# Low-Energy Physics in Neutrino LArTPCs

**Contributed Paper to Snowmass 2021**

D. Caratelli,[17,*] W. Foreman,[44,*] A. Friedland,[99,*] S. Gardiner,[33,*] I. Gil-Botella,[21,*] G. Karagiorgi,[26,*] M. Kirby,[33,*] G. Lehmann Miotto,[20,*] B. R. Littlejohn,[44,*] M. Mooney,[27,*] J. Reichenbacher,[100,*] A. Sousa,[23,*] K. Scholberg,[29,*] J. Yu,[108,*] T. Yang,[33,*]

S. Andringa,[69,†] J. Asaadi,[108,†] T. J. C. Bezerra,[104,†] F. Capozzi,[43,†] F. Cavanna,[33,†] E. Church,[87,†] A. Himmel,[33,†] T. Junk,[33,†] J. Klein,[89,†] I. Lepetic,[96,†] S. Li,[33,†] P. Sala,[47,†] H. Schellman,[85,33,†] M. Sorel,[43,†] J. Wang,[100,†] M. H. L. S. Wang,[33,†] W. Wu,[33,†] J. Zennamo,[33,†]

M. A. Acero,[4] M. R. Adames,[90] H. Amar,[43] D. A. Andrade,[44] C. Andreopoulos,[67] A. M. Ankowski,[99] M. A. Arroyave,[31] V. Aushev,[63] M. A. Ayala-Torres,[25] P. Baldi,[16] C. Backhouse,[110] A. B. Balantekin,[118] W. A. Barkhouse,[80] P. Barham Alzás,[20,21] J. L. Barrow,[73,106] J. B. R. Battat,[117] M. C. Q. Bazetto,[111] J. F. Beacom,[84] B. Behera,[27] G. Bellettini,[92] J. Berger,[27] A. T. Bezerra,[112] J. Bian,[16] B. Bilki,[6,53] B. Bles,[82] T. Bolton,[62] L. Bomben,[48,51] M. Bonesini,[48] C. Bonilla-Diaz,[79] F. Boran,[6] A. N. Borkum,[104] N. Bostan,[81] D. Brailsford,[65] A. Branca,[48] G. Brunetti,[48] T. Cai,[121] A. Chappell,[116] N. Charitonidis,[20] P. H. P. Cintra,[111] E. Conley,[29] T. E. Coan,[101] P. Cova,[88] L. M. Cremaldi,[78] J. I. Crespo-Anadón,[21] C. Cuesta,[21] R. Dallavalle,[33] G. S. Davies,[78] S. De,[1] P. Dedin Neto,[111] M. Delgado,[48] N. Delmonte,[88] P. B. Denton,[8] A. De Roeck,[20] R. Dharmapalan,[39] Z. Djurcic,[2] F. Dolek,[6] S. Doran,[55] R. Dorrill,[44] K. E. Duffy,[86] B. Dutta,[107] O. Dvornikov,[39] S. Edayath,[55] J. J. Evans,[72] A. C. Ezeribe,[98] A. Falcone,[48] M. Fani,[68] J. Felix,[37] Y. Feng,[55] L. Fields,[81] P. Filip,[46] G. Fiorillo,[49] D. Franco,[120] D. Garcia-Gamez,[35] A. Giri,[45] O. Gogota,[63] S. Gollapinni,[68] M. Goodman,[2] E. Gramellini,[33] R. Gran,[77] P. Granger,[56] C. Grant,[7] S. E. Greenberg,[14] M. Groh,[27] R. Guenette,[72] D. Guffanti,[48] D. A. Harris,[121] A. Hatzikoutelis,[97] K. M. Heeger,[120] M. Hernandez Morquecho,[44] K. Herner,[33] J. Ho,[38] P C. Holanda,[111] N. Ilic,[109] C. M. Jackson,[87] W. Jang,[108] H.-Th. Janka,[74] J. H. Jo,[120] F. R. Joaquim,[54,69] R. S. Jones,[98] N. Jovančević,[82] Y.-J. Jwa,[26] D. Kalra,[26] D. M. Kaplan,[44] I. Katsioulas,[12] E. Kearns,[7] K. J. Kelly,[20] E. Kemp,[111] W. Ketchum,[33] A. Kish,[33] L. W. Koerner,[40] T. Kosc,[36] K. Kothekar,[10] I. Kreslo,[5] S. Kubota,[38] V. A. Kudryavtsev,[98] P. Kumar,[98] T. Kutter,[70] J. Kvasnicka,[46] I. Lazanu,[11] T. LeCompte,[99] Y. Li,[8] Y. Liu,[22] M. Lokajicek,[46] W. C. Louis,[68] K. B. Luk,[14] X. Luo,[17] P. A. N. Machado,[33] I. M. Machulin,[64] K. Mahn,[75] M. Man,[109] R. C. Mandujano,[16] J. Maneira,[69,32] A. Marchionni,[33] D. Marfatia,[39] F. Marinho,[113] C. Mariani,[119] C. M. Marshall,[95] F. Martínez López,[94] D. A. Martinez Caicedo,[100] A. Mastbaum,[96] M. Matheny,[81] N. McConkey,[72] P. Mehta,[58] O. E. B. Messer,[83] A. Minotti,[48] O. G. Miranda,[25] P. Mishra,[52] I. Mocioiu,[89] A. Mogan,[27] R. Mohanta,[52] T. Mohayai,[33] C. Montanari,[33] L. M. Montano Zetina,[25] A. F. Moor,[18] D. Moretti,[48] C. A. Moura,[115] L. M. Mualem,[13] J. Nachtman,[53] S. Narita,[57] A. Navrer-Agasson,[72] M. Nebot-Guinot,[30] J. Newby,[83] J. Nikolov,[82] J. A. Nowak,[65] J. P. Ochoa-Ricoux,[16] E. O'Connor,[102] Y. Onel,[53] Y. Onishchuk,[63] G. D. Orebi Gann,[14,66] V. Pandey,[34] E. G. Parozzi,[20] S. Parveen,[58] M. Parvu,[11] R. B. Patterson,[13] L. Paulucci,[115] V. Pec,[46] S. J. M. Peeters,[104] F. Pompa,[43] N. Poonthottathil,[55] S. S. Poudel,[87] F. Psihas,[33] A. Rafique,[2] B. J. Ramson,[33] J. S. Real,[36] A. Rikalo,[82] M. Ross-Lonergan,[26] B. Russell,[66] S. Sacerdoti,[3] N. Sahu,[45] D. A. Sanders,[78] D. Santoro,[88] M. V. Santos,[111] C. R. Senise Jr.,[114] P. N. Shanahan,[33] H. R Sharma,[61] R. K. Sharma,[91] W. Shi,[103] S. Shin,[59] J. Singh,[71] J. Singh,[71] L. Singh,[24] P. Singh,[94] V. Singh,[24,9] M. Soderberg,[105] S. Söldner-Rembold,[72] J. Soto-Otón,[21] K. Spurgeon,[105] A. F. Steklain,[90] F. Stocker,[5] T. Stokes,[70] J. Strait,[33] M. Strait,[76] T. Strauss,[33] L. Suter,[33] R. Svoboda,[15] A. M. Szelc,[30] M. Szydagis,[1] E. Tarpara,[8] E. Tatar,[41] F. Terranova,[48] G. Testera,[50] N. Chithirasree,[92] N. Todorović,[82] A. Tonazzo,[3] M. Torti,[48] F. Tortorici,[19] M. Toups,[33] D. Q. Tran,[40] M. Travar,[82] Y.-D. Tsai,[16] Y.-T. Tsai,[99] S. Z. Tu,[60] J. Urheim,[42] H. Utaegbulam,[105] S. Valder,[104] G. A. Valdiviesso,[112] R. Valentim,[114] S. Vergani,[18] B. Viren,[8] A. Vraničar,[82] B. Wang,[101] D. Waters,[110] P. Weatherly,[28] M. Weber,[5] H. Wei,[70] S. Westerdale,[93] L.H. Whitehead,[18] D. Whittington,[105] A. Wilkinson,[110] R. J. Wilson,[27] M. Worcester,[8] K. Wresilo,[18] B. Yaeggy,[23] G. Yang,[14] J. Zalesak,[46] B. Zamorano,[35] and J. Zuklin[46]

[1] *University at Albany, State University of New New York (SUNY), Albany, NY USA*

[2] *Argonne National Laboratory, IL, USA*

[3] *Université de Paris, CNRS, Astroparticule et Cosmologie, F-75013 Paris, France*

[4] *Universidad del Atlántico, Puerto Colombia, Atlántico, Colombia*

[5] *Universtät Bern, Bern, Switzerland*

[6] *Beykent University, Istanbul, Turkey*

[7] *Boston University, Boston, MA, USA*

[8] *Brookhaven National Laboratory, NY, USA*

[9]Banaras Hindu University, Varanasi, INDIA
[10]University of Bristol, Bristol BS8 1TL, United Kingdom
[11]University of Bucharest, Bucharest, Romania
[12]University of Birmingham, Birmingham B15 2TT, United Kingdom
[13]California Institute of Technology, Pasadena, CA, USA
[14]University of California, Berkeley, Berkeley, CA USA
[15]University of California, Davis, Davis, CA USA
[16]University of California, Irvine, CA, USA
[17]University of California, Santa Barbara, Santa Barbara, CA USA
[18]University of Cambridge, Cambridge CB3 0HE, United Kingdom
[19]University and INFN Sezione di Catania, Catania, Italy
[20]CERN, 1 Esplanade des Particules, Geneva, Switzerland
[21]CIEMAT, Centro de Investigaciones Energéticas, Medioambientales y Tecnólogicas, Madrid, Spain
[22]University of Chicago, Chicago, IL, USA
[23]University of Cincinnati, Cincinnati, OH, USA
[24]Central University of South Bihar, Gaya, INDIA
[25]Centro de Investigación y de Estudios Avanzados del Instituto Politécnico Nacional (Cinvestav), Mexico City, Mexico
[26]Columbia University, New York, NY, USA
[27]Colorado State University, Fort Collins, CO, USA
[28]Drexel University, Philadelphia, PA, USA
[29]Duke University, Durham, NC, USA
[30]University of Edinburgh, Edinburgh EH8 9YL, United Kingdom
[31]Universidad EIA, Envigado, Antioquia, Colombia
[32]Faculdade de Ciências da Universidade de Lisboa, Lisboa, Portugal
[33]Fermi National Accelerator Laboratory, Batavia, IL, USA
[34]University of Florida, Gainesville, FL 32611, USA
[35]University of Granada & CAFPE, Granada, Spain
[36]University Grenoble Alpes, CNRS, Grenoble INP, LPSC-IN2P3, 38000 Grenoble, France
[37]Universidad de Guanajuato, Leon GTO, MX
[38]Harvard University, Cambridge, MA, USA
[39]University of Hawaii, Honolulu, HI, USA
[40]University of Houston, Houston, TX, USA
[41]Idaho State University, Pocatello, ID, USA
[42]Indiana University, Bloomington, IN, USA
[43]Instituto de Física Corpuscular (IFIC), CSIC & Universitat de València, Paterna, Spain
[44]Illinois Institute of Technology, Chicago, IL, USA
[45]Indian Institute of Technology Hyderabad, Kandi, Sangareddy, India
[46]Institute of Physics, Czech Academy of Sciences, 182 00 Prague 8, Czech Republic
[47]Istituto Nazionale di Fisica Nucleare, Sezione di Milano, Milano, Italy
[48]Istituto Nazionale di Fisica Nucleare, Sezione di Milano Bicocca, Milano, Italy
[49]Università degli Studi di Napoli "Federico II" & Istituto Nazionale di Fisica Nucleare, Sezione di Napoli, Napoli, Italy
[50]Istituto Nazionale di Fisica Nucleare, Sezione di Genova, Genova, Italy
[51]Università degli Studi dell'Insubria, Dipartimento di Scienza e Alta Tecnologia, Como, Italy
[52]University of Hyderabad, Hyderabad - 500046, India
[53]University of Iowa, Iowa City, IA, USA
[54]Instituto Superior Técnico, Lisbon, Portugal
[55]Iowa State University, Ames, IA, USA
[56]IRFU, CEA Saclay, Gif-sur-Yvette, France
[57]Iwate University, Morioka, Iwate, Japan
[58]Jawaharlal Nehru University, New Delhi 110067, India
[59]Jeonbuk National University, Jeonju, Jeonbuk, Korea
[60]Jackson State University, Jackson, MS, USA
[61]University of Jammu, J&K, India
[62]Kansas State University, Manhattan, KS, USA
[63]Taras Shevchenko National University of Kyiv, 01601 Kyiv, Ukraine
[64]National Research Center Kurchatov Institute, Moscow, Russian Federation
[65]Lancaster University, Lancaster LA1 4YW, United Kingdom
[66]Lawrence Berkeley National Laboratory, Berkeley, CA USA
[67]University of Liverpool, Liverpool L69 7ZE, United Kingdom
[68]Los Alamos National Laboratory, Los Alamos, NM, USA
[69]Laboratório de Instrumentação e Física Experimental de Partículas, Lisboa and Coimbra, Portugal
[70]Louisiana State University, Baton Rouge, LA, USA
[71]University of Lucknow, Lucknow, Uttar Pradesh, India
[72]University of Manchester, Manchester M13 9PL, United Kingdom

[73]Massachusetts Institute of Technology, Cambridge, MA, USA
[74]Max Planck Institute for Astrophysics, Garching, Germany
[75]Michigan State University, East Lansing, MI, USA
[76]University of Minnesota, Minneapolis, MN, USA
[77]University of Minnesota Duluth, Duluth, MN, USA
[78]University of Mississippi, University, MS, USA
[79]Universidad Católica del Norte, Antofagasta, Chile
[80]University of North Dakota, Grand Forks, ND, USA
[81]University of Notre Dame, Notre Dame, IN USA
[82]University of Novi Sad, Faculty of Sciences, Novi Sad, Serbia
[83]Oak Ridge National Laboratory, Oak Ridge, TN, USA
[84]Ohio State University, Columbus, OH, USA
[85]Oregon State University, Corvallis, OR, USA
[86]University of Oxford, Oxford, UK
[87]Pacific Northwest National Laboratory, Richland, WA, USA
[88]University of Parma, Department of Engineering and Architecture, Parma, Italy
[89]Pennsylvania State University, University Park, PA, USA
[90]Universidade Tecnológica Federal do Paraná, Curitiba, PR, Brazil
[91]Punjab Agricultural University, Ludhiana, India
[92]University and INFN Sezione di Pisa, Pisa, Italy
[93]Princeton University, Princeton, NJ, USA
[94]Queen Mary University of London, London E1 4NS, United Kingdom
[95]University of Rochester, Rochester, NY, USA
[96]Rutgers University, Piscataway, NJ, USA
[97]San José State University, San José, CA, USA
[98]University of Sheffield, Department of Physics and Astronomy, Sheffield S3 7RH, United Kingdom
[99]SLAC National Accelerator Laboratory, Stanford University, Menlo Park, CA, USA
[100]South Dakota School of Mines and Technology, Rapid City, SD, USA
[101]Southern Methodist University, Dallas, TX, USA
[102]Stockholm University, Stockholm, Sweden
[103]Stony Brook University, Stony Brook, NY, USA
[104]University of Sussex, Falmer, East Sussex, UK
[105]Syracuse University, Syracuse, NY, USA
[106]Tel-Aviv University, Tel-Aviv, Israel
[107]Department of Physics and Astronomy, Texas A& M University, College Station, TX, USA
[108]University of Texas, Arlington, TX, USA
[109]University of Toronto, Toronto, Canada
[110]University College London, London WC1E 6BT, United Kingdom
[111]Universidade Estadual de Campinas (UNICAMP), Campinas, SP, Brazil
[112]Universidade Federal de Alfenas (Unifal-MG), Poços de Caldas, MG, Brazil
[113]Universidade Federal de São Carlos (UFSCar), Araras, SP, Brazil
[114]Universidade Federal de São Paulo (UNIFESP), Diadema, SP, Brazil
[115]Universidade Federal do ABC (UFABC), Santo André, SP, Brazil
[116]University of Warwick, Coventry, UK
[117]Wellesley College, Physics Department, Wellesley, MA, USA
[118]University of Wisconsin, Madison, WI USA
[119]Center for Neutrino Physics, Virginia Tech, Blacksburg, Virginia, USA
[120]Yale University, New Haven, CT, USA
[121]York University, Toronto, ON, Canada

[*] Editor.
[†] Primary Contributor.

# Contents

# 1 Executive Summary

## 1.1 Key Takeaways

In this white paper, we outline some of the scientific opportunities and challenges related to detection and reconstruction of low-energy (less than 100 MeV) signatures in liquid argon time-projection chamber (LArTPC) detectors. Key takeaways, summarized concisely below, are of broad relevance to practically all Snowmass 2021 Neutrino Frontier Topical Groups. To emphasize this, we also identify the Topical Groups most relevant to each key takeaway.

- LArTPCs have unique sensitivity to a range of physics and astrophysics signatures via detection of event features at and below the few tens of MeV range. Supernova burst and steady-state solar neutrinos are particularly intriguing signals that manifest themselves solely via MeV-scale signatures. [**NF01, NF03, NF04, NF05**]

- Low-energy signatures are an integral part of GeV-scale accelerator neutrino interaction final states, and their reconstruction can enhance the oscillation physics sensitivities of LArTPC experiments. [**NF01, NF02, NF06, NF09**]

- BSM signals from accelerator and natural sources also generate diverse signatures in the low-energy range, and reconstruction of these signatures can increase the breadth of BSM scenarios accessible in LArTPC-based searches. [**NF02, NF03**]

- Neutrino interaction cross sections and other nuclear physics processes in argon relevant to sub-hundred-MeV LArTPC signatures are poorly understood. Improved theory and experimental measurements are needed. Pion decay-at-rest sources and charged particle and neutron test beams are ideal facilities for experimentally improving this understanding. [**NF06, NF09**]

- There are specific calibration needs in the low-energy range, as well as specific needs for control and understanding of radiological and cosmogenic backgrounds. [**NF10**]

- Novel ideas for future LArTPC technology that enhance low-energy capabilities should be explored. These include novel charge enhancement and readout systems, enhanced photon detection, low radioactivity argon, and xenon doping. [**NF10**]

- Low-energy signatures, whether steady-state or part of a supernova burst or larger GeV-scale event topology, have specific triggering, DAQ and reconstruction requirements that must be addressed outside the scope of conventional GeV-scale data collection and analysis pathways. [**NF10**]

## 1.2 Future Needs Summary

Beyond describing the physics opportunities afforded by low-energy LArTPC signatures (Section 3), we identify over the course of Sections 4 through 7 a range of crucial future studies, measurements or R&D items that offer the promise of maximizing the reach and impact of LArTPC low-energy physics. To enable easier reference to these pressing needs, we summarize them below, in order of the section in which they appear:

**Section 4: Modeling Challenges for Low-Energy LArTPC Physics**

- 4.1: Using a decay-at-rest neutrino source, measure neutrino-argon interaction cross-sections and final states below 100 MeV.
- 4.2: Complete and improve modeling of low-energy nuclear physics processes in existing GeV-scale neutrino interaction generator software packages, such as GENIE, FLUKA, and NuWro.
- 4.3: Increase comprehensiveness of low-energy neutrino interaction generators, like MARLEY.
- 4.4: Using MicroBooNE, ICARUS, and SBND datasets, measure properties of final-state neutron activity in GeV-scale neutrino interactions.
- 4.5: Using a neutron test beam, measure inclusive and exclusive inelastic scattering cross-sections and final-state particle content of neutron interactions on argon
- 4.6: Using protoDUNE and LArIAT, measure final-state low-energy activity associated with inelastic charged particle scattering on argon.
- 4.7: With all experimental datasets above, perform experimental benchmarking of particle transport and associated low-energy activity in standard simulation toolkits, such as FLUKA and Geant4.

**Section 5: Detector Parameters**

- 5.1: Optimize wire-based charge readout systems to achieve the best possible physics-limited signal-to-noise characteristics.
- 5.2: Develop low-noise, low-power pixel-based LArTPC charge readout systems.
- 5.3: Study the possibility of proportional amplification of charge signals in single-phase LArTPCs.
- 5.4: Study the introduction of photosensitive dopants into LArTPCs and their impact on LArTPC charge collection.

- 5.5: Maximize the magnitude and uniformity of light collection in large LArTPCs through development of improved large-area light collectors and wavelength-shifting light reflectors.
- 5.6: Study the introduction of xenon into LArTPCs and its impact on LArTPC light collection.
- 5.7: Develop novel LArTPC calibration techniques using internal, naturally-occuring low-energy physics processes.
- 5.9: Develop calibration systems specifically designed for the purpose of benchmarking LArTPC detector response at low energies.
- 5.9: Understand the impact of intrinsic backgrounds on physics capabilities and develop strategies to reduce them with low-background materials.
- 5.10: Understand the impact of external backgrounds and develop shielding strategies.
- 5.11: Develop detector designs capable of achieving extremely low background levels.

**Section 6: LArTPC Reconstruction at Low Energies**

- 6.1: Develop improved tools for identification of weak signatures above the noise threshold in collected ionization charge waveforms.
- 6.2: Using LArSoft, develop robust, experiment-agnostic reconstructed LArTPC data objects for storing and analyzing MeV-scale charge signatures in a uniform manner comparable to tracks and showers.
- 6.3: Using ICARUS and SBND data and DUNE simulated data, explore and develop meaningful reconstruction capabilities and tools for MeV-scale scintillation light signatures in LArTPCs.
- 6.4: Develop machine learning tools tailored for use on low-energy LArTPC activity.

**Section 7: Data Aquisition and Processing Considerations**

- 7.1: Develop supernova burst and steady-state triggering requirements capable of enabling efficient full-DUNE-module readout of interesting low-energy physics signatures while obeying data transfer limitations.
- 7.2: Develop and test realizable data selection schemes tailored for readout of steady-state low-energy signatures, such as TPC-only or ROI-only data storage.
- 7.3: Develop and test tools for performing MeV-scale physics analyses using only information from trigger primitives.

- 7.4: Develop readout hardware and machine learning tools enabling low-latency analysis of MeV-scale features in DUNE supernova neutrino burst trigger registers.

# 2 Introduction

Liquid argon time-projection chamber (LArTPC) detectors have unique and powerful properties for neutrino physics and beyond-the-standard-model (BSM) searches. A LArTPC provides precise digital readout of charged particle trajectories, enabling a detailed picture of the aftermath of neutrino and BSM particle interactions. Charged-particle interaction products in the detector's liquid argon bulk produce scintillation photons and tracks of ionized electrons with a density proportional to each product's energy deposition density, $dE/dx$. By applying an electric field to the argon bulk, electrons can be drifted across meter-long distances over millisecond timescales through the sea of argon atoms to wire planes strung in front of a light collection system along the sides of the TPC. Resulting times and amplitudes of electron-induced charge signals and photon-induced light collection system signals can be used to reconstruct product trajectories to near-mm-scale in 3D.

An international LArTPC program including several small and large-scale LArTPC detectors is well underway. In Europe in the 1990s and 2000s, large LArTPC technology and its application to particle physics was pioneered by the ICARUS experiment [1]. In the United States, the >ton-scale, late 2000's ArgoNeuT experiment was the first to take LArTPC data in a Fermilab neutrino beamline [2], while more recently, the ∼100-ton-scale MicroBooNE experiment broadly demonstrated the reality of the large LArTPCs' precision physics capabilities [3]. This program will continue through the decade with the start-up of the >100-ton-scale ICARUS and SBND experiments in the BNB beamline at Fermilab [4] and the continued operation of the half-kiloton protoDUNE LArTPC prototype detector at CERN [5]. It will culminate at this decade's end with the start-up of the the Deep Underground Neutrino Experiment, which comprises a LArTPC-including near detector facility [6] and four 10-kton underground LArTPC modules [7].

A major part of the physics program of existing and future LArTPC experiments involves fine-grained reconstruction of GeV-scale neutrino interactions for short- and long-baseline neutrino oscillation experiments and for neutrino cross-section studies. However, these detectors have a broad dynamic range that offers exciting opportunities at its low-energy end. The range below about 100 MeV is inhabited by numerous interesting physics and astrophysics targets. MeV-scale signals are important as substantial components of GeV-scale events, and the ability to reconstruct them will enhance oscillation physics and BSM search capabilites. Perhaps the most prominent MeV-scale LArTPC signal category is the supernova burst: a nearby Galactic core collapse will produce a sharp, brilliant flash of neutrinos of energies up to a few tens of MeV within a few tens of seconds. Other steady-state signals, such as solar neutrinos, are also of great interest, and, like supernova neutrinos and some BSM particle interaction signatures, are primarily visible in neutrino LArTPCs via the MeV-scale signatures they produce.

While MeV-scale signals carry special and valuable information, their detection and reconstruction are also associated with challenges specific to the low-energy regime. Aspects of neutrino interaction and neutron production relevant to the sub-hundred MeV regime in argon are quite poorly understood/demonstrated. Steady-state low-energy signals are especially vulnerable to radiological backround contamination. Low-energy events will have specific readout and calibration needs due to their low levels of ionization charge and scintillation light production. Many commonly-used LArTPC reconstruction tools are intended for use on high-energy signals and are inapplicable to MeV-scale scenarios. Data acquisition and computing implementations must be configured to save as much information as possible from rare supernova burst events, and process it with low latency.

The four-day low-energy Physics in Liquid Argon (LEPLAr) workshop was held November 30 to December 4, 2020, virtually, with the first two days open to the physics community, and the last two days held as a Deep Underground Neutrino Experiment (DUNE) collaboration workshop. The focus was on events with less than about 100 MeV of energy deposition in LArTPC experiments. Physics topics included supernova and solar neutrinos, beyond-the-standard-model (BSM) signals, and ∼GeV-scale physics for which ∼MeV-scale event components matter. The goals of the workshop were:

- To identify physics opportunities in the <100 MeV regime that can be addressed by DUNE (and similar large LArTPCs) and related challenges for the different technical working groups, including ancillary measurements.
- To develop a standard set of signal and background assumptions, and identify knowledge gaps and possible experimental/theoretical remedies.
- To enhance communication and exchange ideas between DUNE technical working groups for addressing low-energy physics related challenges.
- To share experience with other LArTPC experiments.

The first half of the workshop was dedicated to exploring theoretical models for low-energy physics signatures in the DUNE far detector (FD) and the status and availability of simulation tools (from generators to detector propagation) for low-energy interaction modeling. The second half of the workshop was focused on relevant technical developments in DUNE necessary for the pursuit of as wide a range of low-energy physics goals as possible.

In this white paper, we roughly follow the organization of the workshop, starting with discussion of the physics topics that can be explored via detection and reconstruction of low-energy physics signals, and the nature of the signatures, in Section 3. We then describe the state of knowledge and issues at each experimental stage from interaction to energy loss to signal detection, data acquisition, event reconstruction and computing. Section 4 describes challenges in modeling of underlying neutrino-argon interactions, and in simulation of particle transport and energy loss in argon. Section 5 describes detector hardware requirements for efficient detection, as well as background requirements. Section 6 describes challenges for event reconstruction. Section 7 describes challenges for data acquisition and computing.

# 3 Low-Energy Physics and Neutrino LArTPC Physics Goals

Many of DUNE's primary physics goals, such as long-baseline oscillation measurements with GeV beam neutrinos [8, 9], involve detection of high-energy (>100 MeV) charged particle signatures. Particles in this energy range present themselves in LArTPC event displays as extended track-like and shower-like topologies encompassing dozens if not hundreds of contiguous affected charge collection elements. In contrast, physics signatures below 100 MeV will generate signals on a far more limited number of charge and photon collection elements. This means that some aspects of detector design and event reconstruction schemes directed towards DUNE's higher-energy physics goals are insufficient for performance of low-energy physics in DUNE. Given the broad array of potential physics that relies on these low-energy features, it is important to consider future improvements that would aid in solidly securing these low-energy capabilities in DUNE.

Figure 1 roughly illustrates the relevant energy and spatial scales involved in performing a broad set of potential low-energy physics measurements in DUNE, including solar and supernova neutrino detection, neutrinoless double beta decay, and sterile neutrino searches. First, the energy of the primary electron produced by low-energy physics processes relevant to each of these low-energy physics goals are pictured. For example, primary electrons produced by charged-current absorption of supernova-produced $\nu_e$ on argon are expected to primarily have energies between roughly 5 and 30 MeV [10, 11], while radiogenic $^{39}$Ar beta decay electrons, which can be used to calibrate aspects of DUNE detector response [12] and perform searches for keV-scale sterile neutrinos [13], entirely inhabit the sub-MeV regime.

In LArTPC event displays, all of these electrons will produce small, topologically isolated ionization signatures, or 'blips,' comprised of a limited number of contiguous affected charge collection elements [17, 18, 19, 20]. Blips at the lower end of this energy regime are likely to have little or no spatial extent, appearing only as a detectable signal of variable energy deposit in one or two charge detection elements. Meanwhile, blips at the higher end of this regime may span five or more charge collection elements, possibly enabling reconstruction of both an electron's energy and direction. To illustrate this range in spatial extent, Figure 1 also highlights electrons whose straight-line trajectories orthogonal to charge collection elements in DUNE would roughly match the spacing of one or five of those elements. In the following section, we will describe the expected physics processes in DUNE that are accompanied or entirely defined by this blip content and summarize the wealth of physics that can be accessed by precision low-energy reconstruction and analysis in LArTPCs.

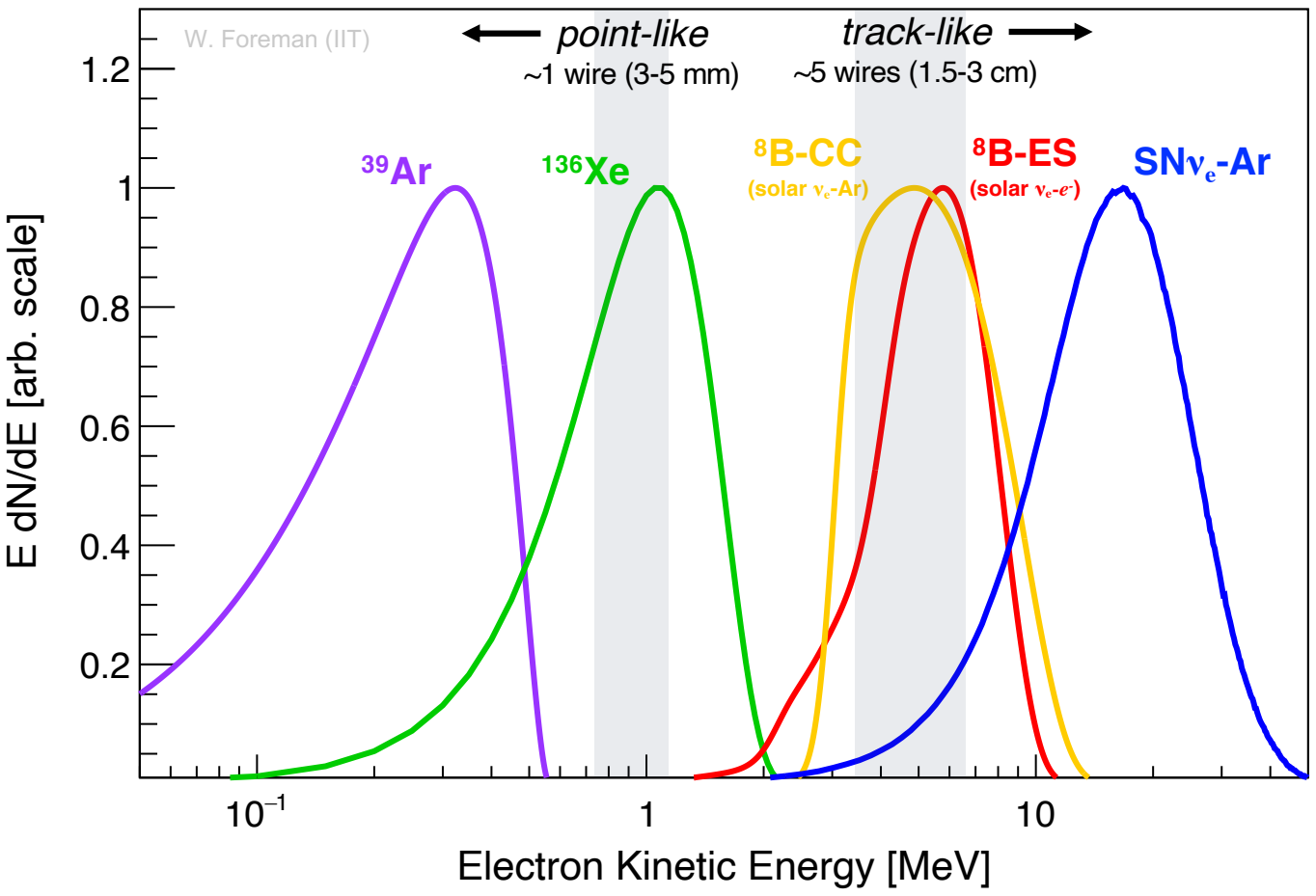

Figure 1: Energy spectrum of primary electrons produced by selected potential low-energy physics processes of interest in DUNE; spectra, from [14, 15, 16, 11]. To emphasize the spatial scales of LArTPC event display features produced by these electrons, energies corresponding to roughly one and five charge collection elements of orthogonal straight-line range in DUNE are also highlighted.

## 3.1 Low-Energy Signatures in High-Energy Neutrino Events

Long-baseline oscillation experiments such as DUNE use artificial neutrino beams to precisely measure the three-flavor neutrino oscillation pattern and to search for new physics effects. The beam neutrinos in these experiments typically have energies between 0.5 and 5 GeV. Even in events of this energy, detection and identification of MeV-scale charge deposits is essential for achieving optimal energy reconstruction.

### 3.1.1 Description of Low-Energy Features

When a GeV neutrino interacts in liquid argon, it produces long charged-particle tracks, sizable electromagnetic showers, as well as an extended halo of MeV-scale blips. The blips ultimately originate from neutral particles created at the primary vertex, from downstream interactions of propagating hadrons, and from the final stages of electromagnetic showers. In all of these cases, the ionization reflected in a DUNE event display blip is generated by electrons from electromagnetic interactions of MeV-scale $\gamma$-rays, which were produced by the various upstream hadrons or leptons. Reliably assessing the role of the blips in a GeV-scale event display thus requires a robust understanding of the entire event, including accurate neutrino-argon and hadron-argon cross section modeling, as will be described in Section 4.

Concrete examples for a few event topologies are given in Figure 2. The left panel shows a simulated 4 GeV $\nu$-Ar interaction created by combining the neutrino event generator GENIE and the propagation package FLUKA [21]. All dots show MeV-scale blips. As is immediately apparent, a blip halo extends more than a meter from the primary vertex and from the charged-hadron tracks.

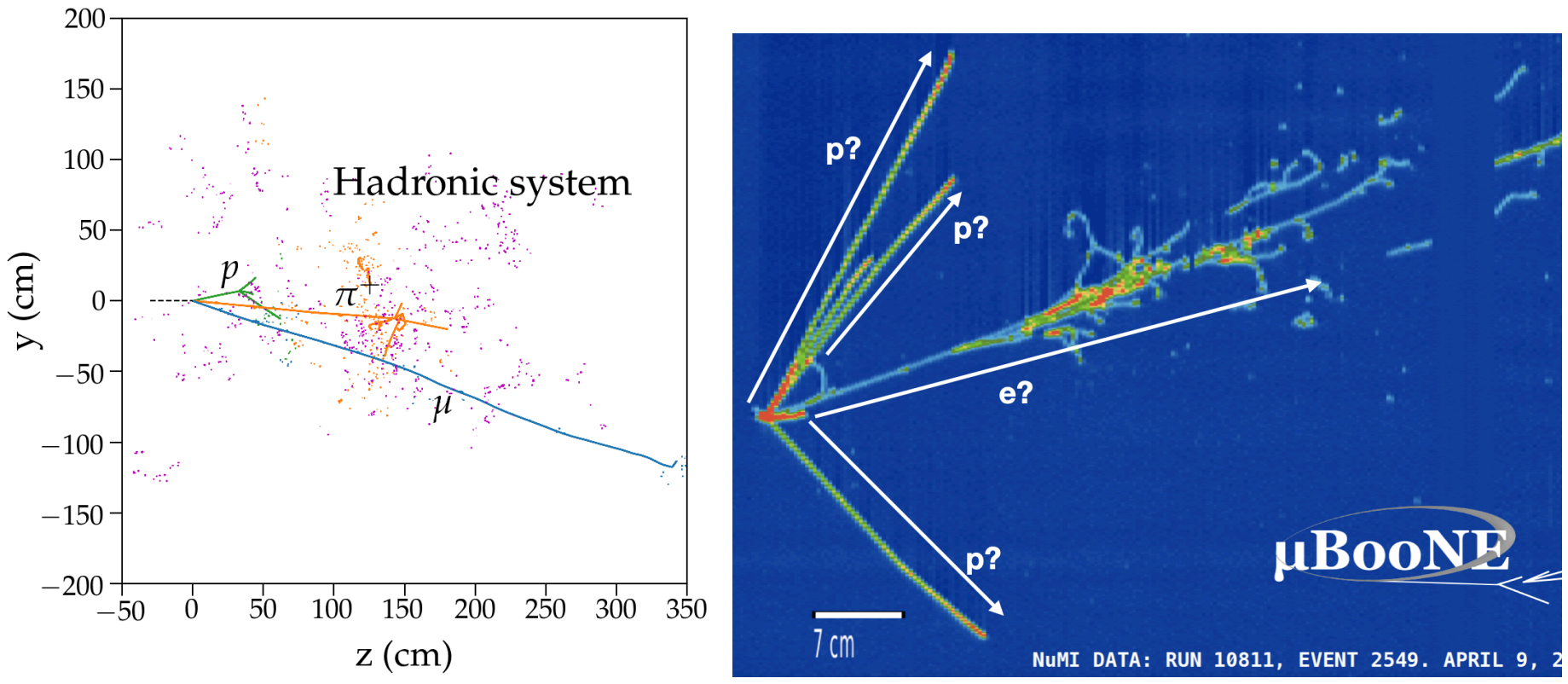

Figure 2: Left: A 4 GeV $\nu_\mu$ event in liquid argon simulated using GENIE and FLUKA. The blue track is created by the final-state muon. The orange color denotes energy originally carried by the prompt charged pion and the blue, by the prompt proton [21]. The magenta energy deposits are caused by neutrons undergoing multiple scatterings. Right: Example of a candidate neutrino interaction in the MicroBooNE detector, displaying electromagnetic activity.

Further examination of truth-level information for the event shows that most of the blips are created by diffusing neutrons. In this particular event, primary neutrons originating from the primary vertex, ranging from 3 to 120 MeV, carry off a total of 170 MeV of kinetic energy. Secondary neutrons created in subsequent hadronic interactions of primary charged hadrons carry further 230 MeV. Much of this energy is then deposited in the form of blips, with those from the primary neutrons and the primary pion, proton, and muon represented in magenta, orange, green, and blue, respectively, in Figure 2. A comparatively smaller fraction (<10 MeV) of the total energy is exhibited as blips produced by de-excitation gammas from the final-state nucleus. Detailed descriptions of blip signatures produced by these hadronic systems in LArTPCs are discussed in detail in Refs. [20, 21, 19, 18]. Blips derived from both final-state neutron production and nuclear de-excitation have been observed in the ArgoNeuT LArTPC [17], and likely in MicroBooNE as well [22].

Apart from any final-state hadronic system, charged- and neutral-current interactions of high-energy accelerator neutrinos can also produce primary or secondary high-energy electrons and gamma rays in a LArTPC event display. Both of these types of particles will generate shower-like event display features as a result of repeated hard scattering and radiative interactions in the liquid argon. A substantial portion of the ionization produced in this shower will be contained within blips. A representative electron-containing event from the MicroBooNE LArTPC is shown for illustration. This event, matching the expected appearance of a charged-current $\nu_e$ interaction with an argon nucleus, contains a high-energy shower. This event illustrates that the identification and inclusion of MeV-scale features into larger topological objects should be an important ingredient in electromagnetic shower reconstruction in LArTPCs. More detailed discussion of electromagnetic shower reconstruction completeness can be found in Refs. [23, 24, 25].

#### 3.1.2 Impact on Oscillation Physics

Oscillation physics measurements depend on reliable estimation of the true energy of neutrinos interacting in the experiment's detectors. As shown in Figure 2, for LArTPCs, reconstructed neutrino energies will depend on the performance of reconstruction algorithms focused on the final-state hadronic and electromagnetic systems in these events. Reconstructed energy estimates for these systems can be impacted by various factors, including smearing from the initial nuclear state and final-state particle mis-identification or non-identification, with aforementioned highly-scattered blip-like energy depositions often not directly associated with the reconstructed vertex. Perhaps more importantly, validations of hadronic energy scale modelling for neutrino-argon interactions, crucial for providing unbiased predictions for oscillated event spectra in DUNE, can only be directly achieved in other blip-reconstructing LArTPC experiments.

Therefore, the ability to efficiently reconstruct the low-energy structure of the hadronic final state of neutrino interactions, both in DUNE and in precursor LArTPC efforts, can have a significant impact on DUNE's physics reach. This is illustrated in Figure 3, which displays results from studies projecting the capability of DUNE in determining the oscillation parameters $\theta_{23}$ and $\Delta m^2_{32}$. In Ref. [15], biases in DUNE's measured values of $\theta_{23}$ and $\Delta m^2_{32}$ are reported for a case where 20% of the hadronic system's proton energy for the full hypothetical neutrino interaction dataset is transferred to un-detected neutrons. In this case, shifts in best-fit parameters are found to be greater than the size of projected 90% confidence level measurement error bands. In this same scenario, measurement of $\delta_{CP}$ would be more mildly biased, producing an effect that may have a modest impact on the long-term CP-violation measurement precision of the experiment. These outcomes underscore the importance of MeV-scale reconstruction in reaching DUNE's centerpiece oscillation physics goals: accurate neutron energy scale estimates and modeling will require reliable blip reconstruction capabilities in DUNE and in other LArTPC experiments.

Reliable electromagnetic energy reconstruction is also an important ingredient in performing DUNE oscillation measurements. high-energy final-state electrons usually carry off the majority of total interacting $\nu_e$ energy, and thus it is essential to precisely reconstruct their energy. Measurement of the small cross-section of the low-energy neutrino-electron elastic scattering process, producing $\sim$ 5000 interactions/year expected in DUNE ND, would provide a strong normalization constraint for the neutrino flux systematic uncertainty, directly impacting oscillation physics measurements, such as those related with the potential discovery of CP violation in the lepton sector. Such an analysis relies on characterization of single very forward low-energy electrons in the final state. Finally, efficient reconstruction of Michel electrons from decay of stopping muons (E$<$ 53 MeV) can assist in muon charge identification in non-magnetized detectors, as a stopping $\mu^+$ will always produce a Michel electron, but a stopping $\mu^-$ will have a $\sim$ 75% chance of being captured by an argon nucleus, for the case of DUNE [26]. Muon charge identification is relevant for mass hierarchy measurements, particularly in the case of atmospheric neutrino interactions, which may provide additional leverage in disentangling new physics from the standard oscillation scenario. It has been shown that for high-energy $\nu_e$ interactions, reconstructed shower energy resolutions can be expected to be close to the few-percent level, and should be fairly insensitive to whether or not MeV-scale blip information is included into reconstructed showers [21]. However, blip inclusion is certain to be a more important consideration for achieving good RMS resolution for low-energy ($<$100 MeV) electrons [19], as they are closer to the critical energy for radiative losses.

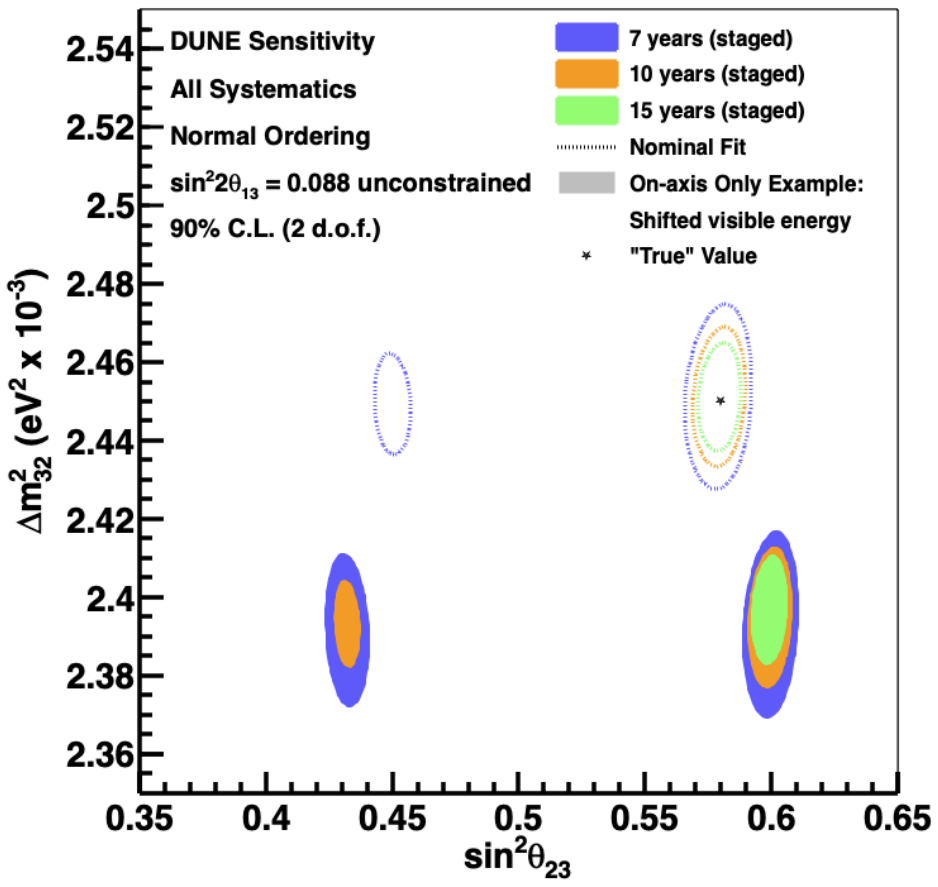

Figure 3: Biases in hypothetical measured DUNE $\theta_{23}$-$\Delta m^2_{32}$ allowed regions caused when 20% of final-state proton energy is shifted to undetected neutrons. Assumed true values of the oscillation parameters are indicated by the stars. From [12].

## 3.2 Low-Energy Signatures in BSM Searches

High intensity proton beams used to generate high flux neutrino facilities, like the LBNF beamline supporting the DUNE experiment, or the BNB beamline supporting the SBN Program, also produce large numbers of photons from Drell-Yan and neutral meson decays. This leads naturally to the suggestion that new chargeless gauge bosons might also be generated in large numbers in these same processes in LBNF and other neutrino beamlines [27]. These gauge bosons may then interact with downstream detectors in this beamline or decay into other dark sector particles which may be detectable. In addition, other dark sector particles, such as axion-like particles (ALPs) can be produced through other coupling processes in the target, such as Primakoff scattering [28, 29, 30].

Observation of these final state dark sector particles can then be performed by detecting the products of their interaction with or decay in a neutrino detector, whether they be electrons, photon pairs, recoiling or de-exciting nuclei, or combinations of heavy charged particles. Long-baseline neutrino experiment near detector facilities or short-baseline neutrino experiments are ideal for detection of beam produced BSM particles, due to the higher flux of incident BSM particles [31]. BSM physics expressed in advantageous neutrino interactions, such as neutrino tridents, are also best accessed in short-baseline detector facilities.

These processes, as well as others, such as annihilation or boosted scattering, may also take place in large numbers in non-terrestrial locations, such as in the Sun [32] or in the galactic halo [33]. The wealth of energetic hidden sector particles generated in these locations can also be observed on Earth in neutrino detectors. Large far detector facilities are more ideal locations for detecting these ambient, lower-rate BSM signatures [34].

A wide selection of BSM searches in neutrino LArTPCs using the hidden portal access methods described above can benefit from the identification of blip activity and subsequent classification of events based on the presence or absence of these features [19]. The remainder of this Section will highlight promising BSM scenarios that can benefit to varying degrees from low-energy blip reconstruction. These scenarios can be roughly grouped into three event topology categories: ones where the BSM signature in question consists purely of high-energy charged particles whose identification can be aided by blip reconstruction, ones in which the signature is entirely represented by low-energy blips, and ones where the signature is partly high-energy and partly low-energy in nature.

### 3.2.1 Blip-Based Particle Identification

Many BSM physics processes discussed as possible points of focus within SBN and DUNE are distinguished by the heavy charged particle combinations they produce [15, 31, 4]. For example, high-energy di-lepton pairs can be expected from Standard Model neutrino trident interactions [35, 36], which have the potential to uncover new physics — such as heavy sterile neutrinos [37], a dark neutrino sector [37], or dark Higgs [38, 39] — if rates diverge from Standard Model predictions. Long-lived dark photons produced following scalar meson decay in the target could similarly manifest themselves via decay to $e^+ + e^-$ or $\mu^+ + \mu^-$ pairs in a detector. Other BSM-specific particle combinations have been discussed: for example, pion-muon pairs can be produced from decays of heavy neutral leptons [40] produced in accelerator neutrino experiments, while pion-pion pairs could be expected from decays of dark Higgs bosons [38] or up-scattered dark neutrinos [41], respectively.

It is expected that the primary backgrounds to muon and charged pion producing BSM searches are different final-state particle combinations produced by common Standard Model neutrino interactions. For example, the primary backgrounds for the two-muon and two-pion channels would likely come from charged pion production in neutirno charged-current interactions. While discrimination of pions and muons is considered to be a challenging task in LArTPCs due to their similar masses and energy loss profiles, it has been shown that they should be distinguishable via the MeV-scale products generated at their points of capture or decay [19]; charge-sign discrimination for pions and muons should also be achievable using these methods. Thus, the pion-muon discrimination capability delivered by analysis of reconstructed blips may be a particularly useful additional tool in reducing backgrounds for these types of BSM analyses.

### 3.2.2 Pure Low-Energy Signatures

Some BSM particle interactions in LArTPCs will manifest themselves purely in the form of MeV-scale blip activity. An obvious example is that of fractionally-charged or millicharged particles (mCP) produced in high intensity proton beams, again via neutral meson decay. These particles scatter with electrons multiple times along their path in a LArTPC, leaving behind blip signatures that can then be traced back to the neutrino beam's target [42], as recently demonstrated by the ArgoNeuT detector on the NuMI beamline [43]; to demonstrate, a candidate mCP event display from this ArgoNeuT search is shown in Figure 4. The ability to identify weak $e^-$ recoils with sub-MeV threshold enables this method to be highly sensitive and virtually background-free. Without MeV-scale thresholds, LArTPC-based mCP searches would struggle to compete with the sensitivities of other experimental approaches.

Other BSM particle interactions in LArTPCs may produce pure but less-distinctive low-energy signatures. For example, interaction of an astrophysical or beam-produced light dark matter (LDM) with an electron in a LArTPC will produce a single energetic recoil electron, while LDM elastic or inelastic interactions with argon nuclei can produce a low-energy nuclear recoil or gamma, neutron, and proton de-excitation products. ALPs can also produce energetic one- or two-photon signals, depending on the precise interaction mechanism. While these products are allowed be higher in energy, the ability of the detectors to identify $e^-$ recoils and de-excitation products with as low an energy threshold as possible greatly enhances LDM and ALP sensitivities. This is illustrated in Figure 5, which shows recoil electron energies from LDM and $\nu_\mu$ elastic scattering [44]. In addition to enhancing the strength of the LDM signal, lower energy sensitivity, as well as varied off-axis 'PRISM-style' run configurations [45, 46], also enables better separation of BSM blip-inducing phenomena from neutrino-generated blip backgrounds; example sensitivity for such a search at the DUNE near detector is also provided in Figure 5.

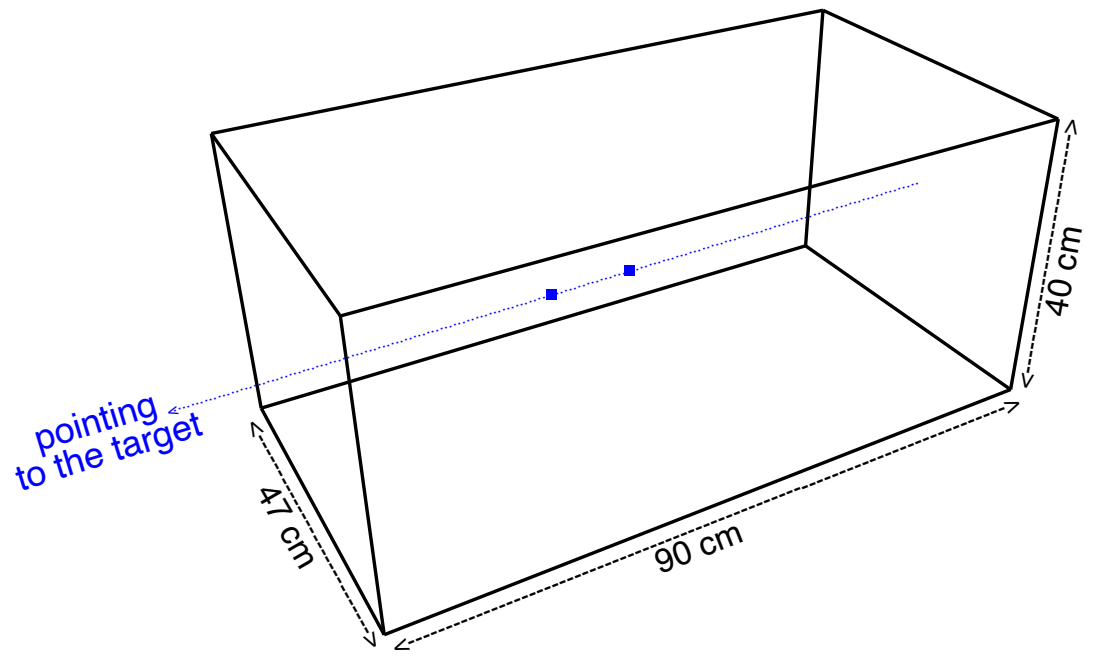

Figure 4: A 3D candidate signal event display from ArgoNeuT's search for millicharged particles produced in the NuMI target [43]. Blue dots represent reconstructed blips that can be connected with a ray pointing upstream back to the NuMI target.

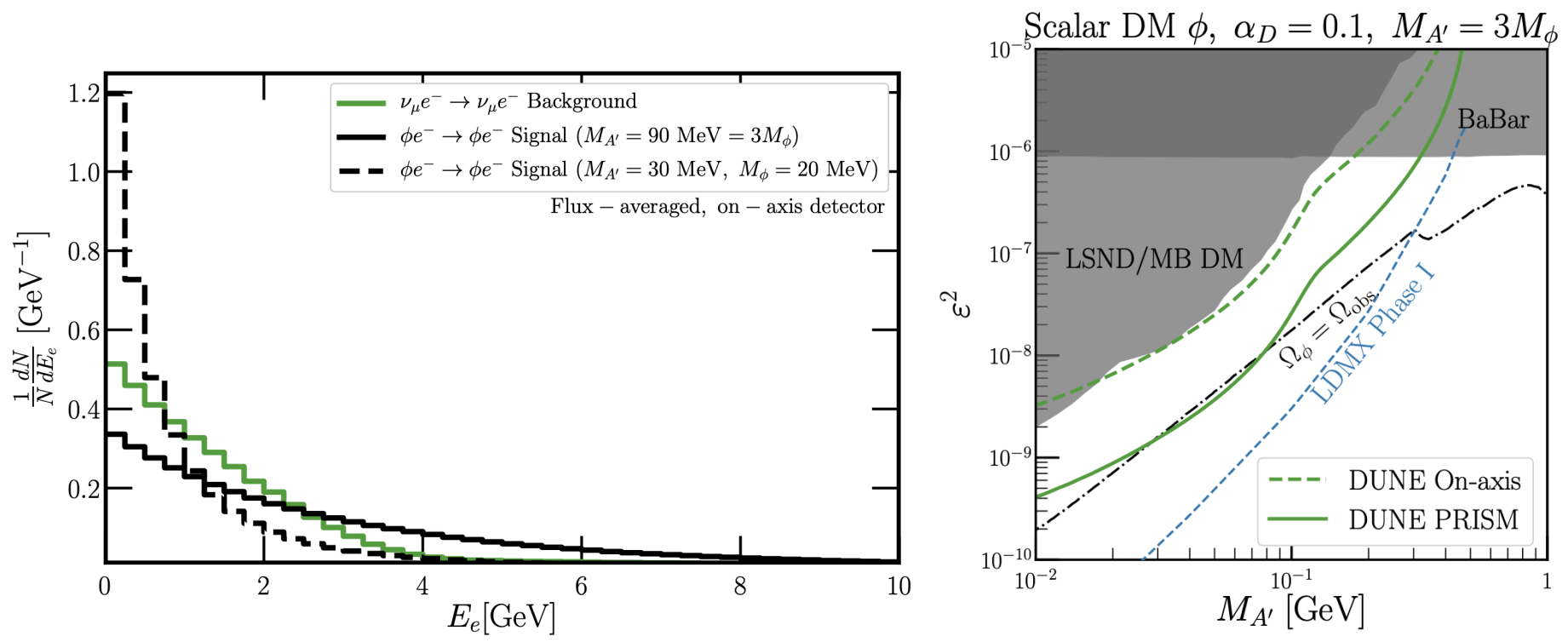

Figure 5: Left: Recoil electron energy distribution from light dark matter (LDM) elastic scattering (dashed, solid black) versus that from $\nu_\mu$ elastic scattering. Right: DUNE LDM sensitivity for a near-detector on-axis (dashed green) versus on-axis plus off-axis (solid green) configuration. Both figures from [44].

### 3.2.3 Mixed-Energy Signatures

Some BSM particle interaction signatures may be separable from backgrounds by the unique character of the blip signals they produce in concert with higher-energy charged particle activity. For example, short-lived hidden sector particles produced inside a LArTPC, such as up-scattered heavy neutral leptons [47, 4], can produce two displaced event vertices, one of which consists of a de-exciting nucleus generating primarily or exclusively blip activity. In this and other cases, identification of these secondary low activity vertices is the only likely method for distinguishing the signal BSM process's most-energetic vertex from a variety of similar neutrino-induced backgrounds. Hidden sector interactions may also be distinguished from neutrino backgrounds by their total lack of associated blip activity. For example, decays of hidden sector particles in LArTPCs, such as heavy neutral leptons, dark photons, or dark Higgs, are not dependent on substantial momentum exchanges with an argon nucleus. These decay vertices, unlike neutrino-argon interactions, will include no neutron and photon products of final-state nuclear de-excitation, resulting in an event with no blip activity near the interaction vertex.

## 3.3 Low-energy Neutrino LArTPC Physics

Many neutrino-related signals in large LArTPCs will manifest themselves primarily or entirely via low-energy and blip-like signatures. To perform physics measurements using signals in this category, such as supernova and solar neutrino interactions and beta decay products, analyses will rely heavily on the identification and reconstruction of MeV-scale LArTPC event display features. In this sub-section, we will briefly describe the appearance of these signals in LArTPCs, and summarize the broad-ranging physics goals that can be pursued with them.

### 3.3.1 Supernova Neutrinos

A nearby core-collapse supernova is a once-per-career event, and will be a prodigious producer of neutrinos in the tens of MeV range. A rich array of high impact nuclear and particle physics can be performed by measuring the energy spectrum, flavor content, and time profile of supernova burst neutrinos. Liquid argon TPCs have unique capabilities for capturing physics and astrophysics from the expected few-tens-of-second burst from a core collapse [48, 11].

The core collapse is expected to produce neutrinos and antineutrinos of all flavors [49]. Given that the supernova burst neutrinos are overwhelmingly below charged-current threshold, muon and tau flavors (plus antineutrinos) are usually lumped together, so that there three effective flavors— $\nu_e$, $\bar{\nu}_e$ and $\nu_x$. The neutrinos are emitted over roughly three physical phases. During collapse, electron neutrinos are emitted at an increasing rate, culminating in a bright "neutronization burst" in the first tens of ms following core bounce. This burst of purely electron-flavor neutrinos is potentially processed by flavor transformation. This is followed by a $\sim$second-long "accretion" phase, over which dynamics of the explosion may manifest themselves in the neutrino flavor and spectrum as a function of time. Subsequently, all flavors cool over a few tens of seconds.

Given the flavor-pure and relatively model-insensitive nature of the neutronization burst signal, it offers a particularly pure view of neutrino oscillations, particularly the promise of cleanly measuring the neutrino mass hierarchy [50, 51]. The shape and absolute scale of the measured spectrum over all times provides insight into the progenitor, explosion and proto-neutron star evolution. The time profile during accretion can provide details about the dynamics of this phase (for example, standing accretion shock instability (SASI) oscillations) [52, 53], while the duration or cessation of the signal at long times (>5 s) can provide insight into the ultimate fate of the remnant [54]. At certain times, the neutrino number density may be high enough that nonlinear effects (e.g., "collective oscillations") resulting in exotic flavor transitions, may imprint themselves on the observable signal.

In a LArTPC, supernova neutrinos are primarily detected via $\nu_e$ absorption interactions on argon nuclei [11], a complementary signal to those primarily accessed in other operating large neutrino detectors, such as Super-Kamiokande, IceCube, and scintillation detectors [49]. These relatively flavor-pure interactions will produce signals similar to the representative simulated event display in the top panel of Fig 6: main features include a short track caused by the final state low-energy electron, as well as accompanying blips from nuclear deexcitation, inelastic collisions of final-state neutrons, and bremsstralung radiation produced by the final-state electron. A $\nu_e$ energy resolution of roughly 20% or better is expected over most of the relevant $\nu_e$ energy range [11]. As demonstrated in Ref. [19], the resolution of calorimetrically reconstructed supernova neutrino energies is optimized when the full event topology – both final-state electron and all associated blips – is incorporated.

Sub-dominant interaction channels, such as $\overset{(-)}{\nu}_x$-e elastic scattering, neutral current $\overset{(-)}{\nu}_x$-argon scattering, and charged current $\overline{\nu}_e$-argon scattering should also be detectable in modest quantities for a supernova explosion located in the Milky Way. Final-state particle content differs greatly between many of these channels and the dominant $\nu_e$ charged-current absorption channel. This is clearly illustrated in Figure 6 by comparing the aforementioned top event display to the bottom event display image containing the single electron product of a $\nu_e$-e scattering interaction. As discussed and quantified in Ref. [19], the two interaction channels are clearly distinguished by the presence or absence of final-state blips. Neutral-current scatters from argon nuclei should be distinguishable from charged-current absorption by the lack of a dominant leading final-state electron track.

The neutral-current CEvNS process is the dominant neutrino interaction cross section in the few-tens-of-MeV regime. For a supernova burst, one expects about 30-100 more CEvNS than CC$\nu_e$ events, depending on the neutrino spectrum. However, the deposited energy per interaction is very tiny, up to tens of keV in the same neutrino energy range. Photons created by recoils also tend to be quenched. Individual CEvNS interactions are observable in dark-matter-style argon detectors, either single or dual phase, and nuclear-vs-electronic recoil selections can be used [55]. However, in large LArTPCs, it is a challenging ex-perimental task to record information when the signal amounts to just a handful of CEvNS-induced photons per interaction.

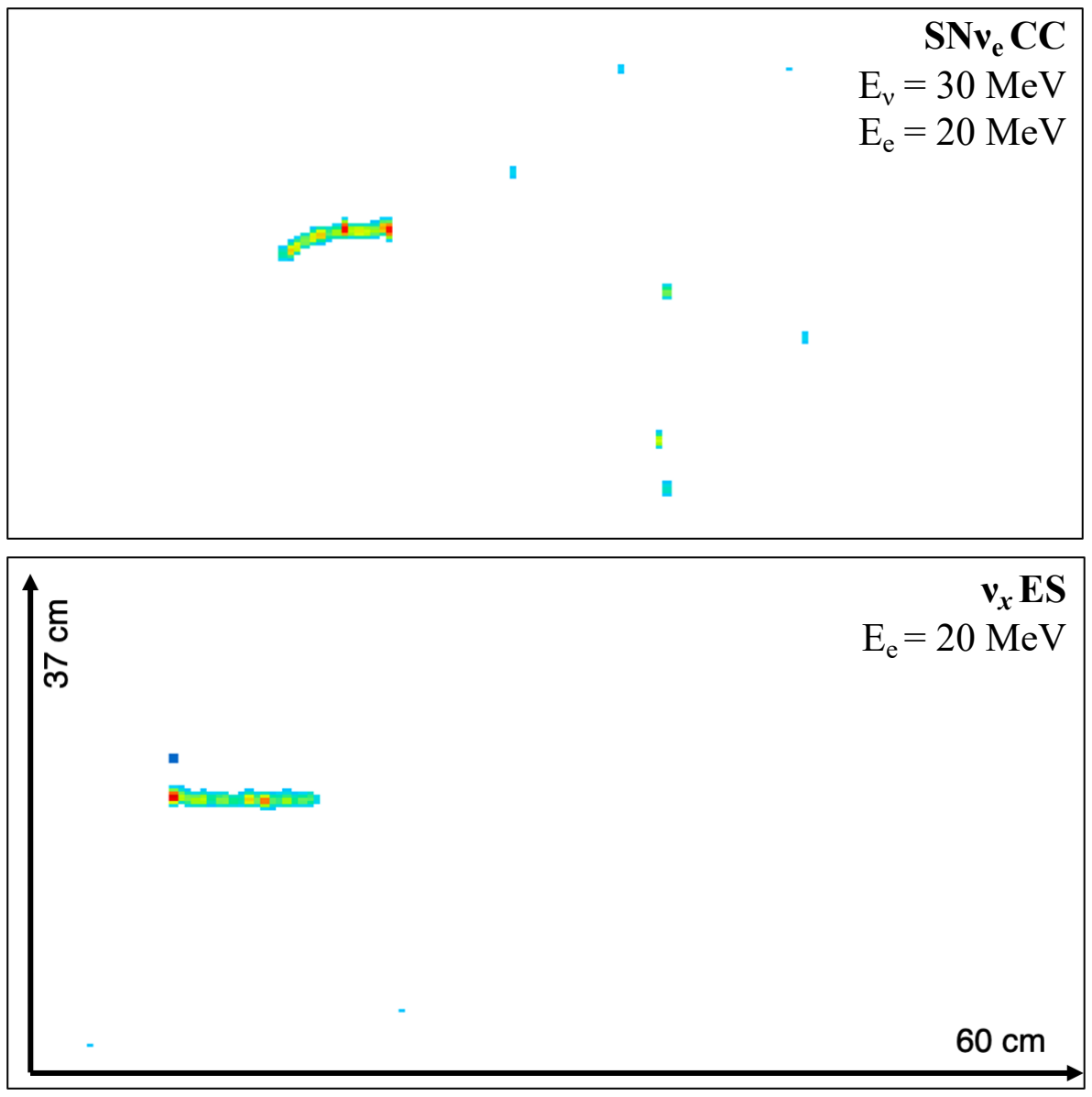

Figure 6: Simulated LArTPC event displays from a MARLEY-generated 30 MeV $\nu_e$ charged current absorption interaction (top) and a single 20 MeV scattered electron from $\nu_e$-e scattering (bottom). From [19].

Nevertheless, given the very large number of interactions from a supernova burst, a statistical "glow" of CEvNS photons could be observable [56]. Neutral-current processes, like CEvNS, are especially valuable due to their sensitivity to all flavors of a supernova burst.

### 3.3.2 Solar Neutrinos

Solar neutrinos are lower in energy than those from supernovae, meaning that even the leading final-state electron track may be blip-like in appearance. Thus, in this case, identification of solar neutrino interaction candidates requires robust, reliable blip reconstruction capabilities. As the associated gamma-related blip activity makes up a larger fraction of the total event energy for solar neutrinos, their incorporation into calorimetric reconstruction is even more crucial than in the supernova neutrino case.

If solar neutrinos can be identified and their energies can be reliably reconstructed, massive underground LArTPCs, like DUNE, may be capable of performing world-leading measurements of fundamental neutrino properties [57, 16, 58, 15]. They can measure the solar mass splitting $\Delta m^2_{12}$ and mixing angle $\theta_{12}$ with higher precision than the previous best measurements from SNO and Super-K [16], enabling more stringent tests of the low significance tension between terrestrial and solar measurements of these parameters. By using its energy spectrum measurement capabilities to powerfully probe the shape of the transition region between MSW and vacuum oscillations, DUNE can also place uniquely strong limits on a variety of processes, including non-standard neutrino-matter interactions [59, 60] and neutrino decay [61, 62]. DUNE may also be able to provide first measurements of the hep neutrino flux thanks to its particular sensitivity to $\nu_e$ interactions [16]. DUNE can make a statement about solar metallicity if a low-background module is built (see Section 5.5.2).

#### 3.3.3 Other low-energy Neutrino Physics

- **Neutrinoless Double Beta Decay in Large LArTPCs**

  Searches for neutrinoless double-beta decay have recently moved into an era of tonne-scale measurements, providing access to effective Majorana masses on the scale of 10 meV. This reach would give access to the parameter space allowed by the inverted mass ordering of the neutrino mass states. To extend this reach into the normal mass ordering regime, the scale of the experiments will need to grow by at least two orders of magnitude. If a large underground LArTPC, such as one DUNE FD module, were to be doped to 2% by mole fraction with 90% enriched $^{136}$Xe, it would be able to leverage more than 300 tonnes of candidate isotope. Beyond doping, utilizing $^{42}$Ar-depleted argon, enhanced external shielding, and enhanced light collection efficiency to greater than 50% (see Section 5.2.2) would further strengthen the power of this detector for a neutrinoless double-beta decay search application. If these modifications are combined into a $^{136}$Xe-doped DUNE FD module, searches for neutrinoless double-beta decay could reach effective Majorana masses in the range of 2-4 meV, well within the phase-space allowed only by the normal mass ordering [63].

- **Search for Heavy Sterile Neutrino Masses Using $^{39}$Ar Beta Decays**

  Another MeV-scale physics measurement that can potentially be carried out at DUNE and other large on-surface or underground LArTPCs is a heavy sterile neutrino search performed by detecting "kinks" in beta decay spectra, providing a handle on $|U_{e4}|^2$. These features are normally present in beta decay spectra due to mixing between the three standard model neutrinos [64], but primarily lead to spectral distortions very close to the end point. A heavy sterile neutrino search can be carried out in large LArTPCs by specifically using $^{39}$Ar beta decays. For example, the large size of the DUNE FD and use of atmospheric argon ($^{39}$Ar beta decay rate of 1 Bq/kg [65]) will lead to $\mathcal{O}(10^{16})$ decays within the detector over the lifetime of the experiment, providing abundant statistics for this measurement. Reconstruction of $^{39}$Ar beta decays has been previously carried out at MicroBooNE, demonstrating that low thresholds (roughly 100 keV) and good energy resolution from low TPC noise levels (roughly 50 keV) are achievable in large LArTPC detectors [66]. The low thresholds and good energy resolution associated with ionization signals, combined with the $^{39}$Ar $Q$ value of 565 keV, allows for a search for sterile neutrinos in the 20 keV to 450 keV mass range.

# 4 Modeling Challenges for Low-Energy LArTPC Physics

Challenges in the simulation and modeling of low-energy interactions in LAr have broad ranging impacts on many aspects of LArTPC experimental physics goals. These modeling limitations fall into two general categories: those related to the probability and products of low-energy and high-energy neutrino interactions in LAr, and those related to the production and transport of neutrons in LAr. The purpose of this section is to describe the current status of modeling of these aspects, existing datasets used to constrain this modeling, and what direct LAr-based experimental measurements could be made to reduce uncertainty or improve robustness of this modeling.

## 4.1 Neutrino-Argon Cross section Physics

The theoretical description of neutrino-nucleus interactions, including those with the argon nucleus in particular, has received increased attention in recent years due to the high precision needed for the success of future oscillation analyses [67]. Aspects of neutrino-nucleus scattering relevant for low-energy physics in liquid-argon-based detectors include both cross sections for neutrinos below ~100 MeV as well as MeV-scale nuclear activity induced by higher-energy neutrinos. These two topics are considered separately in the following subsections.

### 4.1.1 Low-energy Neutrino Interactions and Generators

At energies of tens of MeV and below, only four interaction modes are available to (anti)neutrinos striking an argon atom:

- Neutrino-electron elastic scattering, which has a small cross section but is readily calculable to high precision in the Standard Model;
- Inelastic neutral-current scattering of all flavors of (anti)neutrinos on the argon nucleus;
- Charged-current scattering of $\nu_e$ and $\bar{\nu}_e$ on the argon nucleus; and
- Coherent elastic neutrino-nucleus scattering (CEvNS), a neutral-current process in which the recoiling argon nucleus is left in its ground state.

Neutrino-electron elastic scattering is already modeled precisely enough to meet the needs of foreseeable low-energy experimental analyses. The remainder of this subsection will therefore focus on the inelastic NC and CC modes, which are expected to provide the dominant signal for measurements of tens-of-MeV neutrinos in large underground LArTPCs like DUNE. The CEvNS cross section is the largest of the four considered here. While it is well understood in the standard model, the lack of experimental signatures apart from nuclear recoil presents special challenges. Prospects for using CEvNS signals in liquid-argon-based physics measurements were described in Section 3.

Calculations of inelastic neutrino-nucleus cross sections at energies below 100 MeV typically describe the scattering process in two distinct steps. In the first step, the neutrino participates in a $2 \to 2$ interaction with the target nucleus, leaving it in a final state with a well-defined excitation energy. In the second step, the outgoing nucleus emits zero or more particles in a series of de-excitations until the ground state is reached. The physical justification for splitting the calculation into these two steps is the assumption of *compound nucleus* formation: instead of being directly knocked out, the nucleon originally struck by the neutrino scatters repeatedly within the nuclear medium. This leads to a state of thermal equilibrium in the nuclear system as the energy transferred by the neutrino is shared widely among multiple nucleons. Subsequent de-excitations are thus largely insensitive to the details of the primary neutrino interaction which formed the equilibrated nuclear state.

A substantial literature on calculations of *inclusive* inelastic neutrino-nucleus cross-sections (i.e., those which consider only the first step mentioned above) in the low-energy regime has accumulated over the past several decades. Nearly all calculations for complex nuclei follow variations of a general approach presented by Walecka [68]. This approach neglects the momentum transfer relative to the mediator boson ($W$ or $Z$) mass, leading to a single tree-level diagram in which the scattering amplitude involves the product of a leptonic and a nuclear weak current. The leptonic current is the usual one obtained by representing the neutrino and outgoing lepton in terms of Dirac spinors. For charged-current scattering, corrections for electromagnetic final-state interactions (FSIs) are typically handled using some simple approximations [69]. The nuclear current is written as a sum over single-nucleon terms. This *impulse approximation* neglects any multinucleon contributions although these are understood to be important at higher neutrino energies [70]. The single-nucleon terms are evaluated in coordinate space by applying the free-nucleon weak current operator to the appropriate bound-nucleon wavefunctions. Two series expansions are usually applied and truncated to obtain approximate results. The first is an expansion in inverse powers of the nucleon mass $m_N$, typically kept to $\mathcal{O}(1/m_N)$. The second is a multipole expansion of the nucleon current operator in coordinate space, with higher-order multipoles playing an increasingly important role as the neutrino energy increases.

Brief reviews through 2018 of the literature on inclusive predictions for low-energy neutrino-argon cross sections are given in the supplemental materials of Ref. [16] and in Section 7.1 of Ref. [71]. Since those reviews were prepared, new calculations have been reported in Refs. [72, 73, 10]. The primary difference between the various available models is the strategy used to evaluate the nuclear matrix elements. This is most typically handled using variants of the nuclear shell model [74], the Random Phase Approximation (RPA) [75], or sometimes a combination of the two [76]. A partially data-driven approach, supplemented with an existing quasiparticle RPA calculation [77] at high excitation energies, has also been reported in Ref. [10]. Models can also differ in their handling of forbidden transitions (i.e., the choice of cutoff for the multipole expansion mentioned above) and in whether they adopt an effective (or *quenched*) value of the nucleon axial-vector coupling constant $g_A$.

Modeling of the de-excitation step of low-energy neutrino scattering on argon has only been attempted so far in a single recent publication [10]. As discussed briefly therein, the theoretical techniques adopted for the argon case are similar to the ones used previously for various other target nuclei. For nuclear states that may de-excite by emitting a nucleon or light ion, the Hauser-Feshbach statistical model [78] is employed to describe the competition between these channels and $\gamma$-ray emission. Tables of measured nuclear energy levels and $\gamma$-ray branching ratios (supplemented with theoretical estimates where necessary) are used to handle transitions between bound nuclear states. While these model ingredients are based on highly successful treatments of nucleon- and photon-nucleus reactions at low energies (see, e.g., Ref. [79]), the validity of their application to tens-of-MeV lepton-nucleus scattering remains largely unstudied.

A dedicated neutrino event generator called MARLEY implements the physics model described in Ref. [10] and provides what is currently the only realistic simulation of inelastic neutrino-argon scattering at tens-of-MeV energies. Usage instructions and details about MARLEY's numerical implementation are provided in Ref. [80]. As of the current release (version 1.2.0), MARLEY lacks a calculation of the nuclear matrix elements necessary to simulate the CC $\bar{\nu}_e$ and inelastic NC channels for argon. Including these channels in a future release would be relatively straightforward apart from obtaining the nuclear matrix elements themselves.

#### 4.1.2 Low-Energy Cross-Section Measurements with Pion Decay-At-Rest Sources

Dedicated *measurements* sensitive to nuclear de-excitation products from neutrino-induced interactions in argon and other materials will be required to make further progress. An ideal neutrino source for such measurements is a pion decay-at-rest source [81], for which neutrinos of $\nu_e$, $\bar{\nu}_\mu$, $\nu_\mu$ flavor have energies in the range up to $\sim$52 MeV with well-understood spectrum [82]. The Spallation Neutron Source at Oak Ridge National Laboratory provides a high-power, pulsed instance of such a source; COHERENT has exploited this source for CEvNS measurements [83, 84] and forthcoming measurements are expected for neutrino-induced neutron production on lead and iron targets [85]. Future inelastic NC and $\nu_e$CC measurements on argon will place direct constraints on cross sections and on the distribution of interaction products for processes of interest. Measurements on other materials will furthermore shed light on low-energy neutrino cross section models in general. Specific possibilities for measurements, including single-phase argon scintillation detectors and LArTPCs, as well as other detectors, are described in a Snowmass white papers in progress [86].

#### 4.1.3 High-Energy Interaction Considerations

At the higher energies typical of accelerator neutrinos (multiple hundreds of MeV to ∼10 GeV), neutrino-nucleus interactions generate MeV-scale nuclear activity primarily via the emission of neutrons and de-excitation $\gamma$-rays. Modeling of the neutrino-induced production of these particles is considered in this section, and a discussion of their transport in liquid argon is given in Section 4.2.

Understanding the neutron content (in both multiplicity and kinematics) of the hadronic final state in neutrino-nucleus scattering will be a critical ingredient to achieve the necessary precision for the future accelerator-based oscillation program. Both calorimetric and kinematic techniques for neutrino energy reconstruction [87] must be corrected for missing energy imparted to neutrons, and they are subject to bias in the case of mismodeling. A particular concern for $\delta_{CP}$ measurements, which rely on comparisons of neutrino and antineutrino event rates, is the degree to which neutron production differs between these two channels [67].

Theoretical predictions for neutrino-induced neutron emission in this energy regime are sensitive to multiple nuclear effects which greatly complicate the necessary calculations. Among these are multinucleon knockout contributions to the cross section arising from meson exchange currents, short-range nucleon-nucleon correlations, and hadronic FSIs. Differing models of these effects in neutrino event generators are poorly constrained by currently available data and lead to notable discrepancies, particularly at low neutron energy. A specific example for an argon target can be seen in Figure 7, which displays the neutron multiplicity and kinetic energy distributions predicted by version 3 of the GENIE neutrino event generator [88, 89] for inclusive charged-current $\nu_\mu$ events at 2 GeV. Results obtained using four different models of hadronic final-state interactions are shown, with all other aspects of the generator configuration held constant. The FSI models considered include a single-step (hA2018) and multistep (hN2018) intranuclear cascade developed within GENIE itself [90], as well as the Liège (INCL++) [91] and Bertini treatments, the latter of which was implemented within Geant4 [92].

A second example with significant implications for neutrino energy reconstruction is shown in Figure 8. As a function of true neutrino energy $E_\nu$, both panels show the mean fraction $F_n$ of the leptonic energy transfer $q^0 = E_\nu - E_\ell$ (where $E_\ell$ is the energy of the outgoing lepton) which appears as the kinetic energy of neutrons in the final state. The left panel shows the distributions predicted by NEUT [93], NuWro [94], and several configurations of GENIE v3 for CC inclusive scattering on $^{40}$Ar. The right panel imposes the additional restriction that the final state must contain at least one neutron. The much higher $F_n$ values seen at low $E_\nu$ in the right panel for GENIE versus NEUT/NuWro can be attributed to a key physics difference: the GENIE FSI models used here in-clude a rough approximation for nucleon evaporation (which enhances emission of low-energy neutrons) while the other generators do not.

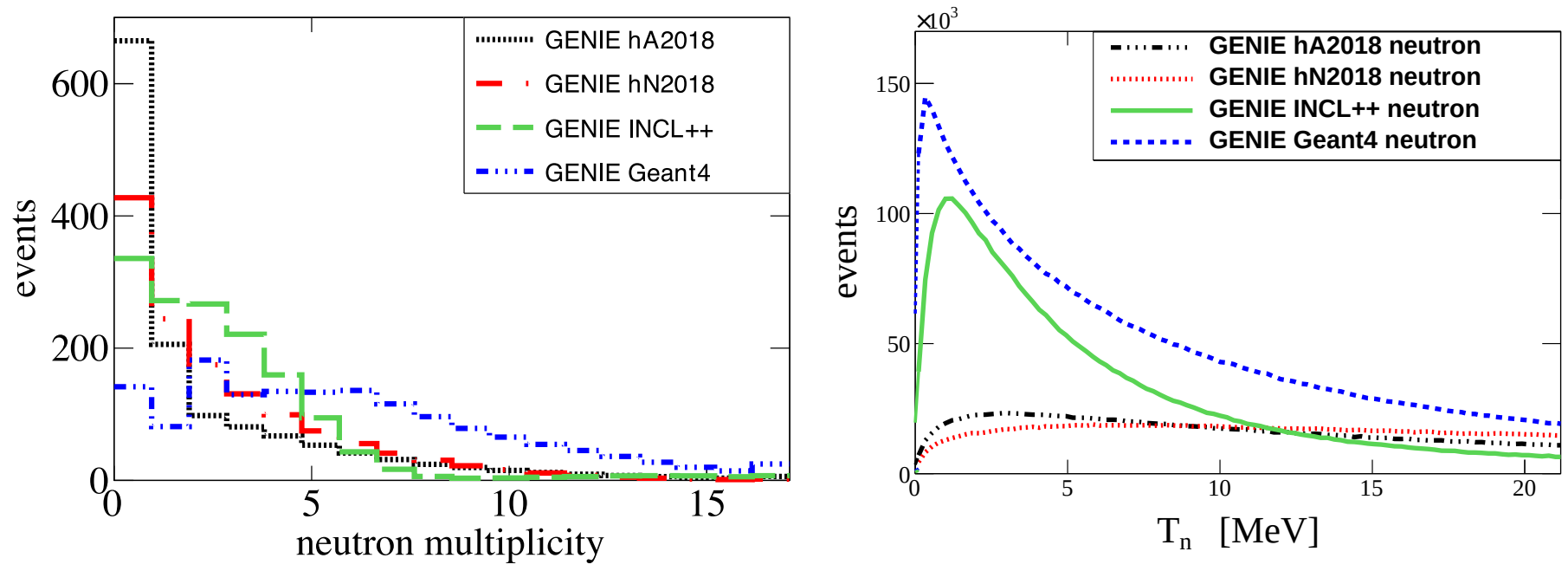

Figure 7: Neutron multiplicities (left) and kinetic energies (right, low-energy region shown) predicted by GENIE v3 simulations of inclusive charged-current $\nu_\mu$ scattering on $^{40}$Ar at 2 GeV. All generator configuration details are the same between the different histograms except for the choice of model for hadronic final-state interactions. Figures from Ref. [88].

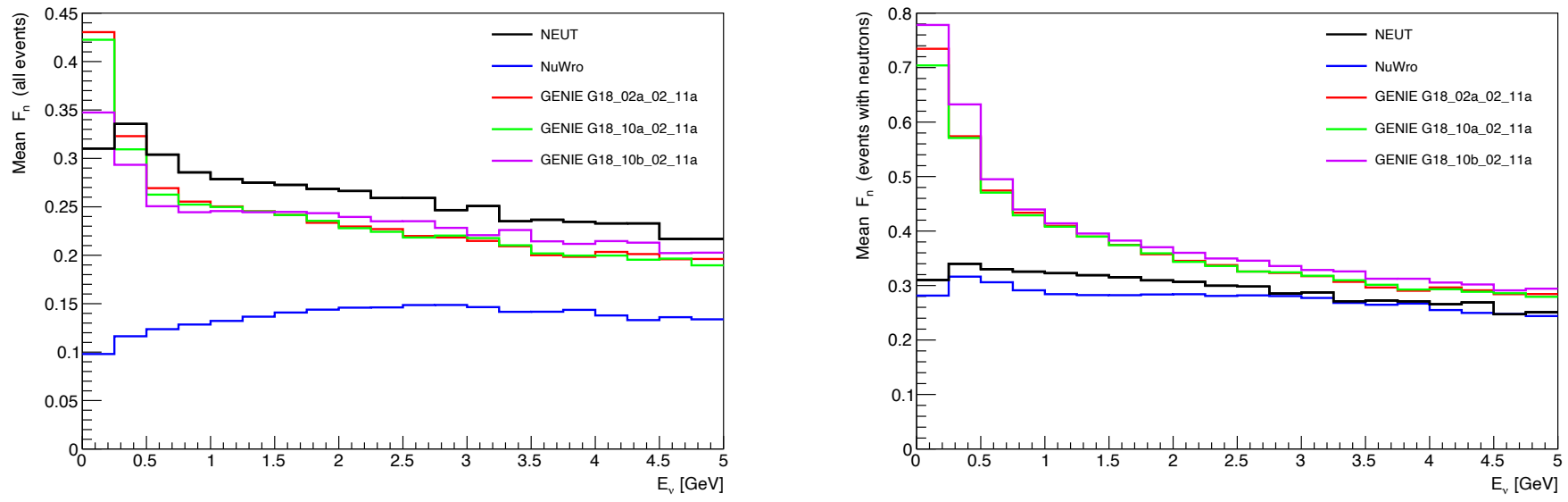

Figure 8: Mean fraction $F_n$ of the leptonic energy transfer converted into the kinetic energy of final-state neutrons. The left-hand plot shows this quantity as a function of neutrino energy for inclusive charged-current $\nu_\mu$ scattering on $^{40}$Ar. The right-hand plot shows the corresponding distributions when events with zero final-state neutrons are excluded. Predictions from the NEUT (black) and NuWro (blue) generators are shown together with three different configurations of GENIE v3 (red, green, violet). Figures from Ref. [95].

The first detailed experimental investigation of neutron production by accelerator neutrinos was recently reported for charged-current $\bar{\nu}_\mu$ interactions on hydrocarbon by the MINERvA experiment [96], and follow-up analyses are ongoing. For water Cherenkov detectors, both a dedicated experiment (ANNIE) [97] and an upgrade to Super-Kamiokande (involving the addition of Gd to enhance the neutron capture signal) are anticipated to shed new light on this important topic. Despite a promising first demonstration of neutrino-induced neutron sensitivity by ArgoNeuT [17], no comparable effort yet exists for LArTPCs. However, the emerging low-energy reconstruction techniques described later in this document may enable the first precision measurements.

At present, the only complete treatment of nuclear de-excitations officially available in a neutrino generator designed for GeV energies is the PEANUT model [98, 99] included in FLUKA. Due to the code's lack of an open-source license, however, its use in experimental production environments has been somewhat limited. Following simulation of the primary neutrino interaction and an intranuclear cascade step similar to other high-energy neutrino generators, PEANUT uses an exciton model to describe preequilibrium nucleon emission. Program execution is then passed to a MARLEY-like compound nucleus simulation which includes discrete $\gamma$-ray emission based on nuclear structure data. In a recent study [17] of MeV-scale detector activity by the ArgoNeuT experiment, the additional de-excitation physics included in FLUKA but not in GENIE was found to be necessary to fully describe data.

Current capabilities for simulating $\gamma$-ray emission (and other low-energy de-excitation processes) in other high-energy neutrino generators are discussed in the introduction to Ref. [80], with the conclusion that realistic handling of $\gamma$-ray lines for nuclei other than $^{16}$O is currently unavailable in any official release of GENIE, GiBUU [100], NEUT, or NuWro. Possible paths toward a full open-source simulation include modifying the default behavior of the GENIE interface to the Geant4 Bertini cascade (as mentioned in Ref. [80]), combining the MARLEY de-excitation simulation with the intranuclear cascade of one of the high-energy generators, and creating an interface between an existing neutrino event generator and a low-energy nuclear reaction code. The last of these options has recently been pursued unofficially [101, 102] for GENIE and NuWro using TALYS [103].

While they are far less important to neutrino energy reconstruction than final-state neutrons, nuclear de-excitation $\gamma$-rays induced by accelerator neutrinos may be of interest for other applications. It has recently been pointed out, for example, that de-excitation $\gamma$-rays could provide a powerful new handle for rejecting backgrounds to proton decay in LAr-based searches [104]. Fully pursuing this strategy, however, would require a realistic simulation of both the signal $\gamma$-rays from proton decay and background $\gamma$-rays induced by atmospheric neutrino interactions. Model predictions for the latter could potentially be tested using accelerator neutrino measurements and the low-energy reconstruction techniques described later in this document.

## 4.2 Particle Propagation and Interaction in Liquid Argon

Upon interacting with an argon nucleus in a LArTPC, the kinetic and rest-mass energy of directly and indirectly ionizing radiation can be transferred to product uncharged particles, which can travel macroscopic (many cm) distances prior to subsequent interaction and detection. For example, in beam, atmospheric, solar, and supernova neutrinos, momentum can be transferred to uncharged non-leptonic final-state particles, such as neutrons and photons. As uncharged particles transport through the liquid argon environment of large LArTPCs, like DUNE, many of them deposit a substantial portion of their their energy via production of topologically isolated low-energy ionization features, or blips. It is useful to consider the most common examples, ordered roughly from greatest to smallest fraction of blip-like energy deposition relative to the total visible energy deposited by the particle. For an expanded description of these cases, see Refs. [21, 19, 20].

- *Nuclear de-excitation gamma rays* interact via Compton scattering and the photo-electric effect, with their produced secondary electrons generating blip signatures.
- *Neutron capture on* $^{40}Ar$ generates a cascade of de-excitation gammas with total energy equal to the relevant mass defect. These gammas in turn generate blip signatures.
- *Low-energy (∼<20 MeV) neutrons* primarily interact in neutrino LArTPCs via inelastic ($n$,$\gamma n$) scattering, in which the scattering nucleus produces one or more de-excitation gammas, which in turn generate blips.
- *Low-energy (∼<20 MeV) protons* have short ranges in LAr, and thus have the capability to produce compact event display features with collected ionization charge similar to or greater than that associated with blips from gamma rays. In some cases, such as ($n$,$p$) scattering on argon or neutral current $\nu$-Ar scattering, this proton can be topologically isolated, and thus manifest itself in an event display as a blip.
- *High-energy neutrons* interact via ($n$,$\gamma n$), but more predominantly via $n$- and $p$-producing inelastic scattering. The former process's neutrons can ultimately generate blip-like signatures. For the latter process, high-energy protons predominantly produce easily-identifiable track signatures, while low-energy protons predominantly generate blips.
- *High-energy pions, protons, and muons* predominantly deposit energy via ionization along a primary track, but their inelastic collisions or terminating processes can also produce final-state neutrons or excited nuclei, which in turn produce blip signatures.

Proper modeling of the pathways to blip production represented above relies on accuracy in understanding of three underlying input categories: the cross-sections of various low-energy $n$-Ar scattering processes, the cross-sections of uncharged-particle-generating inelastic processes for protons, pions, neutrons, and muons, and the number and properties of final-state neutrons, gammas, protons, and pions produced in these nuclear interactions. In the remainder of this section, we will provide a more detailed look at these three broad simulation input categories, using descriptions of both the FLUKA [105] and GEANT4 [106] particle transport codes. Where possible, we will emphasize where direct LAr-based constraints on simulation inputs are available or where further such data could be yielded from existing or future experiments.

### 4.2.1 Transport of Low-Energy Neutrons

Cross-sections for $n$-argon interactions of various types are shown in Figure 9 from 1 eV to >10 MeV. This fully spans the range of expected neutron energies generated in solar and supernova neutrino interactions, as well as many of the secondary neutrons generated in beam and atmospheric neutrino interactions. From roughly 1 eV to 1.5 MeV, interactions are dominated by elastic nuclear scattering, a process invisible to LArTPCs with energy detection thresholds in the neighborhood of 100 keVee [19]. Thus, neutron capture is the only viable process for neutron detection at these energies in LArTPCs. Photon-producing inelastic scattering begins to contribute substantially to the total interaction cross-section at roughly 1.5 MeV, with hadron-producing scattering taking over at roughly 10 MeV. Thus, in this energy range, blips can provide a pathway towards identifying and calorimetrically reconstructing primary neutron energy.

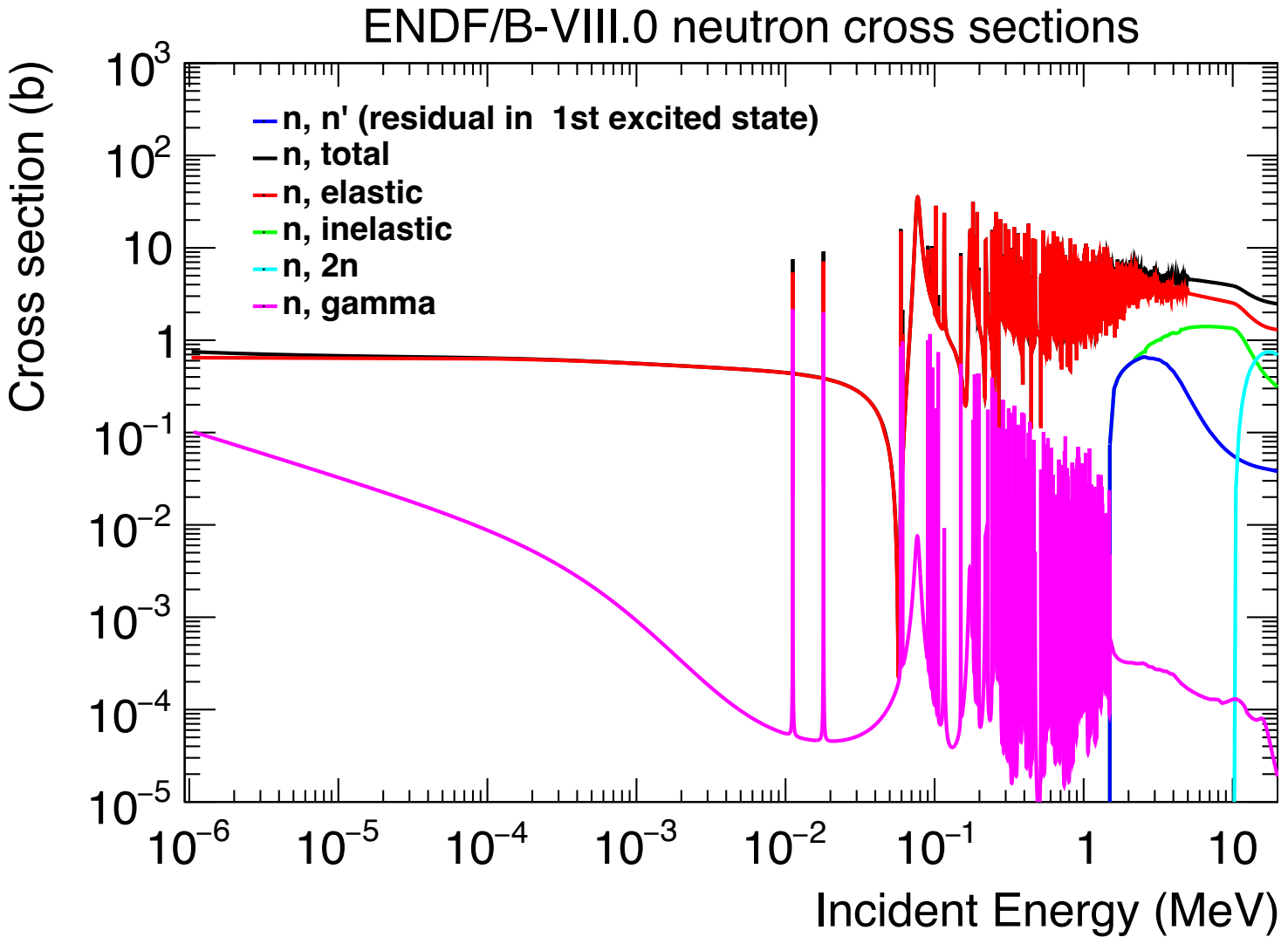

Figure 9: Total and exclusive interaction cross-sections on argon for neutrons ranging in energy from 1 eV to 20 MeV. Capture and elastic scatter processes dominate at and below the 1 MeV scale, while inelastic photon and neutron producing processes dominate above this range.

A variety of uncertainties in existing low-energy neutron-argon cross-section data limit the ability to fully assess LArTPC capabilities with respect to neutron detection at and below the MeV scale. Of particular importance, a deep anti-resonance in the total and elastic scattering cross-sections centered at 57 keV, enables extremely large (>10 m) mean free paths for neutrons in this energy range [107]. The depth of this feature determines whether most neutrino-produced neutrons will be captured in a LArTPC's active volume and subsequently tagged via associated blip detection [19], and whether LAr is an efficient self-shielding material for external neutron backgrounds [108, 11, 109, 54]. Future detailed measurements of this resonance region have been performed and final results expected in the near future [110].

Beyond this feature, the region of Figure 9 from roughly 10 keV to 1.5 MeV exhibits wildly fluctuating scattering and capture cross-section resonances. Moreover, capture and elastic scattering cross-sections begin to become comparable in magnitude over a large range at the figure's lowest energies. These intricacies, combined with the relatively small fractional energy loss of neutrons per argon collision, will generate a complex interplay between elastic and capture processes at low neutron energy. Preliminary large LArTPC simulations suggest that neutron capture time and position distributions are also likely to be size- and geometry-dependent. While some of these low-energy response aspects, such as detailed sub-eV neutron capture cross-sections [111] and resonance spectroscopy [107], have been defined, capture-related response aspects in large LArTPCs are in many ways poorly experimentally demonstrated. Measurements of cosmogenic or calibration source neutron capture positions and times in MicroBooNE, SBND, or protoDUNE may provide some further insights into these aspects of low-energy neutron response for large LArTPCs the near future. The use of a pulsed neutron source in protoDUNE makes its data particularly valuable for this purpose; analysis of this dataset is underway within the collaboration. The high, uniform light collection efficiency of SBND and other future on-surface LArTPCs may also make these datasets particularly amenable for neutron capture analyses. More detailed understanding of low-energy transport may also be accessible through deployment of a LArTPC in a low-energy neutron test beam.

Above 1.46 MeV, corresponding to the favored lowest-energy state of $^{40}$Ar [112], neutrons are capable of inelastic excitation of the nucleus, which de-excites via gamma, neutron, or proton emission. Dominant cross sections for the populating of excited states by neutrons of varying energies between 1 and 30 MeV have been measured fairly precisely in the context of dark matter experiments [113]. However, the interplay between elastic and inelastic collisional energy losses in this energy range has not been accurately probed in previous 'thin target' experiments. This lack of knowledge may result in biased GEANT4 or FLUKA modeling of blip multiplicities, energy spectra, and topological distributions in large LArTPCs.

Future measurements of cosmogenic or neutrino-induced neutrons in the Fermilab SBND Program or protoDUNE, or secondary neutrons produced in LArIAT or protoDUNE test beam data, are capable of providing some level of constraint. However, direct relation of these measurements to cross-sections is limited by the lack of precise understanding of true neutron energies produced by these sources. For this reason, deployment of a large-volume LArTPC at a fast neutron test beam, similar to miniCAPTAIN at the LANSCE facility [114], would be particularly valuable. While further analysis of existing miniCAPTAIN data may offer some value, deployment of a new high performance LArTPC designed and operated with blip detection in mind offers substantially more promise.

### 4.2.2 Transport of High-Energy Hadrons

Looking again at the highest pictured energies of Figure 9, one can see inelastic processes generating free baryons beginning to contribute substantially the total neutron interaction cross-section. This dominance over other inelastic channels continues at higher energies for all hadron types – neutrons, protons, and pions. For neutron and proton interactions on argon nuclei, predictions of total elastic and inelastic cross-sections out to higher energies are pictured in Figure 10. Validations of these cross-sections are essential to ensure proper modeling of the behavior of leading-energy hadrons in high-energy neutrino interactions and BSM processes. Improper knowledge of energetic neutron cross-sections may lead to incorrect understanding of neutronic energy containment within a TPC or biased blip and secondary proton position distributions in event displays; improper proton/pion cross-section knowledge can bias the level of energy deposition by the primary proton/pion with respect to the neutral secondaries it produces.

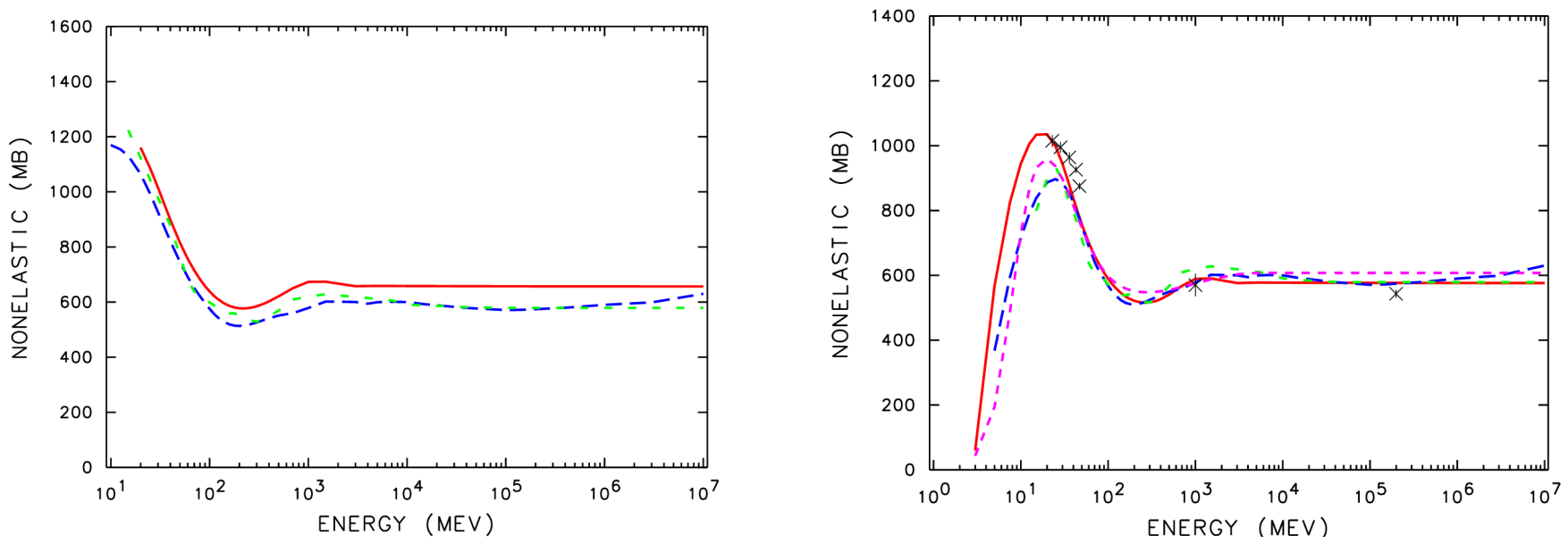

Figure 10: Total neutron (left) and proton (right) inelastic interaction cross-sections on argon versus energy. Lines represent predictions developed by FLUKA (dashed blue) and other nuclear modeling teams, while asterisks and squares represent existing measurements on argon. Data from Ref. [115], with plots reproduced from Ref. [116].

LArTPC experiments currently rely on particle transport simulation codes like such as GEANT4 and FLUKA to model energy depositions produced by high-energy hadronic systems. As an example, for FLUKA, libraries incorporated to predict interaction probabilities include consideration of resonances and quark/parton string models for secondary-producing processes and phase-shift analyses and eikonal approximations for elastic and charge-exchange processes. When needed, these models are fitted to match available experimental data.

Unfortunately, as is visible in Figure 10, there are few argon datasets currently available for model benchmarking. Some historical measurements for protons exist, while no historical measurements are available for neutrons. Given the lack of historical argon data, GEANT4 and FLUKA have instead primarily used proton and neutron scattering data from other targets, such as aluminum and carbon, to constrain models, leading to concerns about the accuracy of this modeling for the specific case of LArTPCs. A full stable of argon-based high-energy cross-section measurements from LArTPCs can foster direct quantification of uncertainties in hadronic energy scales and deposition mechanisms modeled by FLUKA or GEANT, which can then be straightforwardly propagated to higher-level LArTPC physics analyses.

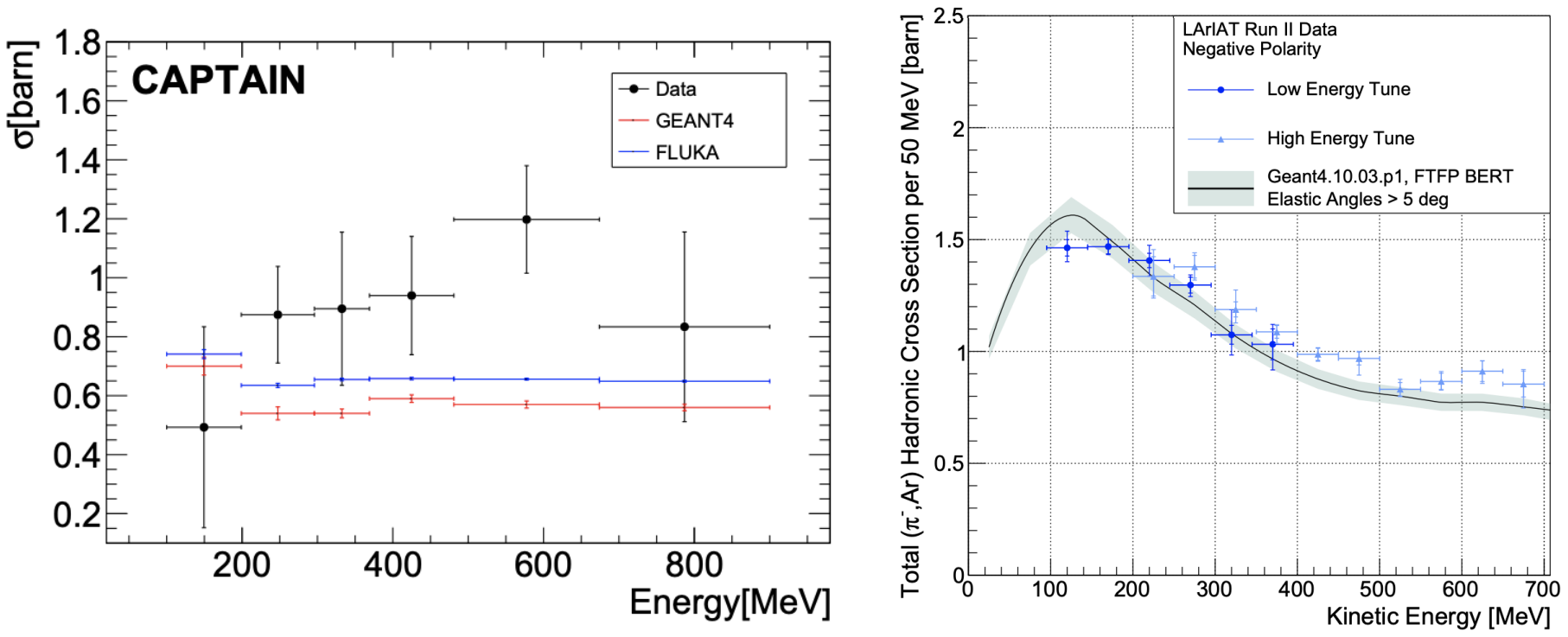

Figure 11: Left: measured inelastic cross-sections on argon versus energy from the miniCAPTAIN experiment [114]. Right: measured inelastic cross-section of pions on Argon versus energy from the LArIAT experiment [117].

Fortunately, recent LArTPC measurements have begun to address current argon data limitations. The miniCAPTAIN experiment at LANSCE has recently provided first measurements of the neutron-argon inelastic scattering cross-section in the 100-800 MeV range using a time-of-flight tagged neutron test beam [114], with roughly 20-50% precision above 200 MeV. The LArIAT collaboration has recently provided inelastic scattering measurements for pions using a charged particle test beam at Fermilab [117], providing consistent cross-section results between low and high-energy beam settings. Both of these results are pictured in Figure 11. Similar measurements for protons and kaons should be achievable in the future using existing LArIAT and protoDUNE test beam datasets. Meanwhile, improvements in the precision of neutron scattering cross-section measurements, as well as extension of the measurement range below roughly 200 MeV, will require a dedicated future experiment with a high performance LArTPC stationed in a high-energy neutron beam facility. In addition, total neutron-argon high-energy inelastic cross-sections may be indirectly probed by measuring blips and isolated proton tracks in the vicinity of hadronic (neutrino) interaction vertices in protoDUNE (neutrino beam LArTPC) event displays. As similarly described in the previous section, however, direct knowledge gleaned from these latter measurements would be limited by uncertainties in other aspects of neutron production and transport in LAr.

#### 4.2.3 Final-State Content From Nuclear Interactions

Even if inclusive elastic and inelastic cross-sections for hadrons in LAr are perfectly known at all energies, proper modeling of the magnitude and distribution of blip-like activity is impossible without knowledge of the final-state products of the involved nuclear interactions. For example, if the previously-discussed inelastic $n$-Ar interactions measured in miniCAPTAIN produced only final-state protons, it is likely that relatively little blip activity is present in miniCAPTAIN event displays; however, if these same interactions produced equal numbers of free neutrons and protons, many blips would be present. Besides multiplicities, final-state particle energies are an equally important aspect of the problem: the total number of $n$-Ar final-state proton tracks observed by CAPTAIN is certainly proportional to the true initial energies of these freed protons. These same issues are relevant to all LArTPC experiments, as mis-modeling of final-state blip energies and multiplicities can bias GeV-scale and MeV-scale neutrino energy assessments, foil blip-based particle and interaction channel identification strategies, and much more.

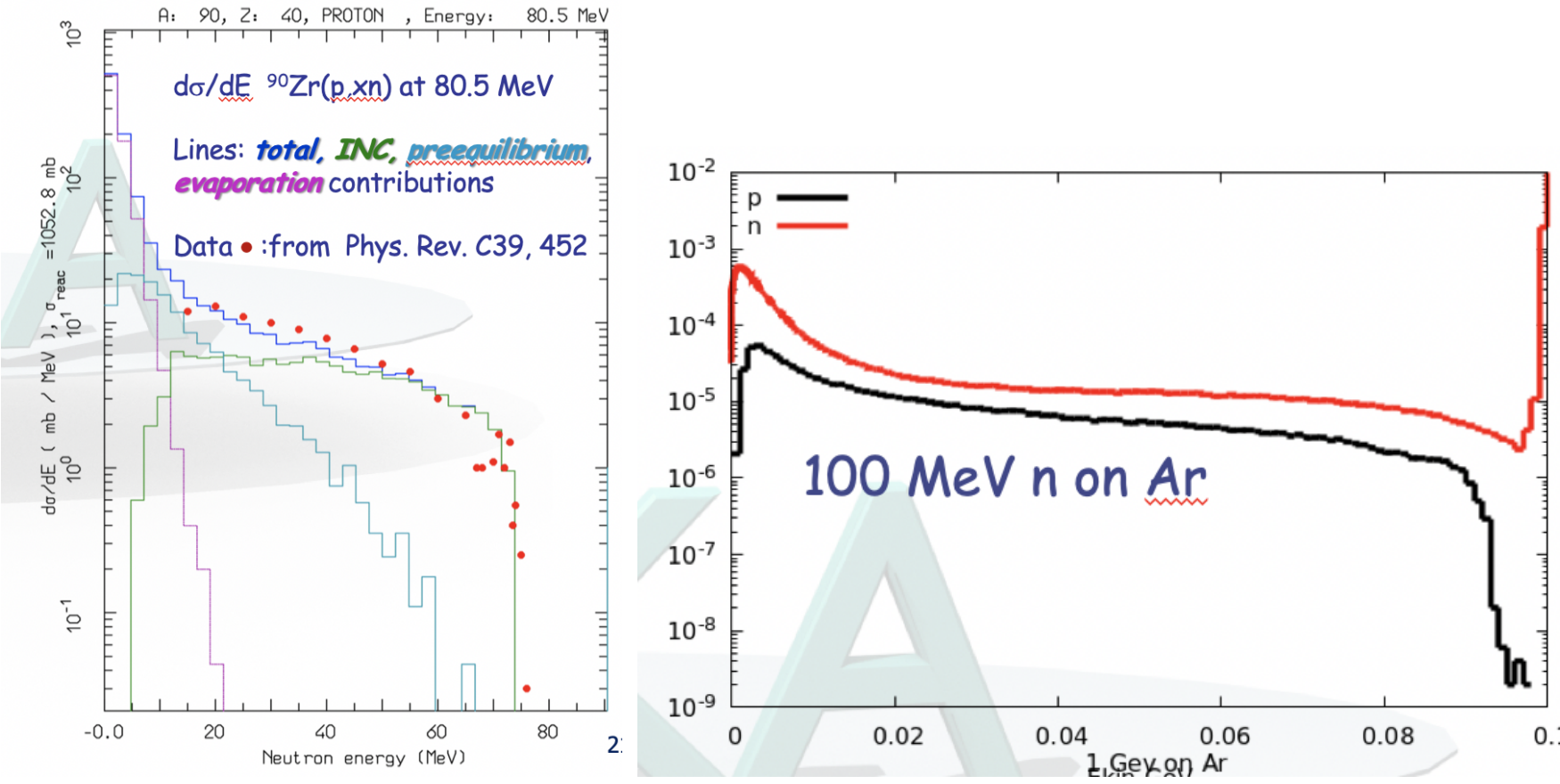

Figure 12: Modeling of final-state neutron properties. Left: total (blue) FLUKA-predicted final state neutron production versus energy for incident 80.5 MeV protons on $^{90}$Zr, with individual contributions from cascade (green), pre-equilibrium (cyan) and evaporation (purple) processes. Right: total FLUKA-predicted final state neutron and proton production versus energy for incident 100 MeV neutrons on argon.

Modeling of final-state content from LAr nuclear interactions in GENAT4 and FLUKA depends on a variety of input physics models, with dominant models differing depending on the energy scales involved. Intra-nuclear cascade effects produce most high-energy (>50 MeV) final-state neutrons, while pre-equilibrium and evaporation processes generate the most content in the medium energy (10-50 MeV) and low-energy (<10 MeV) regimes. The role of these differing contributing processes are well-illustrated in Figure 12, which shows predicted and measured energies of final-state neutrons generated by 80.5 MeV protons interacting on $^{90}$Zr [118]. Regardless of exiting energy, models predict substantially higher numbers of final-state neutrons in LAr nuclear interactions, as demonstrated in Figure 12 by FLUKA for interacting 100 MeV neutrons.

Little to no data exist to directly validate of modeling of final-state particle multiplicities and energies produced by hadronic proton, neutron, pion, and muon nuclear interactions on argon. As with inclusive hadronic interaction cross-sections, confidence for modeling in LAr is largely generated via assessments of other similarly-sized nuclear targets. For example, Figure 13 provides one such demonstration of neutron production versus energy and exiting angle for interactions of 256 MeV protons on aluminum [119]. The only existing measurement of this kind performed in liquid argon, by the ArgoNeuT experiment, is pictured in Fig, 13. This analysis provides a picture of generally accurate FLUKA modeling of the multiplicity, energies, and positions of blips produced by neutrino interaction final-state neutrons and de-excitation photons [17]. While this result represents a watershed moment for low-energy neutrino LArTPC physics, its value for constraining final-state particle production in LAr is limited, given its low statistics and small detector size, and given the lack of knowledge regarding neutrino-produced neutron content.

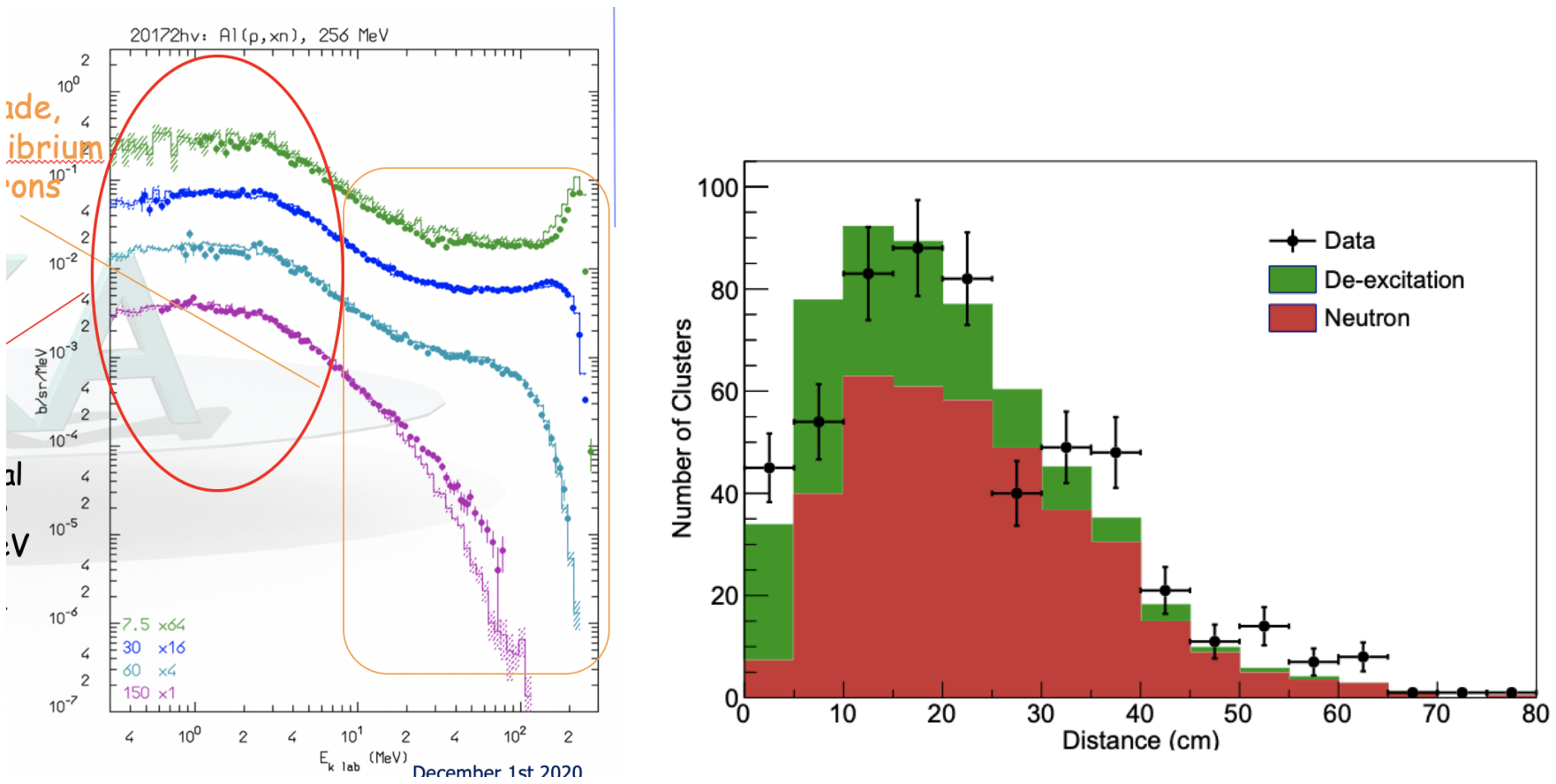

Figure 13: Direct and indirect measurements of final-state neutron properties. Left: Energy and angular distribution of neutrons produced by interactions of 256 MeV protons on aluminum (data from Ref. [119]). Right: position distribution of reconstructed LArTPC blips generated by neutrino interaction final-state neutrons and nuclear de-excitation photons (from Ref. [17].

More direct and precise future constraints can be provided by the protoDUNE and LArIAT test beam experiments, where blips and non-displaced and displaced proton tracks can be identified in LArTPC event displays following the injection of input protons and pions of well-defined energy. Input hadron kinetic energies in these beamlines range from roughly 100 MeV to more than 1 GeV, providing ample sampling of the energy space relevant to GeV-scale neutrino interactions. Nuclear captures at rest of test beam pions and muons provide additional interesting event classes with well-defined input particle kinematics. While explicit determination of high-energy ($\sim$>50 MeV) proton and pion final-state multiplicities and kinematics should be achievable, disambiguation of these two information categories will be difficult for low-energy charged hadrons and for neutrons of all energies. Gaseous argon TPC test beam experiments can aid in addressing the former, while neutron test beam experiments using enhanced LArTPCs should be helpful for the latter. New thin-target measurements on argon at charged particle or neutron test beam experiments using non-LArTPC neutron detectors, similar to those performed in Ref. [119], may also be valuable for more directly constraining properties of final-state neutrons. It should also be noted that, for these LArTPC measurements, determination of only reconstructed quantities, such as total final-state blip, vertex, and displaced proton properties, may be sufficient for reaching LArTPC physics goals; perfect knowledge of true underlying final-state quantities may be unnecessary.

### 4.2.4 Summary of Particle Transport Issues

Improved knowledge of nuclear interactions on argon is required to reliably model the propagation and energy deposition pathways of charged and uncharged particles in LArTPCs. Accurate modeling of these processes is essential for meeting a wide variety of neutrino LArTPC physics goals, from the sub-MeV to the multi-GeV scale. Major improvements in charged particle transport modeling and validation can be achieved via detailed low-energy analysis of test beam data from recent and near-future LArTPC experiments, such as LArIAT and protoDUNE. For low-energy neutron transport, on-surface detectors containing high fluxes of calibration source or cosmic neutrons, such as protoDUNE or Fermilab SBN, may be valuable, as well as dedicated non-LArTPC experiments targeted at specific aspects of argon's repsonse, such as ARTIE and ACED. While charged particle test beams may provide some level of indirect validation of high-energy neutron modeling, dedicated LArTPC-based, GArTPC-based or non-LArTPC measurements at neutron beam facilities would be particularly valuable. These measurements will improve hadronic energy deposition and transport modeling in LArTPCs, either by enabling improved understanding of true underlying interaction/production mechanisms, or by building high-statistics collections of relevant event display topologies for use in future template-based simulation approaches.

# 5 Detector Parameters

## 5.1 General LAr TPC Requirements for Low-Energy Physics

From the standpoint of *requirements*, the low-energy physics program in a LArTPC detector can be broken up into four, distinct classes:

- Rare, time-correlated events like supernova bursts
- low-energy ($< 100$ MeV) signatures in high-energy events
- Rare searches at thresholds that are above all radiological backgrounds ($20 - 100$ MeV)
- Precision measurements well into the radiologial background regime ($<$ 20 MeV)

The first class holds the promise of the biggest physics contribution that could be made by a large, underground LArTPC like DUNE, and fortunately the requirements for being able to see a supernova burst or other time-correlated signal are not overly stringent: the burst profile is its own distinct signature (even if it is unknown) that allows a detection even in the presence of rates of radiological backgrounds or noise that would preclude a measurement of other low-energy signatures, such as solar neutrinos. A low efficiency for the detection of individual *interactions* does not necessarily translate into a low *burst* efficiency. Even a very small number of detected interactions in a short (e.g., 10 s) time window with visible energy above 15 MeV or so would be a clear signature of an interesting burst. What can matter more, however, are fake bursts caused by spallation events from cosmic-ray muons, and thus a high overburden is a requirement for this physics.

Time-correlated events are by definition rare, and can have intrinsically small signals (such as very distant supernovae), and thus one requirement is on detector size: for LArTPCs with signal thresholds above the few-hundred keV scale, a large mass (at least 10 ktonne) of argon is needed to see supernova bursts to the edge of the galaxy. Global timing is also an obvious requirement for these events, as extracting the physics from the worldwide detection of a burst requires both coordinate timing and timing precision at least good enough to resolve burst structure, or measure relative times of neutrino vs. photon emission, or even potentially dispersion in neutrino arrival times. Maintaining high uptime is also necessary; while obviously the probability of missing a burst due to deadtime is only as large as the deadtime, the impact of missing a burst would be incredibly damaging to the worldwide neutrino program. Lastly, a highly efficient—and preferably inclusive or at least somewhat model-independent—burst trigger with a tolerable "fake rate" is needed.

Less critical requirements for bursts would be precision high electron lifetime and a precision $t_0$ for electron-lifetime corrections, and, as stated above, reasonably low radiological backgrounds (particularly above 10 MeV of visible energy deposit) and low front-end noise, including particularly coherent noise which can lead to fake track-like events.

For the second class of physics, low-energy signatures of high-energy events, the requirements are much less stringent. The primary requirement is low front-end channel thresholds, to provide high efficiency for detection of these low-energy signatures and, correspondingly, low front-end noise (again, including coherent noise). Good electron lifetime is also needed so that these small energy deposits survive with enough charge that they cross threshold. Lastly, diffusion needs to be low, so that as the charge drifts to the anode it does not spread out so much that the deposits cannot be distinguished from higher-energy deposits that are part of the rest of the event, or fall below threshold as they spread out across many channels.

For rare searches in the 20-100 MeV regime, the requirements are weakest of the four classes listed above. In this regime, the dominant backgrounds are atmospheric neutrinos, possibly cosmic-ray muons that clip the detector's active volume or stop and are captured, or long-lived kaons produced in the cryostat that charge-exchange and then enter and decay. Thus a large overburden is important, as is an efficient and inclusive trigger. Robustness against unusual instrumental backgrounds—like high voltage streamers—is also important here.

The last category of physics, low-energy signals below 20 MeV and down into the radiological backgrounds, has the strictest requirements. There are many possible radiological background sources, most of which become problematic just below 4 MeV or so ($^{42}$Ar and its daughters; U and Th chain decays including $^{214}$Bi and $^{212}$Bi; and ultimately $^{39}$Ar with its 500 keV endpoint and enormous rate of 1 Bq/kg of argon). External backgrounds, in particular neutrons, are potentially more dangerous because the $\gamma$ cascade from their capture on $^{40}$Ar deposits roughly 6 MeV. While sensitivity to bursts can tolerate a relatively high rate of backgrounds because of the burst signature itself, the physics in this regime—for example, solar $^{8}$B and *hep* neutrinos with rates in the regime of a dozen per day or so—is far below all of the background rates listed above. To do this physics very likely requires going to energies above the neutron captures (whose rate in a 10 ktonne mass might be in the regime of 1-10 Hz), and thus energy resolution, and in turn electron lifetime and diffusion plays an important role so that these backgrounds do not leak into the (higher-energy) signal regime, and can be topologically distinguished from events that include primary electron tracks. Also, because so much has been already done for solar neutrinos, the precision goals are higher, and thus precision knowledge of the energy scale and resolution will be important to make a meaningful contribution. Shielding of neutrons with water or some other hydrogenous material would be a requirement if a solar neutrino program pushing below 5 MeV of visible energy deposit is planned. The trigger for these events will need to be not just inclusive and efficient, but its differential efficiency curve must be *known* with reasonable precision. Lastly, because of the high rate of background events at the trigger level, some online data reduction, such as region-of-interest selection of interaction hits, would be needed so as not to create impossibly large data sets.

## 5.2 Improving TPC Charge Readout

The performance of the single-phase LArTPC design as tested in ProtoDUNE-SP is already quite good for the low-energy physics program. Electrons from $^{39}$Ar decays are plainly visible with strong sculpting of the energy distribution below 200 keV, but with quite useful efficiency above 300 keV. The signal-to-noise ratio in the collection plane is the most important determinant of whether events can be triggered, read out and analyzed to produce useful physics results. The signal-to-noise ratio for minimum-ionizing particles passing perpendicular to the wires is 48.7 with correlated-noise removal (CNR) [120]. Without CNR, the signal-to-noise ratio is 30.9 in the collection plane. Improvements in the electronics, grounding, and investigation of sources of excess noise such as instrumentation in the liquid argon volume and power supplies, is likely to make the noise performance in the first DUNE FD module even better than it was in the first run of ProtoDUNE-SP. A lower intrinsic noise value will allow for lower hit-finding thresholds and lower trigger thresholds.

One feature seen in ProtoDUNE-SP is the presence of very low frequency noise on the induction-plane wires, which was not seen as prominently on the collection-plane wires. Frequency filtering and deconvolution easily suppresses this noise, but such approaches would have to be implemented at the input to the trigger in order to have the induction-plane wires contribute meaningfully to the trigger. Polynomial or sinusoidal fits to the slow variations of the pedestal as a function of time may be sufficient to allow for threshold-based hit-finding and triggering to collect as many low-energy physics events as possible.

Other optimizable parameters are the electronics gain, the shaping time, and deliberate nonlinearities introduced in the preamp stage in order to maximize resolution of low-energy pulses while avoiding saturation for high-energy pulses. A two-gain solution has been proposed for the dual-phase detector readout, due to the large intrinsic gain expected from the LEM [121]. The 12-bit ADC solution required by the DUNE FD is sufficient to detect $^{39}$Ar decays as well as detect 10-MIP signals without saturation. Designs with 14-bit ADCs are being investigated, which would reduce the ADC discretization noise on the low end of the signal strength range, or allow for larger signals without saturation, depending on the choice of preamplifier gain.

Optimizable detector design choices include the drift field, which affects the recombination factor and thus the split between charge and light production, the bias voltages, the wire spacing, the induction-plane wire angles, the number of planes, and the presence or absence of a grid plane. MicroBooNE has shown that the signal-to-noise ratio in its U wire plane is better than that in its V plane, due to the different signal shapes of the two planes owing to the lack of a grid plane [122]. DUNE plans on installing a grid plane in order to protect the electronics from electrostatic discharges during handling and installation. If the U plane performs very differently from the V plane, the detector response will be less homogeneous and isotropic.

The wire bias voltages have an impact on the ratio of peak signal to the RMS sample noise because they change the drift velocity of electrons moving between the wire planes. While the total charge collected by the collection plane wires remains unchanged by changes in the wire bias voltages and similarly, the total of the absolute value of the induced current on induction-plane wires is also unchanged, the time over which the deposits occur is shortened, sharpening the hits if the potential differences between the wire planes are increased. The deconvolution recovers the total charge regardless of how it is spread out, but the contribution due to noise increases as the pulses widen. A downside to increasing the electric fields between the wire planes is increased sensitivity to mechanical oscillations on the wires, causing currents to flow as they are in an external field. Operations are also expected to be more difficult with higher electric fields between the wire planes. Naturally, every effort must be taken to filter noise from the bias voltage power supplies.

The wire spacing for the single-phase horizontal-drift DUNE FD module design was optimized based on the anticipated size of diffusion and the desire to keep the signal-to-noise ratio high. A finer wire pitch improves the spatial resolution, but the spatial resolution is limited by the typical size of the diffusion radius. Finer spacing divides the same detected charge among more wires, but since it increases the total plane capacitance, it increases the total noise and thus reduces the signal-to-noise ratio. We do not anticipate a significant improvement for low-energy physics by changing the wire pitch from its value of approximately 5 mm. The natural spacing between planes is similar to the spacing between wires within a plane so that the range over which significant signals occur from induced charge is similar to the wire spacing. A narrower gap between planes makes hits sharper in time, but does not continue to do so if the inter-plane gap is narrower than the wire pitch, as the planes do not make as effective shields if they are closer together.

The induction-plane wire angles are expected to have only a small impact on low-energy physics when varied within ranges that are optimal for beam physics. The main effect is on the point resolution along the vertical direction. Very low-energy deposits, from radiological decays for example, show up simply as blips which may appear on single wires or which may span pairs of neighboring wires. These have little direction information. Estimating the direction of electron stubs in supernova-burst neutrino interactions however is an important ingredient to the pointing measurement [15]. The distribution of angles of the electrons with respect to their progenitor neutrinos is broad, however [11]. The choice of the $\pm 35.7^\circ$ angles for the induction-plane wires was made so that no induction-plane wire intersected the same collection-plane wire more than once, reducing inter-plane hit-association ambiguities compared to designs with larger wire angles [15].

### 5.2.1 Pixel-Based Readout

A possible avenue to improve both the signal-to-noise ratio and reduce the ambiguities of interpreting three-dimensional events by matching data in multiple two-dimensional views is to use a pixel-based readout. The capacitance of a pixel 5 mm on a side is much less than that of a ProtoDUNE-SP collection-plane wire, and the intrinsic noise would be that much lower. Limitations in the noise performance due to the electronics choices (considerations of cost, power, and size are all important to meet DUNE's goals) are expected to dominate over the intrinsic channel noise for pixel-based LArTPC detectors. Furthermore, wires have an intrinsic limitation in resolving ambiguities, which result in making the event reconstruction difficult in some cases. Finally, the construction, mount-ing, and testing of these large anode plane assemblies to host the wire planes poses substantial engineering challenges. Moreover, the potential for a broken wire poses a significant single point failure (SPF) design in the system. For these reasons, a non-projective readout would have many advantages.

Non-projective readout has been realized in many gas based TPC's but until recently was not considered viable for LArTPC's. This was because the number of readout channels and power consumption requirements on existing LArTPC readout technologies made such an approach prohibitive. The number of pixels for equal spatial resolution will be two or three orders of magnitude higher than the number of corresponding wires, with a corresponding increase in the number of readout channels, data rates and power dissipation. A transformative step forward for future LArTPCs is the recent advancements to build a fully pixelated low power charge readout.

The endeavour to build a low power/low noise pixel-based charge readout for use in LArTPC's has independently inspired two research groups to pursue complimentary approaches to solving this problem. The LArPix [123] and Q-Pix [124] consortia have undertaken the challenge of the research and development necessary to realize a pixel-based readout. Pixel-based readout is also the leading candidate readout option for the proposed GRAMS LArTPC-based compton telescope [125, 126], targeting the detection of MeV gamma rays during a balloon or future satellite deployment of the detector.

A pixelated charge readout system provides a uniform detection efficiency with respect to the readout plane, and native three-dimensional information of physics activities, which bypasses the ambiguities from inter-plane hit-associations. As was shown in Ref. [127], and is represented in Figure 14, the 3D pixel-based readout is found to be superior to the 2D projective one across a wide range of classifications in the high-energy regime. In particular, the identification of electron-neutrino events and the rejection of neutral current $\pi^0$ events, a 3D pixel-based detector significantly outperforms a 2D projective one by about a factor of two.

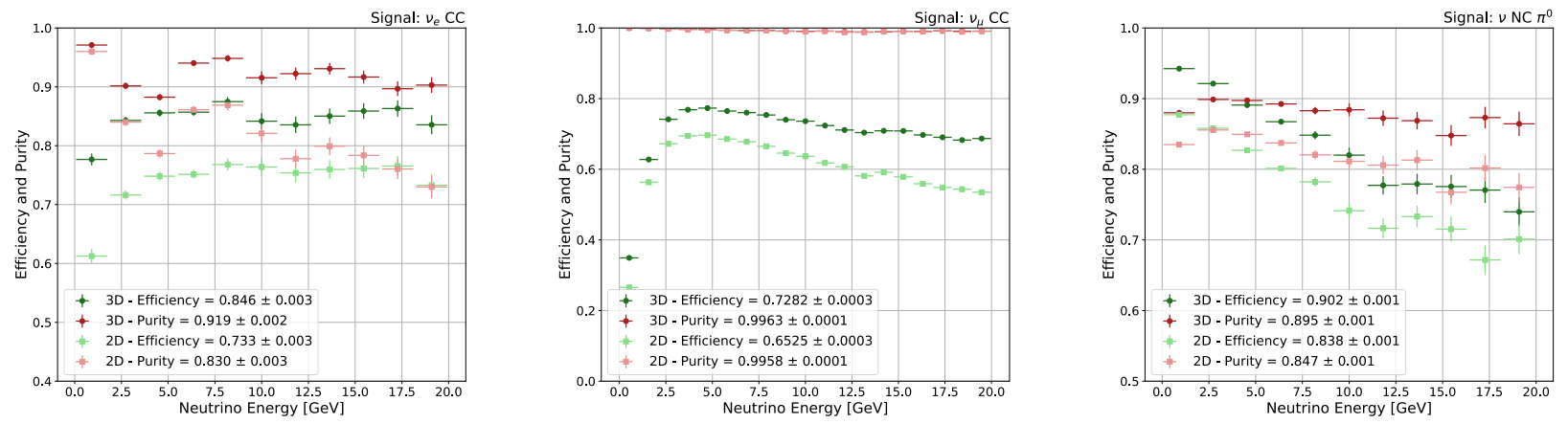

Figure 14: Efficiencies (green) and purities (red) as a function of neutrino energy for inclusive $\nu_e$ CC selection, inclusive $\nu_\mu$ CC selection, and $\nu$ NC $\pi^0$ selection. Results are shown for both 2D (light colors) and 3D (dark colors) taken from Ref. [127].

Data of low-energy particles collected in a pixelated charge readout system will help us study the profile of such particles, thereby improving simulation of particle propagation and developing reconstruction algorithms. Moreover, the intrinsic 3D readout allows for drastic improvement in the detection, reconstruction, and classification of events in the region below 10 MeV of deposited energy, opening the opportunity to enhance the physics opportunities of LArTPCs in regions such as solar, atmospheric, and supernova neutrinos and enhance the beyond standard model reach of these detectors.

For next-generation LArTPC neutrino experiments that will have the potential to make many new discoveries, leveraging the technology of 3D pixel readout to maximize their potential is a crucial task. The challenge to realize LArTPC pixel-based readout is a non-trivial one and efforts from the LArPix and Q-Pix groups are well underway. The pursuit of 3D pixel technology is well motivated by the foreseen physics impact presented here.

### 5.2.2 Photosensitive Dopants

A critical aspect of using LArTPCs to explore low-energy signals is how the energy deposited can be measured precisely. When charged particles cross the liquid argon, the argon is excited from its ground state and ionized. The resulting ionization electrons will recombine with the argon ions without an electric field, forming unstable argon dimer molecules. As the excited argon dimer molecules relax, they release 9.6 eV photons. When in the presence of an electric field, a fraction of the electrons drift to the charge readout before recombining, forming the charge signal. Different electric fields will yield different quantities of electrons and photons but will sum to a fixed number. This effect creates an anticorrelation in the light and charge signals observed in the detector, where the normalization of each channel depends on the applied electric field. At low energies collecting information from both light and charge is key to increasing the energy resolution of a LArTPC. For example, MicroBooNE simulations predict an impressive 5% energy resolution at 1 MeV for electrons utilizing only the charge signal [128]. If we wanted to improve on this, we would need to begin leveraging information from the light signal, as demonstrated by LArIAT [129].

In improving the energy reconstruction of a LArTPC, it is crucial to know how much light is required. The Noble Element Simulation Technique (NEST) collaboration has explored the precision of LArTPC for 1 MeV electrons as a function of charge readout signal-to-noise ratio and the efficiency of collecting light [130]. The results of NEST collaboration, in Figure 15 (left), show that LArTPCs with signal-to-noise ratios near that of ProtoDUNE [120] will need to collect nearly 50%, or 15,000 photons/MeV, of the light created by the energy deposit to achieve 1% energy resolution at 1 MeV. Collecting so much light seems a tall order given that the current DUNE baseline requirement is 0.06% [12] and the best achieved by a large LArTPC (SBND) is 0.5% [131]. The introduction of photosensitive dopants into the liquid argon could provide a cost-effective method to increase the light collection efficiency up to 60% by directly converting isotropic scintillation photons to directional ionization charge at the point of creation.

Photosensitive dopants work by having ionization energies close to the scintillation energy of the detector medium. Then through two processes, Penning transfer or photoionization, they convert the light to charge. In Penning transfer, the dopant interacts with the unstable argon dimer transferring the energy to the dopant, which ionizes. For photoionization, the argon dimer relaxes, releasing a photon which the dopant captures, ionizing the dopant. The exact contribution of which process dominates is an open R&D question. These dopants have been studied previously for applications in LAr collider calorimeters [132] and were doped into a prototype LArTPC by the ICARUS collaboration [133].

In LAr collider calorimeter test-stands photosensitive dopants were introduced to the LAr at ppm levels. The coarse test-stand detectors employed monoenergetic MeV-scale alpha sources. These test stands leveraged alpha particles to create large energy deposits over short distances. Depositing so much energy over short distances leads to a quenching of the ionization process results in alphas creating 15 times more scintillation light than ionization charge when compared to a minimally ionizing particle. The results, shown in Figure 15 (right), demonstrate that, at the DUNE nominal electric field, these dopants can increase the charge collected by nearly an order of magnitude. We can reinterpret this charge collection increase to the "light collection" efficiency. A factor of 9.4 increase in charge collected from an alpha particle is analogous to collecting 60% of the light produced. Figure 15 (left) shows this would enable percent-level energy resolution for 1 MeV electrons.

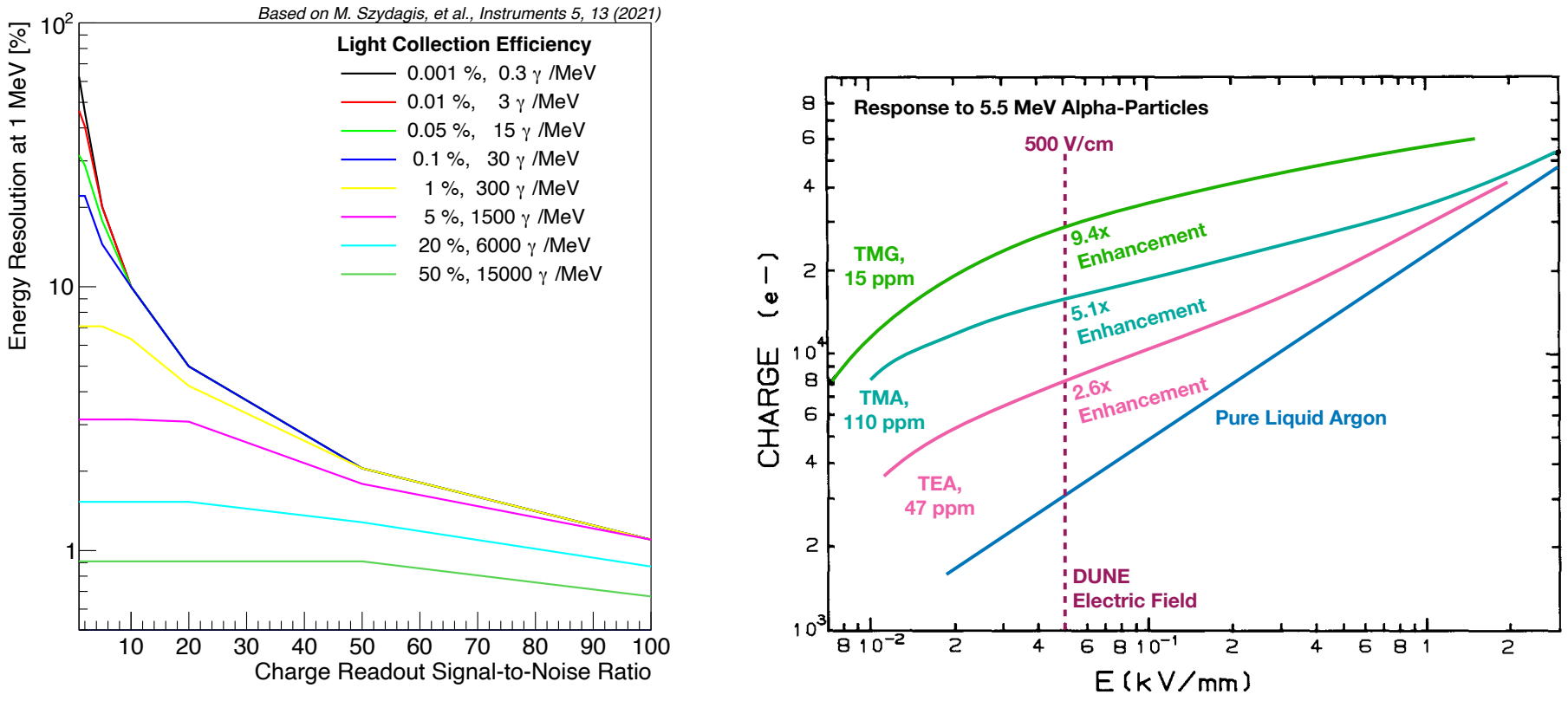

Figure 15: Left: reinterpretation of data presented in Ref. [130]. This shows the theoretical best energy resolution a LArTPC could achieve for a 1 MeV electron as a function of the charge readouts signal-to-noise ratio (x-axis) and the light collection efficiencies (colored lines). This figure is from Ref. [134]. Right: a replotting of data presented in Ref. [132]. This presents the amount of charge observed from a 5.5 MeV $\alpha$ source at various different electric field settings.

Further, ICARUS studied photosensitive dopants for applications in LArT-

PCs targeting GeV-scale physics [133]. ICARUS found that introducing ppm-levels of TMG to liquid argon increased the charge yield from through-going muons by 30% in a 300 V/cm electric field. Further, they found no negative impact on their electron lifetime. They ran stably for roughly 250 days without any observed decrease in the dopant concentrations, implying that their filter did not remove any TMG. Finally, they observed a significantly more linear detector response from stopping cosmic muons and protons. Particles deposit more energy over shorter distances towards the end of stopping particle tracks. This increase in the energy deposited quenches the ionization process, resulting in less charge as more energy is deposited. With the introduction of TMG, ICARUS instead found a significantly more linear response. As particles deposited more energy, more charge was detected by the charge readout. This enhanced linearity could help improve particle identification tools in LArTPCs and improve energy reconstruction for low-energy proton and nuclear recoils.

The conversion of scintillation light to charge introduces a challenge for LArTPCs, which have traditionally used light signals to determine where, in the drift direction, particles passed. Instead of light, the charge signal could be leveraged to estimate the location in the drift direction where the charge originated. Past LArTPCs have demonstrated that using measurements of the charge width (or the amount of charge diffusion) can be used to estimate the drift distance of the charge [135]. Using the charge signals to determine the location of the charge in the drift direction would effectively trade energy reconstruction precision for timing precision. Other possibilities for determining the charge location are discussed later as part of the open R&D possibilities.

A benefit of using photosensitive dopants is that they require no modification of the detector to achieve the stated gains. Photosensitive dopants also provide flexibility for integrating them into large LArTPC physics programs. For example, dopants could be introduced either at the start of data-taking or after the "main" physics program has been completed, giving a "second life" to large LArTPCs without the need to modify the detector.

One application of this technique would be to take one of the 17 kton DUNE FD modules and inject 100 kg of a PS-dopant. This detector could then begin a physics run with enhanced MeV-scale capabilities. Doping a FD module could occur during or after its initial physics run and create a massive MeV-scale sensitive LArTPC situated deep underground. This doped 17 kton LArTPC would enable the detector to forge new ground in the study of solar neutrinos with enhanced precision on $\sin^2\theta_{12}$ [16] and would allow significant improvements in the reconstruction of supernova neutrino energies. Another application would be to combine PS-dopants, $^{42}$Ar depleted liquid argon, and a neutrinoless double-beta decay isotope (such as $^{136}$Xe) into DUNE FD module [134].

The prospects that photosensitive dopants offer to enable new discovery opens a rich R&D program. This includes which photon-conversion process that dominates the light-to charge conversion, photoionization or Penning transfer. If photoionization were to dominate the argon dimer relaxation times, this would lead to characteristic smearing of the charge by 4 mm in the drift direction [136], which would negatively impact the light collection efficiency. By introducing a second dopant specifically, xenon would reduce the scintillation time which would reduce the smearing [136]. A second major R&D question is if light can survive the photon-conversion processes. Enabling light to survive would enable LArTPCs to leverage light for determining the location of the charge in the drift dimension and preserve the timing capabilities of LArTPCs. One concept is to introduce a second dopant (such as xenon) to create residual light. Xenon-doped liquid argon has been shown to shift the energy of a fraction of the photons to lower energies and could enable it to bypass the photosensitive dopants [137]. Finally, past demonstrations of photosensitive dopants were performed in the context of LAr calorimeters and cosmic particles. New test stands should study the improvements these dopants contribute in the context of low-energy electron reconstruction, explorations of the best doping combinations, studies of residual light, and how a photosensitive doped LArTPC would behave in the context of GeV-scale neutrino interactions.

### 5.2.3 Single-Phase Proportional Amplification

Physics measurements with signatures at the MeV-scale and below are one of the major drivers of detector developments in the large-scale noble element LArTPC experimental program. From improved imaging capabilities at the MeV-scale, to the possibility of detecting 10-100 keV energy nuclear recoils (NRs), the benefits of improved detection capabilities can enhance the physics potential for neutrino-oscillation and astrophysical measurements of neutrinos. The LArCADe project aims to investigate the feasibility of reducing detection thresholds for ionization electrons in single-phase LArTPC detectors by enough to enable the detection of nuclear recoils. The program aims to allow for the amplification of drifting electron signals directly in the liquid phase by modifying the geometry of the charge-collecting anode sensors. Nuclear recoils in liquid argon lead to small ionization signals, further reduced by the significant quenching caused by ion recombination and dissipation of energy into atomic excitations. Nuclear recoils of $\mathcal{O}(10)$s of keV, originating from $\mathcal{O}(10$ MeV$)$ CEvNS interactions, are expected to yield 1-100 free electrons, with significant variation in the tails of such distributions depending on the assumed quenching model. These values are up to a factor of 100 smaller than current state of the art detection thresholds in single-phase LArTPCs. In order to amplify ionization charge directly in the liquid phase, strong fields of $> 10^5$ V/cm are necessary. The LArCADe program is exploring the possibility of obtaining stable charge amplification of drifting electrons by shaping the electric field over micron-scale distances in the proximity of the charge collecting anode-planes. The first phase of this R&D effort is employing tungsten tips of micron radii, and has demonstrated preliminary controlled amplification in gaseous argon using a few-cm drift chamber which records ionization charge produced by a pulsed LED source impinging on a photocathode. A second phase, currently underway, aims to use $\mathcal{O}(10$-$100$ nm$)$ tips to obtain amplification in liquid, characterizing stability and potential complications which may arise, such as the formation of argon gas bubbles which can disrupt signal detection. A successful demonstration of this program can lead in the future to the construction of small-scale detectors sensitive to CEvNS interactions in the proximity of intense neutrino beams

## 5.3 Improving Photon Detection

The outstanding successes of the Borexino and KamLAND experiments demonstrate the large potential of liquid-scintillator detectors in low-energy neutrino physics. low-energy threshold, good energy resolution and efficient background discrimination are inherent to the liquid-scintillator technique. A target mass in the tens of kilotonnes would offer a substantial increase in detection sensitivity. At low energies, the variety of detection channels available in liquid scintillator will allow for an energy- and flavor-resolved analysis of the neutrino burst emitted by a galactic supernovae and sensitivity to faint signals of the diffuse supernova neutrino background. Solar metallicity, time-variation in the solar neutrino flux and deviations from MSW-LMA survival probabilities can be investigated based on unprecedented statistics. Low background conditions allow to search for dark matter by observing rare annihilation neutrinos. The traditional existing (and also more recently proposed) scintillator based detectors for UG low-energy neutrino physics implement $4\pi$ photo-sensitive active coverage. Along this line, by positioning large area photon detectors over multiple sides of the active LArTPC volume, the VD PDS Reference design aims to reach uniform LY throughout the volume and high on average, so as to be able to perform calorimetry and space reconstruction (and therefore also trigger with max efficiency) for neutrino events down to a very low threshold. Under this perspective, the FD2-VD Light system could perform those measurements on its own, completely independent and redundant to the charge TPC. At one hand, this represents a notable risk mitigation for physics, guaranteeing highest live time (PDS active also when LArTPC may be off for purity drop/maintenance, HV issues/maintenance, etc.) very relevant for long duration UG operation. On the other hand, given the complementarity of charge and light collected signals, the overall reconstruction capabilities of the FD2-VD detector can be improved when combining the information from the TPC with the PDS. Enhancement in energy resolution is expected (as demonstrated by previous analysis from LArIAT experiment), as well as in position resolution particularly helpful for rejection of radiological background near detector boundaries by volume fiducialization. PDS distinctive features, time resolution and pulse-shape PID, and TPC specific features, like event directionality reconstruction, are expected to provide unprecedented means of overall physics reconstruction when combined information from high performance PDS and TPC is utilized.

### 5.3.1 Xenon Doping

One particularly exciting possibility for improving photon detection is doping the liquid argon with a few parts-per-million of xenon [138]. With a sufficient concentration of xenon, most of the 'late' triplet light released by the dissociation of argon dimers instead comes from xenon dimers, which brings two key advantages: longer wavelength and faster dissociation time. Since Rayleigh scattering depends on $\lambda^{-4}$, a longer wavelength leads to substantially longer scattering lengths: the 128 nm Ar light has a scattering length of approximately a 1 m while the 176 nm Xe light has a scattering length closer to 4 m. This longer scattering length substantially improves the uniformity of the response of the detector, especially in the context of the vertical drift design described above which has PDS coverage on multiple sides of a larger active volume. This improved uniformity will improve both the information which can be extracted from the event (for example, better calorimetry resolution) as well as triggering efficiency by reducing 'dark spots' where events are missed. The faster dissociation time of the Xe dimers also improves triggering on low-energy events by getting more of the available energy from the event into a narrower time window, improving efficiency. Since the 'early' singlet Ar light is mostly not converted, we still can take advantage of the 6 ns decay time from this component for determining event times with high precision. Potentially the mix of different wavelengths could be taken advantage of to do a variant of 'pulse shape discrimination' for excluding background sources like $\alpha$ particles by looking at the relative amount of the two wavelengths, but this would depend on having PDS modules with sensitivity to different wavelengths, an option not yet studied in detail.

## 5.4 Calibrations

### 5.4.1 Calibrating with Low-Energy Signals

It has been proposed [66] to use the reconstructed energy spectrum of $^{39}$Ar beta decays to perform a variety of in-situ and ex-situ measurements of detector effects relevant for particle reconstruction in large LArTPCs, like DUNE. By using the fact that the $^{39}$Ar beta decays are uniformly distributed in the drift direction, one is able to precisely determine the expected reconstructed energy spectrum for a given set of detector response parameters. This can be done independently of using timing information (e.g., from prompt scintillation light). The primary detector response parameters of interest are the electron lifetime and electron-ion recombination factor, and since these two effects impact the shape of the reconstructed energy spectrum in different ways (e.g. recombination shifts the end point, while electron lifetime does not), one is able to use the reconstructed $^{39}$Ar beta decay energy spectrum to constrain these two quantities simultaneously.

The viability of this method has already been demonstrated with MicroBooNE data [66], where $^{39}$Ar beta decays have been observed and their energies reconstructed. Figure 16 illustrates the different possible reconstructed $^{39}$Ar beta decay electron energy spectra one might see after correcting for all other detector effects except for electron lifetime, for $^{39}$Ar beta decays occurring in the single-phase DUNE FD. Also shown in Fig, 16 is the impact of varying the true recombination model from the one assumed in energy reconstruction of the $^{39}$Ar beta decay electron, with infinite electron lifetime. The impact on the reconstructed energy spectrum is very different for the two detector effects, allowing for simultaneous determination of both quantities.

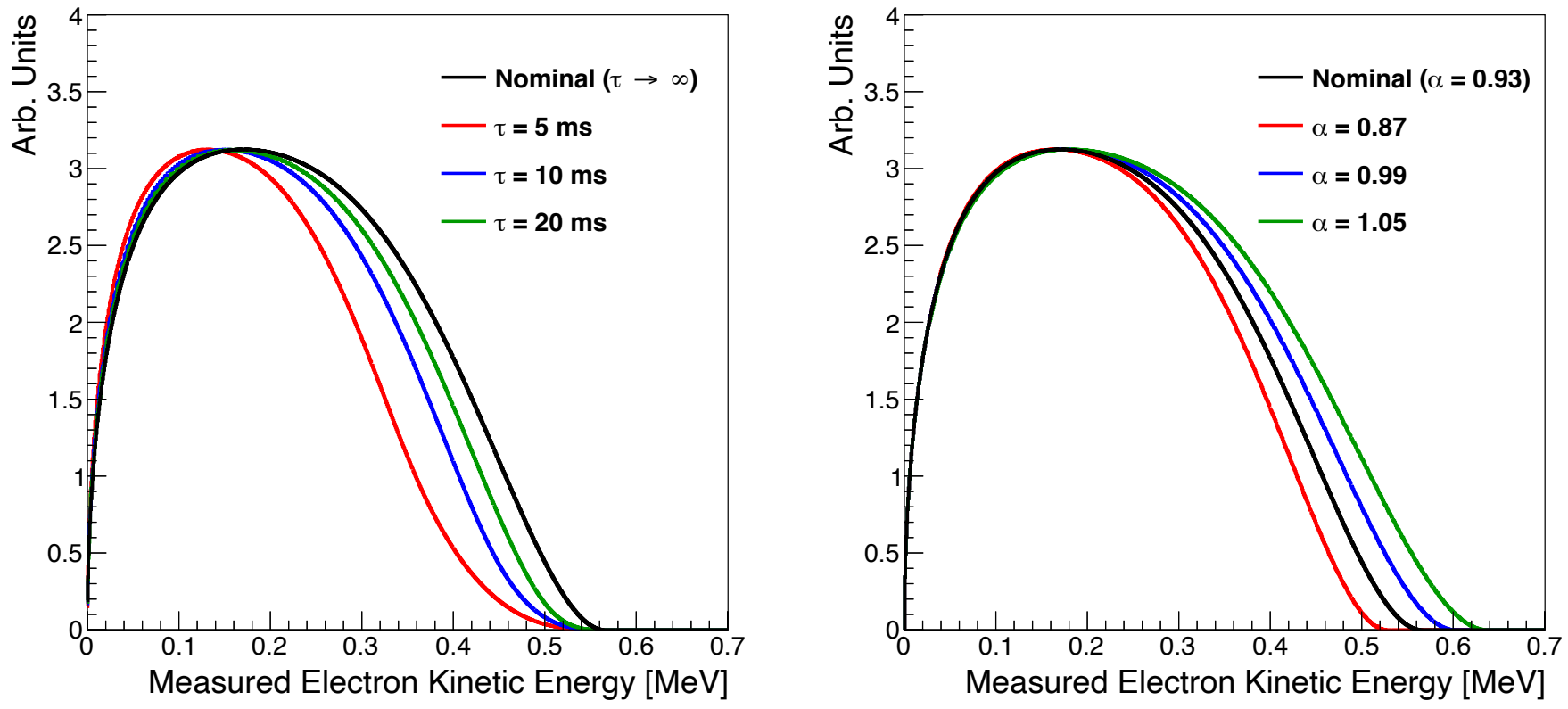

Figure 16: Illustration of the impact of different detector effects on the reconstructed $^{39}$Ar beta decay electron energy spectrum for decays observed in the single-phase DUNE FD. On the left are examples of the reconstructed energy spectrum for various different electron lifetimes, as well as the nominal $^{39}$Ar beta decay spectrum (corresponding to an infinite electron lifetime). On the right are examples of the reconstructed energy spectrum when the true recombination model is different from the one assumed in energy reconstruction (varying the $\alpha$ parameter of the modified Box model [139], $\mathcal{R} = \ln(\alpha + \xi)/\xi$, where $\xi = \beta \frac{dE}{dx} / \rho E_{\mathrm{drift}}$ and with fixed $\beta = 0.212$) and the electron lifetime is infinite. All curves have been normalized to have the same maximal value.

This method is one foreseeable way to obtain a fine-grained (spatially and temporally) electron lifetime measurement in the DUNE FD. It can also provide other necessary calibrations, such as measurements of wire-to-wire response variations and diffusion measurements using the signal shapes associated with the beta decays, and could serve as an online monitor of electric field distortions in the detector by looking at the relative number of decays in the detector near the edges of the LArTPC. These applications are currently being studied using ProtoDUNE data, well in advance of first operations with the DUNE FD.

One important consideration is whether or not the DUNE FD data acquisition system can provide the necessary rate and type of data in order to successfully carry out this calibration at the desired frequency and level of spatial precision. Knowing that the $^{39}$Ar beta decay rate is about 1 Bq/kg in natural (atmospheric) argon, one finds that $\mathcal{O}(50\text{k})$ $^{39}$Ar beta decays are expected in a single 5 ms event readout in an entire 10-kt module. From studies at MicroBooNE, an estimate of the number of decays necessary to carry out a percent-level calibration of electron lifetime is $O(250\text{k})$. This means that in order to make a single measurement of electron lifetime in an entire DUNE FD module, one would only need roughly five readout events. However, one must also allow for the electron lifetime to spatially vary throughout the entire 10-kt module; as a consequence, it may be necessary to collect much more data in order to obtain a precise electron lifetime measurement throughout the detector. Studies of data rates and alternative methods for recording special $^{39}$Ar calibration data are currently in progress. For example, one possibility is to make use of the continuously-streaming "trigger primitive" data for $^{39}$Ar calibrations, assuming trigger primitives are generated with sufficiently low thresholds, and provide sufficient energy resolution (see Section 7.3).

It should be mentioned that $^{39}$Ar beta decays closer to the cathode will be more likely to be below threshold (and thus undetected) in comparison to ones closer to the anode. While this is folded into the electron lifetime measurement and so would not bias the result, it does impact the interpretation of the result; this is because the extracted electron lifetime would be more representative of regions of the detector closer to the anode. Extrapolating this to regions closer to the cathode requires making the assumption that the electron lifetime is constant as a function of the drift coordinate, which may not be the case, though it is more likely to be uniform in the drift coordinate (total drift length of 3.6 m in $x$) than in the other two directions, along which the detector has greater extent (12 m in $y$, 58 m in $z$). In the case that there is variation in $x$, one could make use of an auxiliary measurement using $t_0$-tagged cosmic muon tracks to determine the dependence of electron lifetime on $x$. However, this would require integrating over a much larger period of time in order to obtain the appropriate level of statistics; a pulsed neutron source may be able to provide this auxiliary measurement more rapidly.

### 5.4.2 Calibration Systems for Low-Energy Physics

low-energy calibration systems need to be considered to complement the calibration of low and high level parameters that can not be attained with the desired precision from natural sources alone.

Ionization lasers can provide very high statistics for calibration of low level detector parameters and a higher level calibration of position and direction reconstruction for long tracks, but not for electromagnetic clusters.

For calibration of the low-energy electromagnetic signals typical of solar neutrinos or of neutrinos from a core-collpase supernova we can rely instead on sources of real low-energy particles. Two such sources, inspired by these two channels, are being planned for DUNE, and having at least one of them has been considered essential for the low-energy calibration. These sources should test the trigger models, and set the energy scale and resolution close to the threshold. They should also allow for a more frequent test of low level parameters, contributing to monitoring their evolution in time and/or their uniformity over a substantial volume.

Solar neutrinos can be emulated by single gammas of a few MeV inserted in the detector medium. Supernovae can, instead, be emulated by an intense pulse of gammas spread around the full volume. While the first focuses on having a low rate to avoid pile-up and directly test the efficiency to trigger on a single particle with an energy close to threshold, the second focuses on a high rate of low-energy gammas populating the full volume, and checks the detector response uniformity.

**Radioactive Source Deployment System** A first proposal considered in [12] adapts existing designs for deploying radioactive sources inside the DUNE volume. A well characterized and calibrated source would be deployed from the top next to the TPC, producing gammas of fixed energy at a known position. The baseline design uses $^{252}$Cf neutrons impinging on a nickel target to produce 9 MeV gammas.

This energy is chosen close to the trigger threshold, so that it provides a direct test of the trigger efficiency and the validation of the trigger model at a value in which its gradient as a function of energy is very large. The main requirement is a precise knowledge of the full source activity, which needs to be calibrated before installation.

The yield of escaping neutron and secondary gammas needs also to be precisely characterized before installation. The large Delrin moderator needed for $^{252}$Cf case, leads to a relatively large source of 30 cm and 10 kg. The same deployment system can then be adapted to use other similar size sources, in order to provide other energy values (in principle, it is possible to use a single one or several different ones).

For calibration of the energy scale and resolution it is particularly important to control the pile-up of source events, and so the source activity should not be high (of the order of 200 Hz). Pile-up is further reduced in the active volume, by the geometrical acceptance from a outside position.

The positioning of a single deployment system needs to be chosen to optimize the possibility of calibrating both the charge based and light based detection, for which the distance to the APA is the main factor. The detected energy will naturally be different for events directed towards or away from the APA, which is adequately modeled by the simulation of electromagnetic cascades.

Because the calibration is restricted to only a particular detector region, it is also more sensitive to the basic detector model parameters, which it can monitor along the experiment lifetime. The local electric field can be unambiguously determined from the drift time distributions, and the electron lifetime can be determined from the measured charge distributions. Preliminary studies show that sensitivities better than a few % and 1 ms, respectively, can be achieved with a source located 40 cm from the field cage and 220 cm away from the APA.

**Pulsed Neutron Source** A second proposal detailed in [12] is a new development for DUNE that explores a particular feature of argon to use a pulsed neutron source to cover a large detector volume with electromagnetic signals: low-energy neutrons can travel tens of meters in argon, before being captured; each of the captures in $^{40}$Ar leading to a 6.1 MeV gamma cascade.

Neutrons of 2.5 MeV are obtained from a deuterium-deuterium (DD) generator, moderated in the full source and by the liquid argon itself. The generator pulses can be adjusted to 0.1 ms, while the moderation and capture time are expected at around 0.25 ms. The position of each capture signal can be obtained by comparing the time interval between the generator pulse and the charge detection, with a resolution of the order of 30 cm; or much more precisely, by comparing the time of the charge and light signal. The cascade itself can have a size of 10 cm or 100 cm.

The intensity of the full source should be high to allow neutrons to reach the full TPC volume. DD generators can achieve $10^5$ neutrons per pulse, but the intensity can also be tuned down to avoid localized pile-up closer to the source. The calibration of each full DUNE module can be done in a few hours, while not upsetting normal operations in the other modules. An external trigger can be connected to the neutron generator, to function independently of the single particle or burst trigger; with the trigger models being tested by offline analyses.

The first mode of the calibration uses the full cascade of lower energy gammas as a standard candle of 6.1 MeV, and is dependent on the capability to isolate the cascade from radioactive backgrounds and recognize it as a single physics object. On top of reconstructing a 6.1 MeV cascade as a whole, it can be also possible to use individual gamma lines for detailed studies at even lower energies.

**PNS Data at ProtoDUNE** At the end of the ProtoDUNE-I runs in July 2020, 10 days were devoted to testing the PNS concept, with a DD generator with a minimum pulse rate of 200 Hz (width=0.175 ms), and varying intensities, trigger configurations and electric field configurations of the TPC. The generator was not fully shielded, it had 2.5 MeV neutrons entering the LAr together with some 2.2 MeV gammas from the moderator and the large cosmic-ray background.

The PNS extends the beam test of ProtoDUNE, providing a large sample of neutrons that can be used not only for the design of the calibration but also for other measurements, relevant for the development of neutron identification tools, low-energy electromagnetic clustering, and neutron cross-section measurements. Encouraging results were shortly obtained showing —- despite the large cosmic ray background— a clear excess of signals close to the neutron entrance but also at some distance from the source. These data will be used as a guide for optimization of a new run at ProtoDUNE-II, but also for other low-energy response studies.

## 5.5 Backgrounds

### 5.5.1 Low-energy Background Optimization

By sheer size alone, it is inevitable for ton to multi-kton scale liquid argon detectors like DUNE to have many radioactive materials inside the detector, as well as surrounding the detector such as the rock in the underground cavern walls. Internal radioactivity arises not only from dust deposition and radon daughter plate-out during the construction and installation phase of the detector, but of course from the components that the detector is made of because they contain traces of radioactivity, most notably the detection medium itself. $^{39}$Ar with a half-life of 269 years is a naturally occurring radioactive isotope of argon that one will not reasonably be able to avoid. $^{39}$Ar undergoes a $\beta^-$-decay with an endpoint energy of 565 keV and its energy spectrum is significantly refined by a shape correction due to the forbiddenness of the decay. $^{39}$Ar is accepted to be the dominant background for detectors utilizing natural argon, being fixed by the natural abundance of atmospheric $^{39}$Ar to a specific activity of $(1.01 \pm 0.10)$ Bq/kg in LAr [65]. Inevitably, the radiological requirements for DUNE, or any other ton to multi-kton scale natural liquid argon detector, are to the first order set to guarantee that all other backgrounds remain subdominant to the intrinsic $^{39}$Ar background throughout the active volume of the detector.

Development of a detailed radiological simulation, as DUNE has done within LArSoft, is paramount. The simulation must be sufficiently informed by extensive radiological assay data from all detector and cryostat components, as well as from the surrounding rock, shotcrete and concrete materials. Only with assay-data-driven informed simulations one can ensure that the low-energy physics goals are being met. To the first order the sum of all internal and external background rates has to be subdominant to the intrinsic $^{39}$Ar signal over the detector's active volume.

A particularly critical background is the relatively abundant external radiological neutrons from the surrounding rock, shotcrete and concrete materials. These external neutrons can penetrate the cryostat, enter the active argon volume of the detector where they can deposit a visible energy of 6.1 MeV, and 9 MeV, respectively, via neutron capture on $^{40}$Ar, and $^{36}$Ar, respectively, that create signals difficult for the DAQ to distinguish from electron-neutrino induced CC interactions in argon. It is therefore, in addition to radiological assay data, just as important to perform chemical composition analysis of the rock, shotcrete, concrete and all cryostat materials, in order to sufficiently inform the simulation for correct propagation of external radiological neutrons throughout all materials. Moreover, knowledge of the chemical composition is essential for estimating the ($\alpha$, n) production yields. For certain elements ($\alpha$, n) reactions can produce neutrons with energies of almost 10 MeV, thus enabling such neutrons to deeply penetrate the structures surrounding the active volume of the detector.

Next-generation neutrino experiments like DUNE and generation-3 dark matter experiments will not only have to be located deep underground to shield cosmic induced backgrounds, but the sheer size of these next generation detectors can bring forth unprecedentedly large excavation costs. Therefore, it will be challenging to have an abundantly large passive and/or active shield around these large-sized detectors and cost effective solutions will have to be found. Crucial for such assessments is the accurate prediction of residual backgrounds that could enter the fiducial volumes of these detectors. Radiological neutrons from the surrounding rock and shot/concrete are hereby most critical, but also neutrons produced in the detector materials themselves, such as steel structures, insulating foam layers, internal cables, electronics components, etc. or the target material. It is relatively straightforward to assess neutron production yields from spontaneous fission of e.g., radiological $^{238}$U concentrations in the rock or detector materials. But to date, it is still difficult to ascertain from $^{238}$U, $^{226}$Ra and $^{232}$Th concentrations the precise ($\alpha$, n) production yields and neutron energy spectra that are induced by $\alpha$-ray energies of up to about 9 MeV. These $\alpha$ particles arise from $\alpha$-decays in the early and late $^{238}$U decay chain, and the $^{232}$Th decay chain, respectively. The uncertainties in the ($\alpha$, n) production yields stem mostly from a lack of measurements and/or uncertainties in the existing measurements of ($\alpha$, n) cross sections on many relevant target isotopes. More precise measurements of ($\alpha$, n) cross sections in the $\alpha$-ray energy range of up to 10 MeV on certain critical target isotopes would greatly mitigate the uncertainty on radiological neutron backgrounds for next generation dark matter and neutrino experiments like DUNE, and thus could in turn greatly help saving costs [140].

Intrinsic $^{42}$Ar in natural argon has a half-life of 33 years and is another important addition to the radiological model although its specific activity is four orders of magnitude lower than that of intrinsic $^{39}$Ar. $^{42}$Ar in natural argon has a specific activity of roughly 50 $\mu$Bq/kg [141, 142]. It is still notable due to the high total visible energy of the subsequent $^{42}$K $\beta^{-}$-decay with an endpoint energy of 3.5 MeV and a correlated de-excitation $\gamma$-cascade. $^{42}$K has a half-life of 12.4 hours and about half of the produced $^{42}$K atoms are positively ionized after the $^{42}$Ar $\beta^{-}$-decay with an endpoint energy of 599 keV. Thus, about half of the produced $^{42}$K atoms will migrate to the cathode where they decay. The relatively high total visible energy of the $^{42}$K $\beta^{-}$-decay starts to push up into the boundary where it could be a critical background for solar neutrinos and low-energy supernova triggers. As such, its accurate inclusion in the radiological simulation is very important.

Intrinsic $^{85}$Kr with a half-life of 10.8 years undergoes a $\beta^{-}$-decay with an endpoint energy of 687 keV, very similar to that of $^{39}$Ar. For all intents and purposes, it is a very difficult to recover contaminant for detectors employing ton to multi-kton masses of liquid argon extracted from atmosphere. The measured specific activity of $^{85}$Kr in natural argon is $(0.115 \pm 0.093)$ Bq/kg [65]. However, it appears that depending on the commercial vendor of liquid argon, it can be three times larger, thus potentially amounting to about a third of the background rate of intrinsic $^{39}$Ar.

The type of signal from backgrounds is not the only concern one needs to address within a radiological simulation. The location of a background can also play a large part in how critically that background impacts the detector performance for low-energy physics. This is why one needs to also simulate e.g. $^{210}$Po on the photon detectors (PD) themselves. $^{210}$Po is a part of the $^{222}$Rn decay chain. Specifically, it is a daughter of $^{210}$Pb which is known to stem from radon daughter plate-out on materials. It can be a notable background for DUNE as the FD will be assembled, installed and filled underground at Sanford Lab over the period of at least two years per module. Underground there is a significantly elevated level of several hundred Bq/$m^3$ of $^{222}$Rn in the mine air. The radon daughter $^{210}$Pb has a half life of 22.3 years, hence a $^{210}$Pb contamination inside the active volume of the detector would result in a contamination lasting the entire lifetime of the experiment. A subsequent $^{210}$Po $\alpha$ decay on the surface of a PD can produce a 5.3 MeV $\alpha$ particle depositing its energy in the liquid argon right in front of the PD. In liquid argon light production from $\alpha$ particles is barely quenched. This can result in a large flash right into the PD, which can in turn create a strong signal on a single PD. This is potentially an issue for any optical triggering scheme intending to run on hits. The issue could be easily mitigated with smart triggers, but simple triggers are preferred, as they cost less and can be implemented earlier in the read-out chain, thus reducing the overall rate of the detector. This makes $^{210}$Po on the light collectors a necessary addition to the radiological simulation for DUNE and and generation-3 dark matter experiments. Radon reduction systems that reduce the amount of radon in the air during assembly, installation and filling are desirable.

$^{222}$Rn itself is part of the late $^{238}$U decay chain and it can emanate into the liquid argon from detector materials, most noticeably filter materials in the re-circulation and purification chain. As a noble gas, $^{222}$Rn can easily diffuse through most materials resulting in some ingress to the active volume of the detector. $\alpha$ particles from $^{222}$Rn and its decay chain can then produce critical neutrons directly in the liquid argon volume via ($\alpha$, n) reactions. Moreover, $\alpha$ particles from $^{222}$Rn and its decay chain can produce critical $\gamma$-rays directly in the liquid argon volume with an energy of about 15 MeV via ($\alpha$, $\gamma$) reactions, although with orders of magnitude smaller cross section than for the ($\alpha$, n) reactions in argon. It is therefore paramount to control the radon emanation, especially from filter materials, by conducting both extensive $\gamma$-ray assays that screen for $^{222}$Rn's progenitor $^{226}$Ra in filter materials and by performing additionally cold emanation measurements of these filter materials to assess how much of the $^{222}$Rn produced in the bulk material can actually diffuse out into the liquid argon. Once in the liquid argon, radon and its daughters will migrate throughout the detector, which needs to be modeled with the inclusion of all half-lives, polarities, drift-velocities as well as time and spatial correlations of all daughters in the $^{222}$Rn decay chain. The measured radon emanation rates and the migration model of its daughters need to be implemented in the radiological simulation of next-generation neutrino experiments like DUNE and generation-3 dark matter experiments, in order to get a sufficiently complete background assessment for the low-energy physics performance of ton to multi-kton scale LArTPCs.

### 5.5.2 Achieving a Low Background DUNE Module

With controls over radiopurity and some modifications to a detector similar to the DUNE FD Vertical Drift (VD) design we find that it is possible to increase sensitivity to low-energy physics in a third or fourth FD module. In particular, sensitivity to supernova and solar neutrinos can be enhanced with improved MeV-scale reach, and a WIMP dark matter search becomes possible.

Low-background steel and shielding can lower neutron backgrounds to necessary levels. An underground source, from discussions with industry, would seem to be able to produce the needed volume of $^{39,42}$Ar-suppressed argon at reasonable cost. Radon controls can be developed to limit the $^{222}$Rn background. Altogether, along with dense photon detection instrumentation in an inner fiducial volume, such a detector allows to get to significantly lower energy thresholds than the baseline DUNE FD modules.

Improved photon detection also allows to avail of argon's characteristic pulse shape discrimination for nuclear recoil detection at yet lower thresholds for gamma rejection. Beyond this, in combination with charge readout from the VD-baseline CRPs, few percent energy resolutions at 2-3 MeV and higher can be achieved [129] to pursue even further advantages for interesting physics.

With the above improvements in hand, we show in [143] the following results, among others. A competitive, neutrinoless double beta decay search with $^{136}$Xe loading, as discussed in Section 3.3.3 appears feasible. The coherent elastic neutrino nucleus scattering of supernova neutrinos per Section 3.3.1 is observable above background for core collapses at 10 kpc. Solar oscillation parameter searches of the type discussed in Section 3.3.2 can be significantly tightened over the baseline DUNE program due to a fuller sensitivity to the $^{8}$B spectrum. A definitive claim on the high or low metallicity solution in the CNO solar neutrino process mentioned in section 5.3 appears viable in this module. Furthermore, with an O(75-100) keV$_r$ threshold, which we show to be achievable with significant photodetector coverage, sensitivity to Weakly-Interacting Massive Particle (WIMP) Dark Matter becomes competitive with the planned world program in such a detector [144].

We perform studies –= mostly in a standalone Geant4 simulation –= to show the physics reach of a low-background moudle for the topics mentioned here. See companion Snowmass white paper [143] for further details.

# 6 LArTPC Reconstruction at Low Energies

## 6.1 Low Level Charge Signal Reconstruction

LArTPC detector offers millimeter-scale spatial resolution and excellent calorimetric capabilities in the detection of particles traversing in the liquid argon and the measurement of their properties. The capability for detecting low-energy activity is affected by the ∼23.6 eV mean energy to ionize an electron in liquid argon, the high ionization electron collection efficiency, and the low level of noise achievable in modern electronics readouts. With its low noise electronic readouts and good signal-to-noise ratio (SNR), LArTPC makes the low-energy reconstruction feasible and provides great opportunities for rich low-energy physics. Beyond the fundamental capability, the threshold for detecting low-energy activity mainly depends on the raw waveform processing algorithms used in the event reconstruction. These include noise filtering, signal processing, and the detection of localization of signals in the waveforms.

Low level LArTPC reconstruction starts with data preparation including evaluation of pedestals, charge calibration, noise filtering, and signal processing, etc. low noise electronics readouts are critical to properly extracting the ionization electrons. Noise filtering is a key step towards high quality signal processing. In LArTPCs, common noise contains the coherent noise, tails, and "sticky code" noise, etc. Coherent noise found across neighboring wire channels on each wire plane at the same time tick mainly due to power supply line noise, digital noise from the same electronics board or nearby boards, or some external interference. The tails, which usually occur on the collection plane, originate from the capacitive coupling discharge in the ADCs. Standard tools to remove different noise have been developed and are available [2, 145, 120]. Figure 17 shows example event displays from a collection plane of ProtoDUNE-SP before and after noise removal. To understand the noise features better as well as improve noise simulation, we also investigate the noise frequency distribution and develop a realistic data-driven noise model, which accounts for both the mean value of each noise frequency component and the fluctuation around that mean [146].

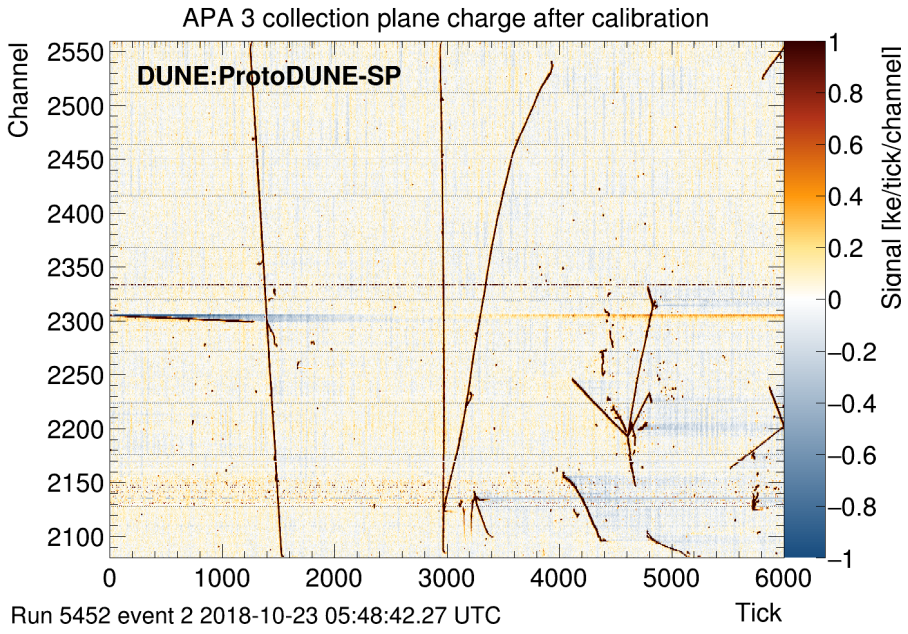

(a) After pedestal subtraction and calibration.

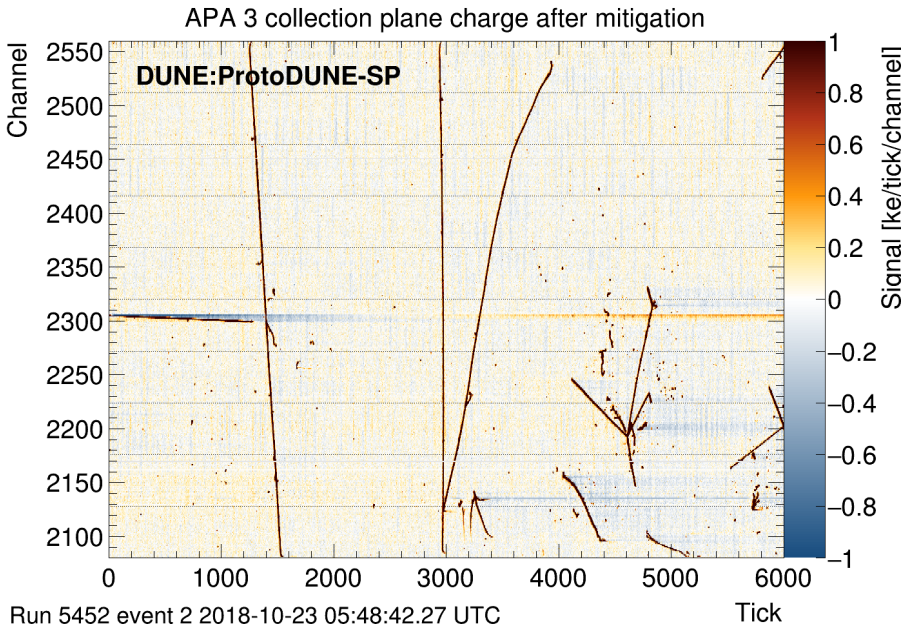

(b) After ADC sticky code and timing mitigation.

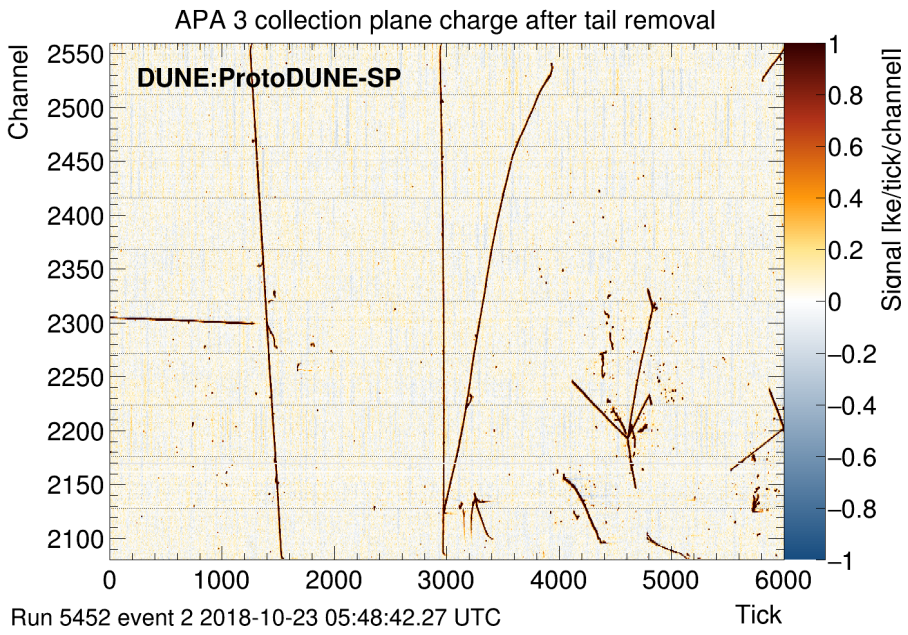

(c) After tail removal.

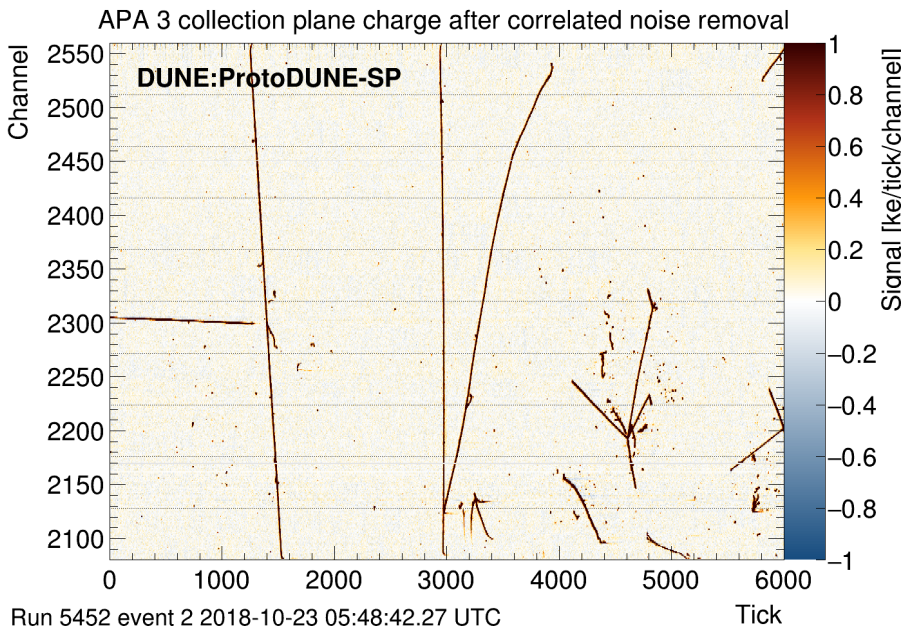

(d) After correlated noise removal.

Figure 17: Example event displays for a collection plane showing background reduction in successive stages of data processing [120].

The LArTPC raw digitized signal waveform is a convolution of arriving electron distribution, field response, and electronics response. Deconvolution will remove the field and electronics responses and convert digitized waveform to ionization electron distribution. The output of deconvolution is a standard signal shape, such as a Gaussian shape, which can be used for the next-stage hit reconstruction. Details of signal processing can be found at Ref. [122, 147, 120].

Understanding and optimizing the signal and noise discrimination capabilities of LArTPCs is crucial in performing charge/energy reconstruction, which is especially critical for low-energy physics. The threshold for extracting small signals is largely determined by the SNR. For example, DUNE is developing its electronics to achieve a wide range of physics goals. The SNR measured at ICEBERG used for DUNE cold electronics test can help DUNE benchmark and quantify the relative improvement in different electronics designs [148]. Figure 18 shows the average values of angle-corrected SNR before and after the noise filtering of the three planes using cosmic-ray muons from MicroBooNE [145], ProtoDUNE-SP [120], and ICEBERG's test of COTS-ADC FEMBs [149]. The detection of localization of signals within wire waveforms, region of interest (ROI), usually is considered prior to any high-level event reconstruction, which preserves the potential for maximizing signal detection efficiency in the initial stages of reconstruction and is essential for achieving the overall high efficiency required in low-energy physics studies. Traditionally, the waveform ROI finder is based on an over-threshold algorithm that selects signal candidates with pulse heights above a predefined threshold. This method has the disadvantage of discarding signals below certain energies. Recently, a novel deep-learning approach based on the application of a simple one-dimensional convolutional neural network (1D-CNN) has been developed, which shows a very encouraging ability to extract small signals and offers great potential for low-energy physics [150, 146]. Figure 19 shows the comparisons of the ROI efficiencies of the two ROI finders for the induction plane and the collection plane from ArgoNeuT.

## 6.2 High-Level Blip Reconstruction

A method to reconstruct low-energy (sub-MeV-scale) activity produced by de-excitation photons and inelastic neutron scatters in neutrino interactions is discussed here, using data from ArgoNeuT. In particular, this study focuses on reconstructing Compton-scattered electrons from the previously mentioned photons. A more complete description of this study, its methods and its application on LArTPC data can be found in [17].

Reconstruction is performed in two steps. First, events are sent into an automated reconstruction based on LArSoft which scans wire signals and finds hits. After this is done, a second reconstruction procedure determines which hits may be due to photons and reconstructs those hits. To identify candidate hits, a series of cuts are applied to hits in an event. A minimum peak height cut is applied first to remove hits due to electronics noise. To remove protons, a maximum peak height cut is applied. After applying a fiducial cut, a track cut is applied which removes hits identified by LArSoft as belonging to a track.

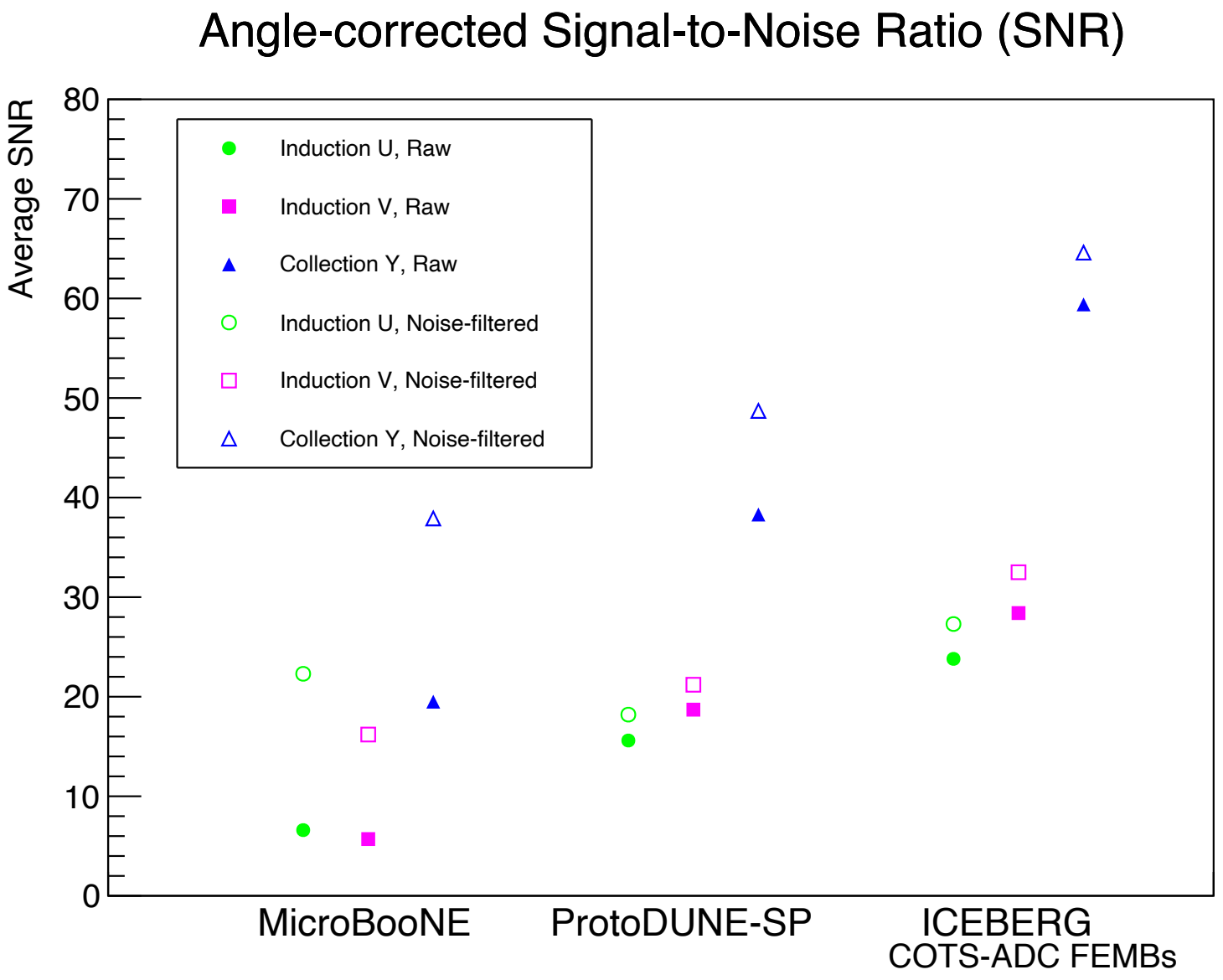

Figure 18: Average values of angle-corrected signal-to-noise ratio before and after the noise filtering of the three planes using cosmic-ray muons for the characterization from MicroBooNE [145], ProtoDUNE-SP [120], and ICEBERG's test of COTS-ADC FEMBs [149].

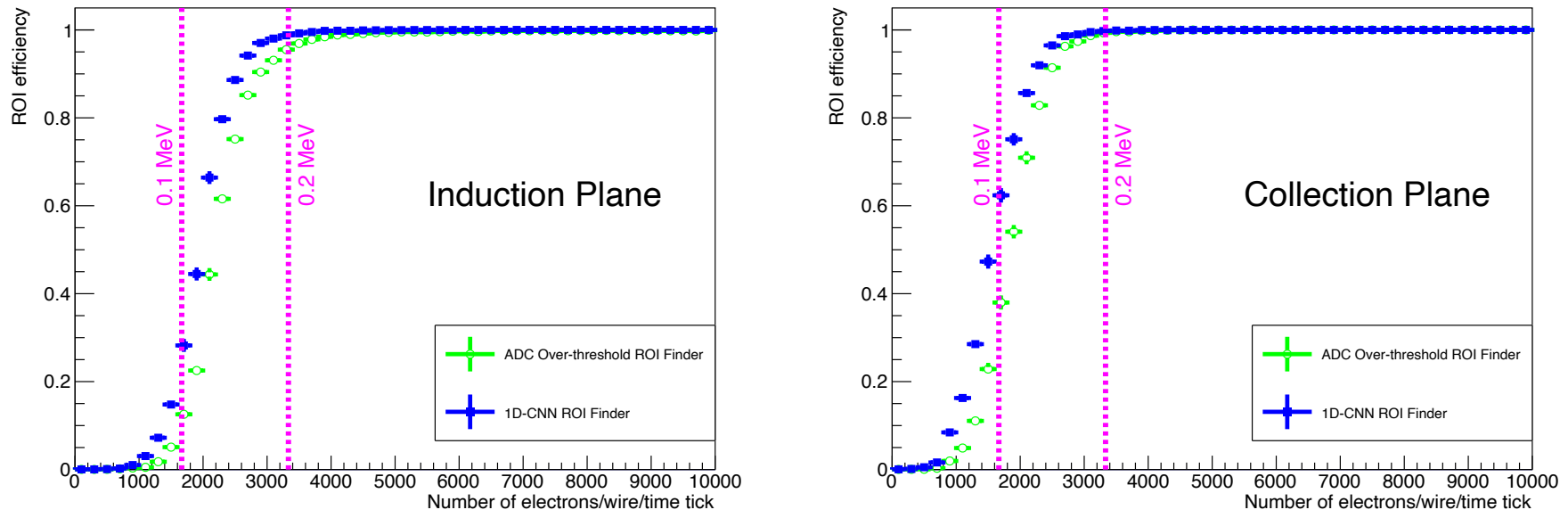

(a) Induction Plane.

(b) Collection Plane.

Figure 19: Comparison of the ROI efficiencies of the two ROI finders for the induction plane and the collection plane from ArgoNeuT [146].

Finally, hits associated with track activity are removed with a cone cut which removes hits within a specified distance from a track.

Once all cuts are applied, the remaining hits are grouped into clusters and reconstructed. A cluster is defined as a collection of one or more hits on adjacent wires within a specified time period. Calorimetric reconstruction is then performed on the clusters. To do this, the sum of the integral of the charge for all collection plane hits in a cluster is considered. A calibration constant and lifetime correction is applied to the total charge. To account for recombination, $dE/dx$ must be known. However, electrons with energies less than 1 MeV will travel less than the wire spacing, making a $dx$ calculation difficult. One can use the NIST table of electron track lengths at various energies (ESTAR) [151] to determine $dx$ and the amount of energy deposited. For each row in the NIST table, we divide the energy by the range, giving $dE/dx$. This is inserted into the inverted Modified Box Model to give $dQ_{\text{coll}}/dx$, where $Q_{\text{coll}}$ is the charge collected. By multiplying by $dx$, we obtain the collected charge. We can then plot energy or range versus collected charge and fit. We use the fit to convert between collected charge and deposited energy. Figure 20 shows the application of this method to a sample of simulated electrons in ArgoNeuT ranging from 0 to 5 MeV. In ArgoNeuT data, this method yielded an energy resolution of 24% at 0.5 MeV and an energy resolution of 14% at 0.8 MeV. More comparisons to ArgoNeuT data can be found in [17].

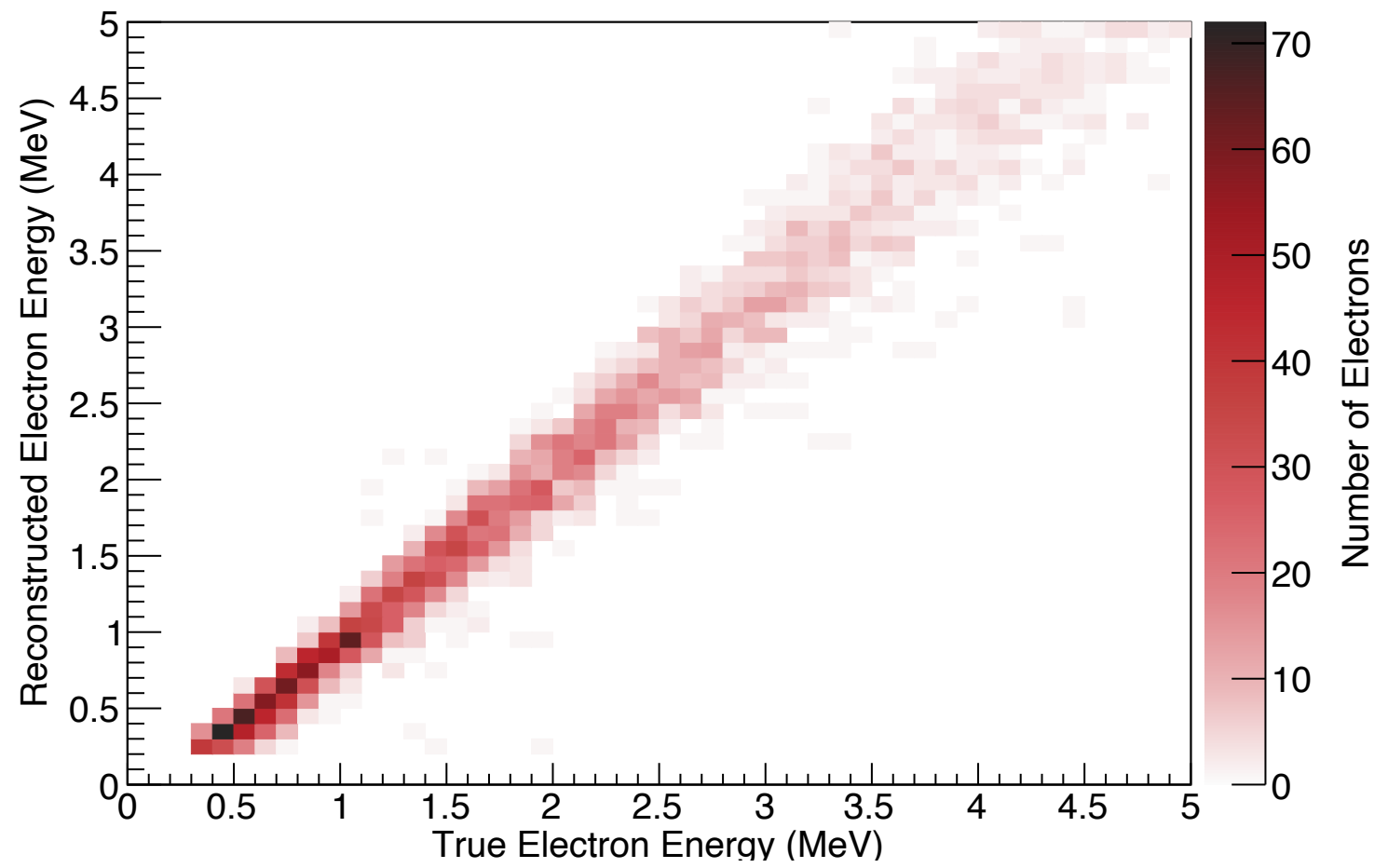

Figure 20: Reconstructed energy vs true electron energy using the charge method for a sample of simulated electrons with energies between 0 and 5 MeV. Events where the electron was not detectable are excluded. Figure from [17].

While ArgoNeuT has successfully demonstrated the ability to reconstruct MeV-scale activity, more work needs to be done. Currently, LArTPC experiments must develop their own, experiment-specific MeV-scale reconstruction software; no standardized software package exists. Further, detector calibra-tion should be performed at these energies, e.g., with a radioactive source. In addition, detector effects, e.g., diffusion, will play a role and have not been fully studied at these energies. Finally, improvements in electronics and signal processing algorithms will lower thresholds even further and improve energy resolution.

## 6.3 Scintillation Light Signal Reconstruction

Scintillation light typically plays an ancillary role in LArTPCs by providing a prompt signal for real-time trigger decisions – though it has the potential to augment physics analyses as well. The scintillation time profile can be indicative of the mean ionization density, allowing for some level of particle discrimination. Sufficiently high light yield (LY) can aid in calorimetry, particularly in low-energy EM showers where charge-based methods alone are prone to electron recombination uncertainties.

A demonstration of optical reconstruction at low energies was carried out by the LArIAT Collaboration. LArIAT's novel use of reflective foils coated in a film of wavelength-shifting TPB, arranged to cover the four field cage walls, allowed it to achieve a relatively large and uniform LY throughout its volume, averaging approximately 18 PE/MeV [152]. Scintillation light in LArIAT was used to collect and analyze a sample of Michel electrons from stopping cosmic muons [129]. Michel electrons are a useful proxy for the low-energy EM showers expected in supernovae neutrino interaction final-states, which DUNE will be designed to capture.

The ability to identify the rising edge of very small pulses becomes more challenging at lower energies. In the LArIAT analysis mentioned above, the gradient of signals from two PMTs was thresholded in order to maintain sensitivity to small pulses, even those that ride on top of low frequency baseline fluctuations or overlap with the late (triplet-component) scintillation from the decaying muon. Signals identified as belonging to the Michel electron were integrated to get the total light arriving at the PMT, subtracting off the estimated contribution of the late-arriving light from the muon pulse that overlapped with the electron pulse. For smaller optical signals, counting individual photoelectrons may produce better results than this kind of direct pulse integration.

Charge and light were combined in LArIAT to measure the total energy deposited by the electron within the active volume by assuming charge-light complementarity, $Q + L = E/W_{\mathrm{ph}}$, where $Q$ and $L$ are the total ionization electrons and scintillation photons, respectively, and $W_{\mathrm{ph}} = 19.5$ eV is the average energy needed to produce an electron or photon. A more involved method using an event-by-event maximum log-likelihood fit to the measured $Q$ and $L$ was also used to measure the deposited energy. This likelihood method was found to be essential for combining these two quantities in such a way that ensures the optimal energy resolution is achieved, though it requires knowledge of the detector resolution (via Monte Carlo simulations) on $Q$ and $L$ independently. Comparisons between data and Monte Carlo showed that the combination of charge and light together improved the energy resolution, though only by a small amount due to LArIAT's relatively high signal-to-noise ratio (SNR) of ≈50. By varying wire SNR and LY in the simulation, the enhancement to energy resolution for low-energy electrons was estimated for larger LArTPCs employing reflective TPB foils in a similar way. For a wire readout SNR $\simeq 30$, for example, the energy resolution for electrons below 40 MeV is improved by ≈ 10%, ≈ 20%, and ≈ 40% over traditional charge-only calorimetry for average LYs of 10 PE/MeV, 20 PE/MeV, and 100 PE/MeV, respectively.

| Experiment | Average light yield (PE/MeV) | Uniform light collection? |
|---|---|---|
| MicroBooNE | $\sim 5$ | no |
| LArIAT | $\sim 18$ | yes |
| pDUNE-SP | 1.9 at 3.3m | no |
| SBND | $\sim 80$ ($>$ 50 min) | yes |
| DUNE: Vertical Drift | $\sim 38$ ($>$ 16.5 min) | yes |

Table 1: Light yields for several existing and planned LArTPC experiments.

Calculating $L$ from the total integrated PEs requires interfacing with the charge-collection system to determine the location of the source of scintillation. Knowledge of the optical visibility throughout the active volume of the detector is then needed to determine the corrective factor accounting for the limited number of photons that actually make it to a photodetector. In LArIAT, voxelated 3D maps of the photon visibility throughout the volume (one map corresponding to each of the two PMTs) were convolved with reconstructed 3D space-points within the electron shower to obtain this corrective factor. These same maps were also used at the simulation stage. However, for larger volumes and more numerous photodetectors, such a method is impractical due to high memory requirements during data processing. Instead, a more robust method of calculating optical photon visibilty in LArTPCs based only on the relative position between the photodetector and point of scintillation was developed for the Short-Baseline Near Detector (SBND) [153]. Some version of this "semi-analytic" method will likely be employed in DUNE. In either case, a more uniform light yield makes this geometric correction procedure easier and less prone to uncertainty.

The role light may play in reconstructing smaller, isolated, blip-like energy depositions ($<$ 5 MeV) in LArTPCs is unexplored, though SBND will allow us to study this capability in detail. With SBND's large number of photodetectors, combined with its use of passive TPB-coated reflector foils enmeshed in the cathode plane, it is expected to achieve a relatively uniform LY of about 80 PE/MeV — an unprecedented level of light collection for a LArTPC. Plans for the DUNE Vertical Drift module suggest a similarly large LY of 38 PE/MeV, with reflectors and xenon doping used to improve uniformity of response. For comparison, Table 1 lists the expected (or measured) LY in several LArTPCs.

# 7 Data Acquisition and Processing Considerations

A maximally-sensitive search for stochastic low-energy processes in a large LArTPC requires a continually-running and fully-active detector, and in particular a deadtime-less data acquisition system that is capable of selecting potential events of interest with the highest possible (nearly 100%) efficiency. Depending on overall data rates, significant data reduction may also be necessary. In the case of the DUNE FD, the overall data rates associated with readout of low-energy physics processes suggests for the need for several orders-of-magnitude of data reduction, while maintaining high sensitivity to signals of interest, including neutrino interactions from the sun or from a galactic supernova burst. Because of the nature of low-energy signals of interest—being nearly indistinguishable from ambient radioactivity signals or random electronics noise—an intelligent data selection (trigger) scheme is needed in order to sift through the data and preferentially select signals that will maximize sensitivity to low-energy physics processes. This section will begin by providing an overview of the design of triggering, data acquisition (DAQ), and computing infrastructure for DUNE, which is the primary example of a large underground LArTPC experiment striving to overcome these data reduction challenges. It will then describe triggering and data selection methods, applicable to any large LArTPC, that can be employed in trigger and DAQ systems to enable low-energy physics; again, the application of these techniques will be provided as an illustrating example.

## 7.1 LArTPC Trigger and DAQ System Description

LArTPC Trigger and DAQ systems, referred to as TDAQ in DUNE, are responsible for receiving, processing, and recording data from the experiment's LArTPC. Like other particle physics experiments, the design is driven by data rate limits, throughput considerations, and stringent up-time requirements. For DUNE, the TDAQ system:

- provides timing and synchronization to the detector electronics and calibration devices;
- receives and buffers data streaming from the TPC and the photon detection system (PDS);
- extracts information from the data at a local level to subsequently form trigger decisions;
- builds trigger records, defined as a collection from selected detector space-time volumes corresponding to a trigger decision;
- carries out additional data reduction and compression as needed; and
- relays trigger records to offline permanent storage and computing.

The DUNE TDAQ has been subdivided into a set of subsystems with well-defined roles and interfaces, which are currently being developed in parallel. A conceptual overview of a TDAQ system, described in further detail in Ref. [148, 12], is shown in Figure 21. The Trigger and Data Filter are in charge of the selection and compression of data. The Dataflow subsystem provides the communication layer to exchange data and implements the data collection functionality, i.e., the logic for building trigger records as well as the organization of data into files. The Readout receives the data streams from the TPC and PDS, processes them to extract information for the trigger and buffers data while the trigger is forming a decision. The Timing subsystem is in charge of distributing the clock, synchronizing the detector modules, as well as time-stamping hardware signals that may be used for triggering, such as calibration pulses. Finally, all subsystems rely on the functionality provided by the Control Configuration and Monitoring (CCM) and Data Quality Monitoring (DQM), which forms the glue of the overall TDAQ, transforming the set of components into a coherent system.

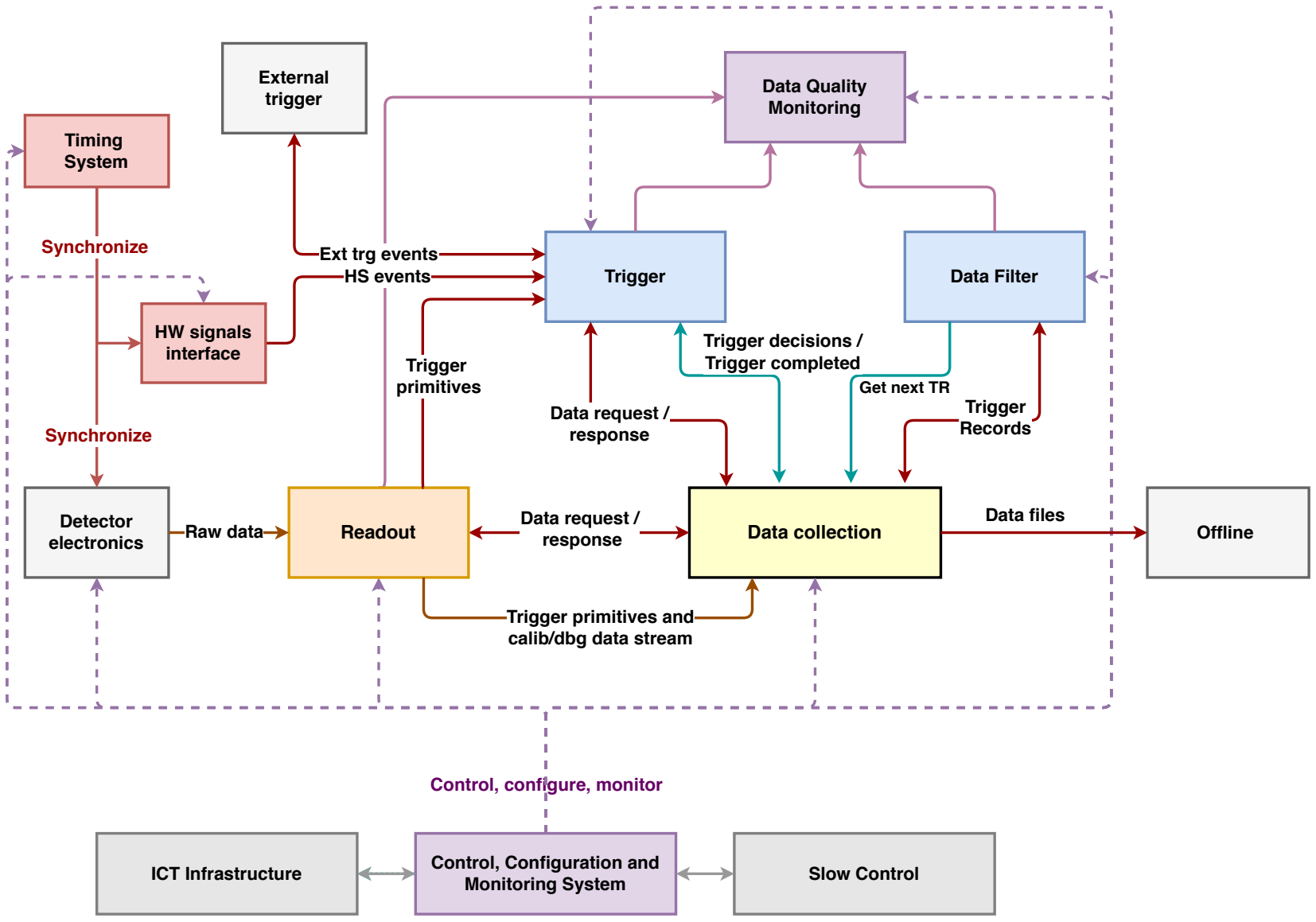

Figure 21: Conceptual overview of the TDAQ system functionality for a single FD module. External systems are depicted in grey. Adapted from Ref. [12].

The TDAQ for the DUNE FD is mainly composed of commercial-off-the-shelf (COTS) components. TDAQ systems for the different DUNE FD modules differ only in minor details so as to support the electronics and the data selection criteria for each. A high performance Ethernet network interconnects all the elements and allows them to operate as a single, distributed system. At the output of the TDAQ the high bandwidth Wide Area Network (WAN) allows the transfer of data from the SURF to FNAL.

To meet DUNE neutrino and BSM physics goals, its TDAQ system must acquire data from both charge and light collection systems. Ionization charge measurement by the TPC for any given activity in the FD requires a nominal recording of data over a time window determined by the drift speed of the ionization electrons in LAr and the detector dimension along the drift direction (up to 6.5 m). Given a target drift electric field of 450 V/cm, the time window of TPC data to be recorded for any given activity of interest is on the order of 4.5 ms; the activity associated with beam, cosmic rays, and atmospheric neutrinos is localized in space and particularly in time, to roughly this order. On the other hand, multiple neutrino interactions, making up the activity of a nearby supernova burst (SNB), extend over the entirety of the detector and last between 10 and 100 s. For either case, the detector data must flow from the electronics output links, on top of the cryostat, through the TDAQ system, undergoing data selection, and to permanent storage at Fermilab. Two stages of data selection, applying data compression algorithms and selection of data only for interesting detector regions and time windows, will allow the needed reduction of the overall data volumes from $\sim$ 50 EB/year produced by each FD module to $\sim$ 30 PB/year, for all FD modules. These inherent computing limitations are defined further in Section 7.2, while implemented data selection methods will be discussed in Section 7.3. While the former discussion is more narrowly focused on the DUNE experimental implementation, the latter is fairly relevant to the design of any large LArTPC experiment's triggering/DAQ systems.

## 7.2 DUNE Computing Considerations

The DUNE Offline Computing Consortium is responsible for the storage, processing, and dissemination of data and simulation for both far and near detectors. To understand the resources needed to accomplish these goals, a DUNE computing model was developed that predicts storage, processing, and network usage beginning with ProtoDUNE operations and extending through full DUNE experiment operations [154]. The computing model puts particular emphasis on storage and file distribution due to the large size of individual trigger records. In particular, prompt and coherent processing of SNB trigger records with sizes on the scale of 150 TB per FD module (for 100 s worth of unbiased TPC data corresponding to a SNB) puts unique challenges on the provisioning of adequate processing resources.

To predict the storage and CPU needs for DUNE, estimates were made upon the experience from operating ProtoDUNE [5, 120]. During the ProtoDUNE I beam run, the 7x7x7$m^3$ Single Phase detector produced uncompressed data that was 178 MB (compressed 71 MB) for each trigger record for 6 APAs and a readout time of 3 ms. This data volume can be scaled to match a predicted far detector module with 25 times as many APAs and a readout window of 5.4 ms. The ProtoDUNE compression factor of 2.5 can be used as a conservative estimate even though occupancy of the ProtoDUNE SP, located on the surface, is considerably different than FD modules operating at 4850 feet. Estimates of near detector (ND) data volumes, expected to be much less than 1 PB per year, are far lower than that of the FD [6].

DUNE plans keep two archival copies of the raw data and one archival copy of derived and simulated data located at Fermilab and other storage sites around the world. Derived datasets will be produced from data twice per year, and once per year for simulation with two copies of each reconstruction pass kept active on disk for two years. From these parameters and the operations schedule for ProtoDUNE and DUNE, predictions for tape and disk storage were calculated. As an example, disk storage predictions are shown in Figure 22; disk storage begins to be dominated by DUNE FD data at the end of the current decade following the start of FD data-taking. The full processing time for a single ProtoDUNE-SP or ProtoDUNE-DP trigger record was approximately 600 seconds, and is used as an estimate for CPU resources needs specified in HEPSpec-06 hours and is shown in Figure 22; estimates for CPU production processing do not exceed 10,000 cores DC until after 2030.

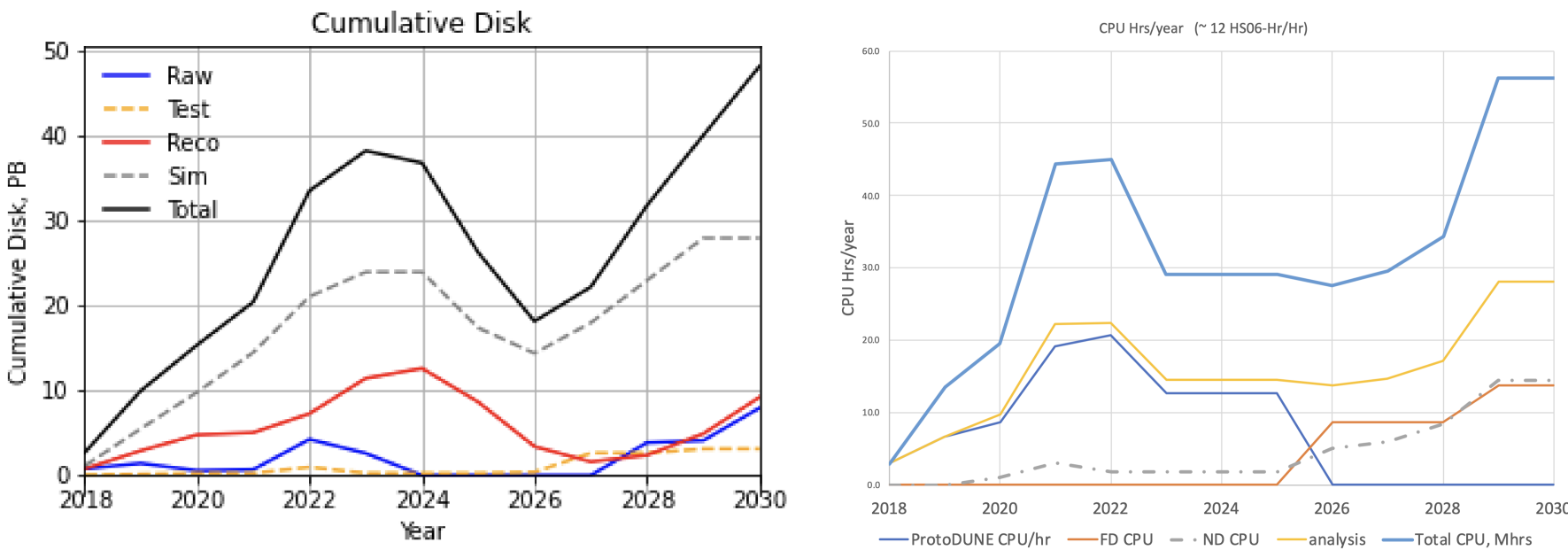

Figure 22: Left: Estimated disk storage for ProtoDUNE and DUNE operations until the year 2030. Right: Estimated CPU needs for ProtoDUNE and DUNE operations given in HEPSpec-06 hours until the year 2030. From Ref. [155]

The estimated DUNE resource needs are modest when compared with data volumes and processing resources for High Luminosity LHC operations, and, along with the success of ProtoDUNE I processing, this gives confidence that there will not be significant issues accomplishing the computing and physics goals of DUNE that align with those scales of challenges. As only 5,000-10,000 beam and atmospheric neutrino interaction events are expected in the full 40 kT FD per year, higher-energy calibration samples, including cosmic muons, are likely to contribute much more substantially to overall data rates. With limited re-calibration but full reprocessing of accumulated beam data occurring with a six-month cadence, the current computing model should be successful.

There are significant additional computing challenges introduced when considering accommodation of low-energy physics goals. Stochastic low-energy events, such as those generated by solar neutrino interactions and related backgrounds of similar energy, are expected to occur in DUNE at very high rates. Attempts to efficiently trigger full-module readout on these signals may thus overwhelm the 30 PB per year data storage limits of the experiment. This limitation emphasizes the importance of incorporating data selection and reduction capabilities into low-energy triggering schemes, as will be discussed in the following section.

With the estimated data volume of a SNB trigger record being on the order of 150 TB, this trigger type offers multiple distinct challenges for offline data processing. The first challenge is network transfer of the trigger record to a storage element from which it can be redistributed for processing. If a 100 Gb/s network connection is available from the output of the DAQ to a storage element, it will take approximately 4 hours to transfer the complete trigger record to offline storage, and then a similar time is needed to transfer outbound from the storage element to processing nodes. Using the estimates for ProtoDUNE trigger record reconstruction timing, to have the SNB trigger record fully reconstructed within 4 hours would require approximately 130,000 cores running in parallel. In comparison with beam trigger processing, this level of resources is more than an order of magnitude larger, and unavailable without considerable workload management development or access to preemption of high performance computing resources. With the estimated trigger rate for the SNB stream being once per month, the challenge of provisioning this scale of resources will take considerable coordination between DUNE Offline Computing, and some combination of OSG, WLCG, and HPC sites. Even then, precision pointing information on the timeline necessary for optical follow-up observations may be dictated by network bandwidth and storage transfer rates. These limitations emphasize the importance of augmenting DUNE's SNB triggerring capabilities with online reconstruction and/or further data reduction. This is discussed in the following section.

A final important consideration for low-energy physics at DUNE is the reprocessing of low threshold data streams targeting energy depositions near the trigger threshold. While all of the raw data will be maintained on archival storage for the extent of the experiment, the presence of data on tape does not guarantee availability of all of the data. The current DUNE event data model involves reducing the size of processing trigger records through slimming, skimming, and discarding raw waveforms from derived datasets. If low-energy physics analyses require fidelity at the level of raw waveforms, the smaller derived datasets may not suffice. If reprocessing of raw data is desired, the amount of tape drive resources dedicated to this processing becomes considerable. The current estimate of 30 PB of raw data written to tape each year means that after 10 years of data taking there will be 300 PB of raw data on tape. In order to reprocess all of this data within a 6 month campaign would require approximately 100 tapes running at full capacity. As of 2021, this is twice the tape drive capacity of the entire Fermilab Intensity Frontier computing operations.

## 7.3 Trigger and Data Selection Strategies

This section briefly summarizes the strategies that have been devised by the DUNE TDAQ team to maximize the retention of interesting low-energy LArTPC datasets while respecting the data volume that can be permanently stored for the DUNE FD (∼30 PB/y). While presented in the DUNE context here, the strategies described are generally applicable to triggering for any large LArTPC. Indeed, many of the selection methods described are under active development within the MicroBooNE collaboration using its continuous readout data stream [156].

| Event Type | Rate (1/s/10 kT) | Data Rate (PB/y) | | |
|---|---|---|---|---|
| | | Module | TPC | $m^3$ Box |
| Beam or Atmospheric $\nu$ | $2 \times 10^{-5}$ | $10^{-3}$ | - | - |
| Cosmic Muons | $4 \times 10^{-2}$ | 4 | - | - |
| Solar $\nu$ | $3 \times 10^{-4}$ | 0.03 | $2 \times 10^{-4}$ | $10^{-6}$ |
| $n$-Ar Captures | 1 | $10^2$ | 0.6 | $4 \times 10^{-3}$ |
| $^{42}$Ar Decays | 1000 | $10^5$ | 700 | 4 |

Table 2: Data storage required for permanent recording of various event classes in a DUNE far detector module assuming various data selection and reduction strategies. Estimates here are directly scaled from demonstrated full ProtoDUNE event sizes; relative scaling factors are detailed further in Section 7.2. Approximate event rates from Ref. [15].

The goal of data selection is to be as inclusive of physics signals of interest as possible; at high energies (> 100 MeV) it is important to accept all possible events with as high an efficiency and as wide a region-of-interest in channel and time space as possible, regardless of event type. As energies approach 10 MeV, where radiological backgrounds (including neutron captures, $^{42}$Ar, and $^{39}$Ar pileup) become dominant, the system should be semi-inclusive, leveraging the topological capabilities of the TPC data to provide some discrimination for low-energy physics signals.

These data selection considerations and impacts are illustrated in Table 2, which gives yearly data collection rates required to store DUNE FD data for some perfectly-identified event types using different data selection and reduction schemes. Of triggers generated by hundreds-of-MeV-scale energy deposits, trigger rates are entirely dominated by cosmic rays, with rates from atmospheric or beam neutrinos sub-dominant to this by multiple orders of magnitude. If all triggers from this event category result in full readout and storage of all FD data, storage rates will remain well below the 30 PB per year requirement.

This situation changes radically at energies at or below 10 MeV encompassing supernova burst and solar neutrinos, as well as beta decay electrons of interest for various physics studies. If they could be triggered on with perfect purity and efficiency, all solar neutrino interaction event displays could be recorded losslessly for the entire FD with modest (<100 TB) yearly storage requirements. In reality, solar neutrino interactions may be difficult to distinguish from a variety of higher-rate radiological backgrounds. While precise low-energy background rates are not yet precisely known, neutron captures on argon, which lead to a 6.2 MeV $\gamma$ cascade, are expected to occur at a rate of 1-10 Hz in a 10 kT module. At lower summed energy, radiogenic $^{42}$Ar ($Q$=0.60 MeV) and $^{42}$K ($Q$=3.53 MeV) present in the atmospherically-sourced LAr are expected to decay at rates of roughly 1 kBq in a 10 kT module. Readout of a full FD module on these higher-rate event types, as shown in Table 2, is obviously infeasible, but also arguably unnecessary. This issue can be circumvented via implementing data selection/filtering schemes or by enhancing triggering criteria to reduce fake trigger rates while maintaining high efficiency for true interactions of interest; studies directed towards these purposes will be discussed in the following section.

### 7.3.1 Triggering Concepts and Methods

Prior to any triggering or data selection, all digitized TPC data are sent to the Readout subsystem and processed. For each channel, a hit-finding algorithm allows identification of activity above the electronic noise with a threshold well below the $^{39}$Ar beta decay endpoint of roughly 0.5 MeV. Every hit forms the basis of a so-called "trigger primitive," or TP. Similarly, PDS electronics boards send waveform data for any channel that passed an internal threshold; TPs are also formed from these data. The trigger primitives serve two main, and one auxiliary purposes:

- They are the basic elements used to form a trigger decision in the TDAQ system.
- They are stored as unbiased (at the "event" level) summary information that can be used for trigger, calibration, low-energy physics studies, etc.
- They are buffered temporarily to be available in case of external alerts of supernova bursts, and/or for other semi-offline detector monitoring and calibration purposes.

To provide good sensitivity to different track topologies, each TP contains information such as the time-over-threshold of the waveform, its peak, and its total charge, as well as the timestamp of the start of the waveform. To date, TPC trigger performance has been established using collection channel information only; exploitation of induction channel information is under development but is not currently seen as needed to satisfy any of DUNE's triggering requirements.

For triggering purposes, the trigger primitives are the basic elements used by the TDAQ to form a trigger record and initiate the collection and storage of raw waveform data. The data selection system takes trigger primitives generated locally and looks for clusters in time and space. These clusters represent what is called "trigger activity," or a TA. Clusters of TAs are then passed to algorithms downstream which determine whether any particular set of TA clusters should be promoted to a trigger candidate. Trigger candidates then are sent to the trigger decision logic, which apply criteria that include both configuration parameters (e.g., which triggers are accepted in a given data run) as well as dynamic decisions (e.g., whether a TPC trigger candidate came shortly after an existing PDS trigger candidate, and is likely part of the same event). The data selection work flow is shown in Figure 23.

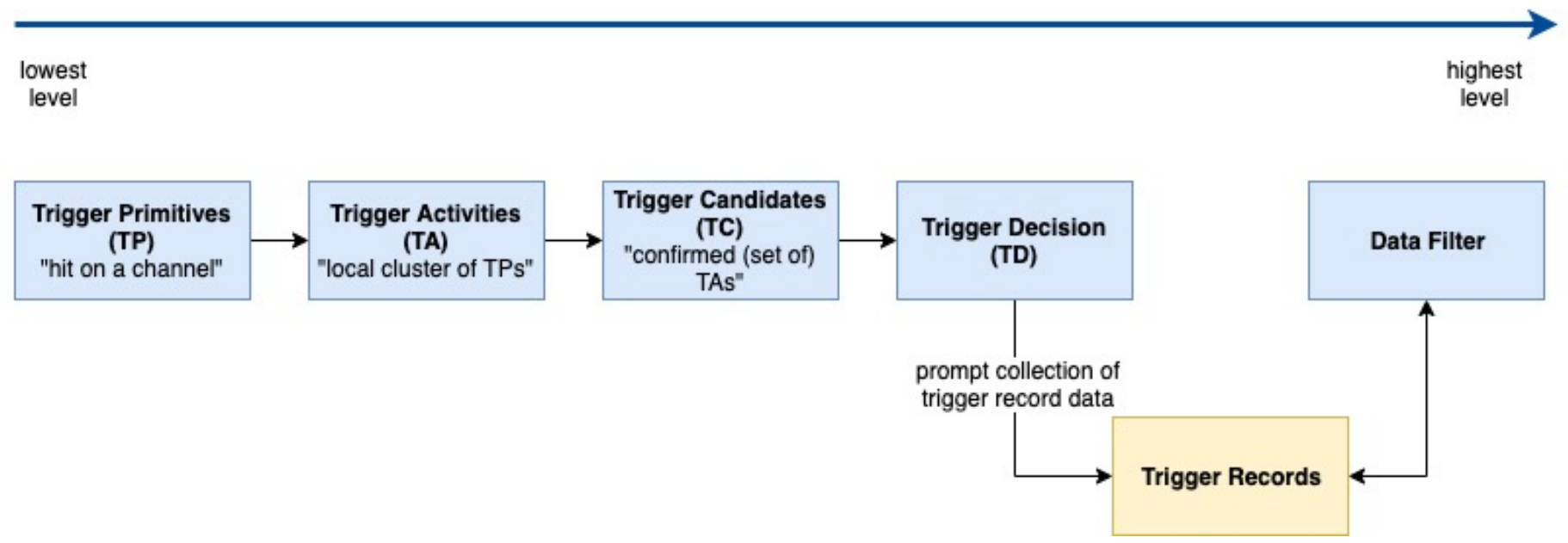

Figure 23: A block diagram representing the data selection work flow for a DUNE FD module. See the text for additional details. Adapted from Ref. [12].

There are two different raw data collection modes foreseen for the DUNE FD:

- A trigger decision based on TA clusters consistent with a single high-energy or low-energy interaction or internal decay, that includes a list of electronics channels to be collected and their associated time window(s). The TDAQ uses this information to collect the relevant raw data from its temporary buffers, form trigger records, and store them persistently.
- A trigger decision based on several TA clusters within a few seconds that are consistent with those of multiple low-energy interactions from a nearby SNB, and inconsistent with the expected fluctuations from background in rate and energy. In this case, a special trigger decision is fired, indicating a SNB candidate. For reference, from only one of the FD modules, the total collected raw data from a SNB, spanning 100 s, will be about ∼150 TB in size. Thus, while the effective burst threshold must be set low enough to satisfy DUNE's requirements on SNB detection efficiency, it is important for this SBN trigger to not fire too frequently on background fluctuations.

Each trigger prompts the collection of data from the Readout and Trigger sub-systems to form a trigger record. In the extreme case of an SNB trigger, data from the whole module is collected over a time window of 100 s. In other cases, data from subsets of the FD module TPC and PDS systems may be collected over much shorter times (<10 ms). Further data filtering on collected and temporarily buffered trigger records can be performed at computing facilities on surface at SURF, as part of the TDAQ's Data Filter. These facilities processes records with the aim of further reducing the data volume to be transferred to FNAL. The following sub-sections will discuss ongoing efforts to tailor trigger criteria for low-energy physics and optimize data selection and reduction within the TDAQ system.

### 7.3.2 Optimizing Triggering Decisions for Low-Energy Physics

With a simple algorithm to form TAs from a stream of TPs, the efficiency of a DUNE FD module for triggering on a core-collapse supernova via neutrino interactions can be evaluated. Specifying that the trigger algorithms form a TA with at least six neighbouring collection channels having TPs within a time tolerance of 10 $\mu$sec, the differential efficiency curve for any given low-energy ($< 100$ MeV) neutrino interaction can be obtained, as shown in Figure 24. The average TA efficiency for any individual supernova neutrino interaction is on the order of 20 to 30% [11]; these efficiencies would also apply to solar neutrino interactions in DUNE.

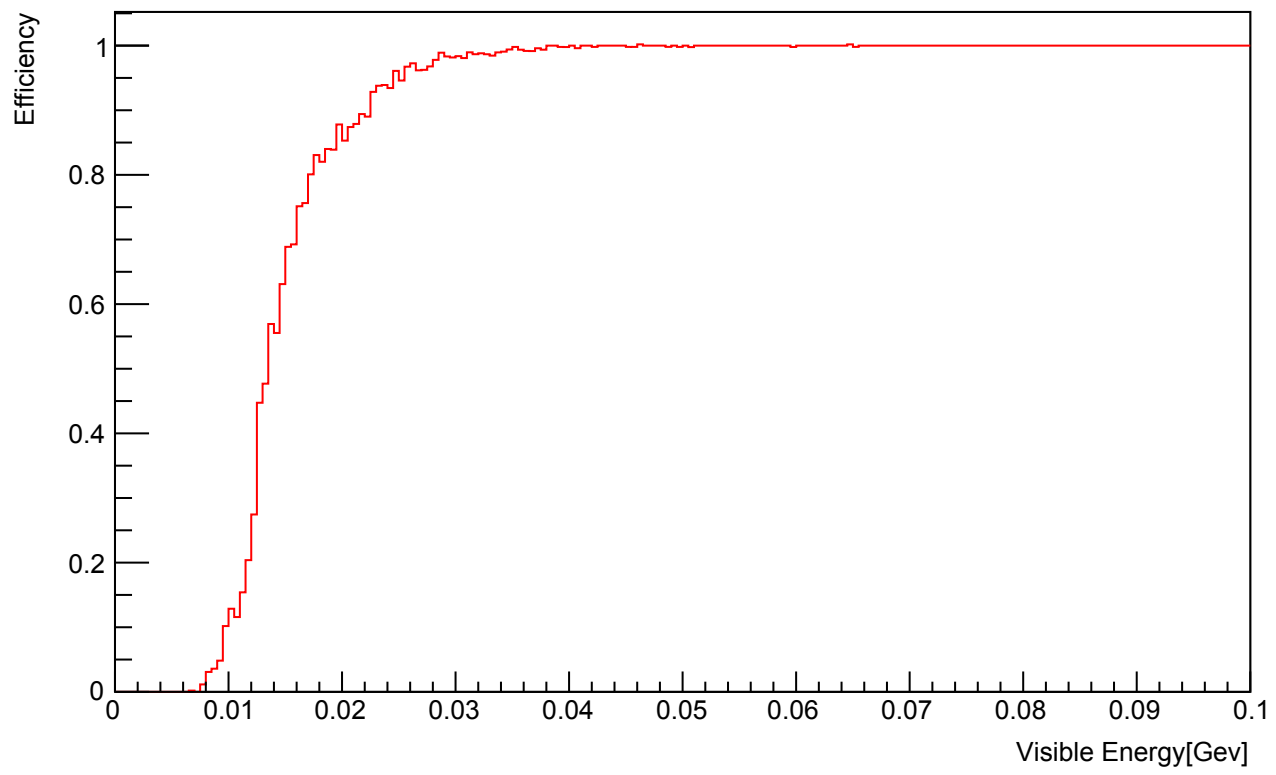

Figure 24: Efficiency for forming a trigger activity object from simulated trigger primitives in a single DUNE supernova neutrino interaction event, as a function of visible energy of the outgoing electron in the interaction [11]. Trigger activity is this study requires at least 6 closely-spaced trigger primitives.

Any low-energy TA, consistent with the ionization deposition due to a single supernova neutrino interaction, serves as input to the SNB trigger. Integrating and considering the multiplicity of such TAs over a 10 second window yields a high trigger efficiency to a supernova burst up to the galactic edge. The resulting fake trigger rate is also at the required level of about one per month when state-of-the-art background models are considered (electronics noise and radiological backgrounds). Decreasing the required number of channels to form a TA from 6 to 4 increases the individual supernova neutrino interaction efficiency to around 70%, and reaches $\sim$100% SNB burst efficiency for any SNB within the Milky Way. The efficiency can also potentially improve further by adding TPs from the induction plane(s). Additionally, the signal from the photosensors gives similar efficiencies [11]. Therefore, the combination of charge and light information can improve the trigger efficiency while reducing the fake trigger rate.

The efficiencies described above are from the most simple "counting" strategy. By considering the ADC information that each TP will contain (summed-waveform digitized-charge), an energy-based approach increases the SNB trigger efficiency and minimizes the fake triggers. By considering the summed ADCs in all TAs in a window of 10 seconds, the distribution of a background-only to a background-and-burst distribution allows for better differentiation. Given that most backgrounds contribute to lower energies than the SNB spectrum, the efficiency gain is significant: 70% (energy-weighted) vs. 6.5% (counting) SNB trigger efficiency is possible for a supernova at the Large Magellanic Cloud [11].

Finally, other online data selection methods, such as ones employing machine learning algorithms, are also being investigated for application in LArTPCs [12], especially at low energies where differences between signal and background become more subtle. In particular, convolutional neural networks (CNNs) have been studied as a way of improving the trigger efficiency of LArTPCs in real-time and online data processing within the TDAQ system [157, 150, 158]. For example, two-dimensional (2D) CNNs have been proposed for application at the TP and/or TA stage of data selection, for example to classify regions of collection wire vs. time data as to whether they contain low-energy (SNB or solar) neutrino interactions [159, 157, 158]. This approach has been shown to yield significantly higher efficiency on individual neutrino interactions, most notably at lower energies, compared to traditional approaches. The individual interaction efficiency is approximately a few % to 20% for neutrino interactions with 5-10 MeV in true neutrino energy, and 70% across all SNB neutrino energies. Once multiplicity considerations (over a 10 second period) are taken into account, a 100% galactic SNB trigger efficiency can be achieved, while meeting the 1/month fake trigger requirement for the future DUNE FD module. An energy-weighted SNB trigger scheme leads to further improvement; specifically, a SNB trigger efficiency of 69% for a supernova at the Large Mangellanic Cloud is possible, while still maintaining the 1/month fake trigger requirement [159]. The increased efficiency at low-energy on individual neutrino interactions (up to 20% at 5-10 MeV) is promising for solar neutrino triggering, especially if low fake trigger rates can be maintained.

### 7.3.3 Data Reduction in Low-Energy Physics Triggering

Even with simplistic trigger requirements unable to distinguish neutrino interactions from low-energy radiogenic backgrounds, manageable data rates should still be achievable by selecting and recording only portions of the module's full charge readout that contain interesting event activity. Data selection and reduction can be performed using differing methods at differing points in the triggering and data acquisition process.

During the generation of a trigger record, the TDAQ system is capable of selecting a subset of electronics channels to be read out and stored. One intuitive method of selection during triggering is to read out all information from the module sub-TPC hosting the TA(s) responsible for the trigger decision, or that TPC and its nearest neighbors. The single-TPC selection scenario is also illustrated in Table 2; the level of data reduction for this case, 150, is based on the total number of TPCs in a 10 kT module. As an example, in one year, such a selection would reduce the total stored data rate triggered by all $n$-Ar captures in a 10 kT module from an unmanageable 100 PB to less than 1 PB.

A greater level of data reduction could be provided by selection and readout only of electronics channels and waveform periods in the immediate vicinity of trigger-inducing TAs. This selection could be performed either online during generation of the trigger decision, or semi-online as part of the Data Filter stage. This case is also illustrated in Table 2 by considering a scheme in which charge collection electronics waveforms corresponding to a $m^3$ cubic volume surrounding the TA of interest are stored. As an example, this scheme would enable complete readout of a 10 kT module's full 1000 Bq activity of radiogenic $^{42}$K decays at a data storage cost of 4 PB per year. Meanwhile, $n$-Ar capture data rates in this scheme would total only roughly 4 TB per year. Such a scheme, if implemented with too tight a volume requirement, risks non-storage of portions of the physics signal in question. However, truth-level studies in LAr confirm that $\sim$$m^3$ volumes are sufficient to achieve excellent energy resolution for supernova and solar neutrino interactions, as well as for isolated gamma-related signals commonly produced by $n$-Ar capture and other $n$-Ar inelastic processes [19]. Ref. [19] also emphasizes that pileup $^{39}$Ar blips contained within these volumes are expected to have only minor impacts on these energy resolutions.

A final promising option for data reduction in collection of low-energy signals is to store only isolated above-threshold signals. Specifically, TPs generated from charge and photon collection systems in the DUNE FD will be stored on disk by the TDAQ system. This data set is very important for carrying out trigger studies but can also be used for calibration purposes, as well as fast data analysis. After compression and minimal clean-up, it is estimated that a few PB/year will be sufficient to store them. It is thus an option for DUNE to not only store the TPs temporarily on months-long timescale for specific studies, but to make them part of the data that will be stored permanently.

This data stream is particularly interesting in view of its role in potentially extending the low-energy physics reach of DUNE. It will also contain summary information for individual interactions with very low visible deposited energy. Depending on the achieved signal to noise ratio on the collection plane (see Section 6 and Figure 18), the trigger primitive threshold is expected to be at or below approximately 250 keV, more than low enough to record charge information for solar and supernova physics, as well as studies of low-energy beta decays for neutrino mass or double beta decay studies.

#### 7.3.4 Augmented Triggers for SNB TDAQ and Computing Needs

As described in Figure 23, in the baseline DUNE SNB trigger, if enough TAs are detected across the entire module, a SNB trigger decision is generated. This causes the previous 10 seconds worth of raw data in the Latency Buffer to be copied into the SNB store and the subsequent incoming raw data lasting up to 100 seconds to be streamed into the same destination. Due to the sheer size of this data ($\sim$150TB/module), it will take at least an hour for the Store Coordinator to copy this data from the underground caverns to the surface for additional processing by the Data Filter, and at least several more hours to eventually send this data to FNAL for archival and offline processing. Within the context of multi-messenger astronomy, one main objective of a SNB trigger is to disseminate useful information, such as pointing, to other experiments and cosmological surveys, with relatively low latency. Waiting hours to provide meaningful information to participants on an alert network is simply too long for a unique astronomical event that is rapidly evolving on much shorter time scales (∼minutes).

The Data Filter sub-system on-site at SURF provides an opportunity for additional data processing which can provide relatively prompt information for multi messenger astronomy. As such, several R&D efforts are being pursued to investigate the possibility of prompt and efficient (in terms of power, computation resources, and time) data processing for this purpose. One approach currently being explored, is to process the data buffered in the SNB Store "in situ" while it is being copied by the Store Coordinator. Since the underlying hardware for the SNB Store will be based on NVMe Solid State Drives (SSDs), it is possible to make use of processing elements such as FPGAs that can access the data stored on the SSDs directly via the PCIe bus, without host intervention. An example of an emerging technology that makes this possible and practical is that of Computational Storage. Implementation of algorithms on FPGAs with this capability would allow extraction of critical information like the source direction from the candidate SNB data, or to execute more sophisticated filters that can reject fake SNB's. The possibility of incorporating ML algorithms (such as CNN's) on FPGAs for this purpose is also currently being explored [158].

# 8 Summary

The key points of this document are summarized in Section 1.1. The future experimental needs that if satisfied will enable us to fully exploit low-energy physics in LArTPCs are summarized in Section 1.2. Among the needs are additional ancillary measurements, improved theoretical and simulation modeling, detector R&D along several directions, improved event reconstruction algorithm development, and development of specialized data acquisition and processing for low-energy events. Addressing these needs will yield rich physics rewards.